\documentclass[10pt,english]{scrartcl}

\usepackage[T1]{fontenc}
\usepackage[latin9]{inputenc}
\usepackage{amstext}
\usepackage{graphicx}
\usepackage[numbers]{natbib}

\makeatletter
\newcommand{\lyxaddress}[1]{
	\par {\raggedright #1
	\vspace{1.4em}
	\noindent\par}
}

\usepackage{etoolbox}

\patchcmd{\@maketitle}
  {\ifx\@empty\@dedicatory}
  {\smallskip
    \begin{center}
    \footnotesize
    \begin{tabular}{c}
      Department of Anatomy \\
      University of Otago Medical School \\
      Dunedin 9012, New Zealand
    \end{tabular}
  \end{center}
  \ifx\@empty\@dedicatory}
  {}{}

\@ifundefined{showcaptionsetup}{}{%
 \PassOptionsToPackage{caption=false}{subfig}}
\usepackage{subfig}
\makeatother

\usepackage{babel}
\begin{document}
\title{\noindent \textrm{\Large{}Infectious diseases, imposing density-dependent
mortality on MHC/HLA variation, can account for balancing selection
and MHC/HLA polymorphism }}
\author{\noindent {\large{}D. P. L. GREEN}}
\date{{\normalsize{}31 DECEMBER 2024}}
\maketitle
\begin{abstract}
The human MHC transplantation loci (HLA-A, -B, -C, -DPB1, -DQB1, -DRB1)
are the most polymorphic in the human genome. It is generally accepted
this polymorphism reflects a rôle in presenting pathogen-derived peptide
to the adaptive immune system. Proposed mechanisms for the polymorphism
such as negative frequency-dependent selection (NFDS) and heterozygote
advantage (HA) currently focus on HLA alleles, not haplotypes. Here,
we propose a model for the polymorphism in which infectious diseases
impose independent density-dependent regulation on HLA haplotypes.
More specifically, a complex pathogen environment drives extensive
host polymorphism through a guild of HLA haplotypes that are specialised
and show incomplete peptide recognition. Separation of haplotype guilds
is maintained by limiting similarity. The outcome is a wide and stable
range of haplotype densities at steady-state in which effective Fisher
fitnesses are zero. Densities, and therefore frequencies, emerge theoretically
as alternative measures of fitness. A catalogue of ranked frequencies
is therefore one of ranked fitnesses. The model is supported by data
from a range of sources including a Caucasian HLA data set compiled
by the US National Marrow Donor Program (NMDP). These provide evidence
of positive selection on the top 350-2000 5-locus HLA haplotypes taken
from an overall NMDP sample set of \ensuremath{\sim}$10^{5}$. High-fitness
haplotypes drive the selection of \ensuremath{\approx}137 high-frequency
HLA alleles spread across the 5 HLA loci under consideration. These
alleles demonstrate positive epistasis and pleiotropy in the formation
of haplotypes. Allelic pleiotropy creates a network of highly inter-related
HLA haplotypes that account for 97\% of the census sample. We suggest
this network has properties of a quasi-species and is itself under
selection. We also suggest this is the origin of balancing selection in
the HLA system.
\end{abstract}

\lyxaddress{\vfill{}
}

\section{\textrm{Introduction}}

\smallskip{}

\noindent The MHC/HLA \emph{super}-locus is a genomic region of about
150-200 loci on the short arm of chromosome 6. It covers approximately
3Mb and includes the six classical transplantation loci: class I (HLA-A,
-B, and -C) and class II (HLA-DP, -DQ, -DR) \citep{Murphy2022}.

HLA molecules are cell surface proteins that are germline encoded
and remain unchanged for the lifetime of the holder. They act with
other cellular components constantly to monitor endogenous and exogenous
proteins of the internal environment of the body. Peptides derived
from these proteins are mounted in peptide-HLA complexes that are
presented to T-cells. T-cells recognising endogenous peptides are
removed during fœtal development, when foreign proteins are absent.
This step limits the T-cell repertoire \emph{post-partum} to a subset
that recognizes only foreign peptides bound to HLA molecules, thereby
detecting pathogen invasion.

The structural feature of HLA molecules that underpins their function
is a pocket of approximately 60 amino-acids derived from exons 2 and
3 for class I alleles, and exon 2 for class II. These define a peptide-binding
site (PBS). HLA class I and II loci show extensive polymorphism in
these pockets. HLA class I and II loci also form numerous haplotypes
that display variation through allelic permutation.

The scale of the polymorphism is large. An analysis of Caucasian five-locus
HLA haplotype frequencies held by the US National Donor Marrow Program
(NMDP) Registry indicates 85,000 haplotypes in a population of \ensuremath{\approx}6.7
million individuals \citep{Slater2015}. Haplotype frequencies range
over nearly six orders of magnitude even in this limited sample. Allele
and haplotype discovery is far from complete. A cardinal feature of
the HLA transplantation is balancing selection, the apparently stable
existence of haplotypes and alleles at intermediate frequencies. A
scan of the human genome for other regions under balancing selection
found no other region, other than the ABO system, that could not be
explained by neutrality \citep{Bubb2006-tx} (although see \citep{Soni2022}
for a more recent assessment).

Observations requiring explanation include:
\begin{itemize}
\item allele and haplotype frequency distributions are characterised by
heavy tails \citep{Slater2015}
\item linkage disequilibrium is positive $(D_{ij}^{'}>0)$ in high-frequency
haplotypes and negative $(D_{ij}^{'}<0)$ among rare haplotypes \citep{Alter2017-df}
\item the Ewens-Watterson test for homozygosity shows excess homozygosity
for common 5-locus haplotypes \citep{Alter2017-df}
\item the distribution of allelic frequencies does not conform to neutral
expectations \citep{Hughes1998-ny}
\item the rate of non-synonymous nucleotide substitution significantly exceeds
the rate of synonymous substitution $(d_{N}/d_{s}>1)$ in codons in
the peptide-binding-region (PBR) \citep{Hughes1998-ny}
\item maintainenance of alleles over long periods of time (Neanderthal/Denisovan
ancestry in HLA class I alleles that has survived since the last introgression
(approximately 45-50,000 years ago) \citep{AbiRached2011,Robinson2017-ze}
and trans-species polymorphism in HLA class II loci that has survived
since the last common ancestor (6-8 mya) \citep{Klein1987,Klein2007,Leffler2013})
\item homogenization of introns relative to exons over evolutionary time
\citep{Hughes1998-ny}
\item HLA class I supertypes \citep{Sette1999-ry,Shen2023-bk}
\item frozen HLA haplotype blocks \citep{Dawkins1999}
\item recent selection on MHC standing variation as shown by inheritance
by descent (IBD) \citep{Albrechtsen2010-zk,Leffler2013}
\item high values of the $f_{adj}^{*}HLA$ metric \citep{Penman2013-eu}
\end{itemize}
\medskip{}

A list of some of the explanations for HLA polymorphism proposed over
the past 50 years or so includes negative frequency-dependent selection
(NFDS), heterozygote advantage (HA), overdominance, rare allele advantage,
divergent allele advantage, pleiotropy, and segregation distortion
balanced by negative selection. Some of these proposed explanations
overlap with others or are broadly synonymous. There is no consensus
theory of HLA polymorphism but the explanations of longest standing
are NFDS, HA, and environmental changes over time and/or space \citep{Neilsen2005},
and Box 3 in \citep{Meyer2018-rm}.

There is an alternative and unexplored explanation for HLA polymorphism
that rests on density-dependent regulation of individual species of
HLA haplotypes by diseases.

The background to the idea is the stable existence of species diversity
among Linnean species under the actions of natural selection/competitive
exclusion on the one hand and limiting similarity/character displacement
on the other. The theoretical basis for haplotype diversity at steady-state
can be be covered by the Lotka-Volterra competion equations, and density-dependent
regulation by disease has its origins in the Verhulst logistic equations
and the equations for disease transmission of Anderson and May \citep{Anderson1979-jn}.

This paper is divided into four sections: (i) this Introduction; (ii)
an outline of relevant theoretical considerations; (iii) an analysis
of the Caucasian HLA haplotype dataset available from the US National
Marrow Donor Program (NMDP) together with some running commentary;
(iv) a discussion of the principal findings.

\noindent \medskip{}

\section{\textrm{Theoretical considerations}}

\subsection{Background}

The broad aim of this paper is to explain the frequency distributions
shown by HLA alleles and haplotypes in the US NMDP Caucasian datasets.
The Caucasian dataset is now sufficiently large to reveal differences
in linkage disequilibrium between high- and low-frequency haplotypes
and excess homozygosity for common 5-locus haplotypes \citep{Alter2017-df}.

Our approach is built on the assumption that HLA polymorphism is an
evolved response to the pathogen environment. We are then faced with
a choice: whether to incorporate population biology and proceed with
density-dependence as a key element, or whether to ignore population
biology and proceed with genetic explanations that are inherently
limited to frequencies.

A recent review of an extensive literature on co-evolution of hosts
and parasites indicates slightly less than half the papers reviewed
incorporate population biology \citep{Buckingham2022}. The majority
that focus on frequency then split into those that follow changes
in allelic or haplotypic frequencies over time, and those that invoke
negative frequency-dependent selection to generate polymorphic populations
through production of genotypes of approximately equal fitness. Density-dependent
models can always generate frequencies by normalisation, but the reverse
is not the case. Moreover, frequency-dependent models make it impossible
to describe a class of non-competitive models of species coexistence
in which haplotype populations are regulated independently. The most
compelling reason for including population biology is the evidence
that disease transmission is density-dependent.

The theoretical approaches used in this paper therefore retain density-dependence
as a key foundation. Other theoretical concepts are drawn from the
literature as the need arises. We start with a brief outline of a
three trophic-level Lotka-Volterra model \citep{Chesson2008,Chesson2018}
and the changes that need to be made to adapt it to coexistence of
different genotypes of the same Linnean species. We examine the Anderson
and May equations for generalised disease transmission \citep{Anderson1979-jn}
since these provide a mathematical approach to top-down forcing. A
key property of these equations is the steady-state of the host that
emerges under certain conditions. Steady-states can be approximated
by logistic functions whose origins in the work of Verhulst \citep{Verhulst1838,Verhulst1845}
are also examined. Finally, we introduce an important theoretical
paper by Neher and Shraiman \citep{Neher2009-ou} that underpins our
approach to the origins of linkage disequilibrium and positive epistasis
in HLA haplotypes.

\subsection{The three trophic-level Lotka-Volterra model of Chesson and Kuang
\citep{Chesson2008}}

This is outlined in Box 1 of \citep{Chesson2018}. In brief, there
are three trophic levels. The focal species, in our case HLA haplotypes,
occupy the middle trophic layer. The resources consumed by human hosts
occupy the lower trophic level and pathogens, behaving as predators,
occupy the upper trophic layer. At this point, we can start making
some simplifying assumptions.

First, we remove the lower trophic level. This can be justified on
a number of grounds. There is an arguable case that historically it
is disease that has regulated human populations for most of human
history and the prior history of primate evolution \citep{May1983-ru}.
Disease has been particularly devastating in generating child mortality,
driven, in part, by the immunological naïvety of children. Second,
we are are seeking an explanation for a major evolutionary adaptation,
the HLA system, that appears unequivocally to be related to defence
against disease and little else. That indicates selective pressure
exercised by diseases over an extended time frame. This is not to
dismiss the importance of famine and starvation as immanent parts
of existence, and there is no doubt about the potentiating interaction
of malnutrition and disease. This could, in principle, lead to resource-based
input into competition between HLA haplotypes. However, to be set
against that is the rapid population expansions following improvements
in public sanitation and supplies of potable water in the latter part
of the 19th century, followed by widespread vaccination in the 20th.
These changes produced major reductions in child mortality.

A second change is not to pursue the resource consumption rates. These
serve little purpose once the lower trophic level is removed. Moreover,
the pathogen environment of the upper trophic layer is not a typical
predator, and we have no easy way of calculating its energy extraction
from the focal species.

Armed with these changes, we end up with a simple system where pathogens
exercise downward pressure on a focal species related to its density,
thereby counterbalancing its incipient tendency to grow. We go one
further, however, by dividing the focal population into a guild of
molecular species, where a guild has the sense used in ecology; a
group of species that exploit the same class of environmental resources
in the same way. The species are molecular variants of a single Linnean
species.

The molecular guild under scrutiny in the current case is the HLA
system of peptide presentation. We propose the effect of the pathogen
environment, as sensed by the HLA system, is to reduce the individual
densities of the major HLA haplotypes and spread their number. Limiting
similarity between haplotypes establishes a set of silos. Variation
within silos is then restricted or eliminated by competitive exclusion.
Major haplotypes within silos move to zero fitness at steady-state.
Effective fitness differences therefore disappear and silos show stable
coexistence.

\subsection{Density-dependent regulation of population by disease}

\noindent Early mathematical models of disease transmission used populations
of fixed size and studied changes in the frequency of different disease
states in the host. Many had their origins in the work of Ronald Ross
on malaria and the compartmental model of Kermack and McKendrick \citep{Kermack1927,Kermack1932,Kermack1933}.
The Kermack-McKendrick model identified three states (Susceptible,
Infectious, Recovered/Removed) and introduced equations to account
for changes in frequency of the three states.

The first work to treat densities of S, I, and R populations as dynamic
variables was that of Anderson and May \citep{Anderson1978a,Anderson1979-jn}.
Their model took the three standard SIR states (Susceptible, Infectious,
Recovered) and assigned them densities $X$, $Y$, and $Z$, respectively.
It proceeded with rate equations that are, effectively, mass action
equations \citep{Anderson1979-jn}. The important figure in \citep{Anderson1979-jn}
is Fig.3, where the infection step is shown as the equivalent of
bi-molecular collision between infected carriers and susceptible individuals.
Equation (14) in their paper \citep{Anderson1979-jn} shows a second-order
rate constant $\beta$ in the denominator. This step creates the density-dependence
of infection, since all other rate constants are first order.

Fig.1 in the current paper is a redrawn version of Fig. 4a in \citep{Anderson1979-jn}.
Anderson and May treat the population in their Fig. 4a as genetically
homogeneous. Curve (a) in Fig.1 of the current paper represents population
growth in the absence of disease; curves (b), (c), (d), and (e) represent
changes in population density in endemic disease where precise behaviour
rests on the duration of immunity. Our Fig.1 shows that host populations
will be capped under certain conditions and is the basis for population
regulation of the host or focal species. If there is no immunity,
individuals are maximally susceptible, and the population may fail
to expand above the threshold, as shown by line (e) to the right of
the arrow head. The threshold population density, $N_{T}$, has its
origin in the Kermack-McKendrick epidemic threshold \citep{Anderson1991}
and is the population density below which infection dies out. The
existence of an epidemic threshold is potentially of fundamental importance
in protecting low-density populations from extinction by disease and
drift.

\begin{figure}
\centering{}\includegraphics[scale=0.7]{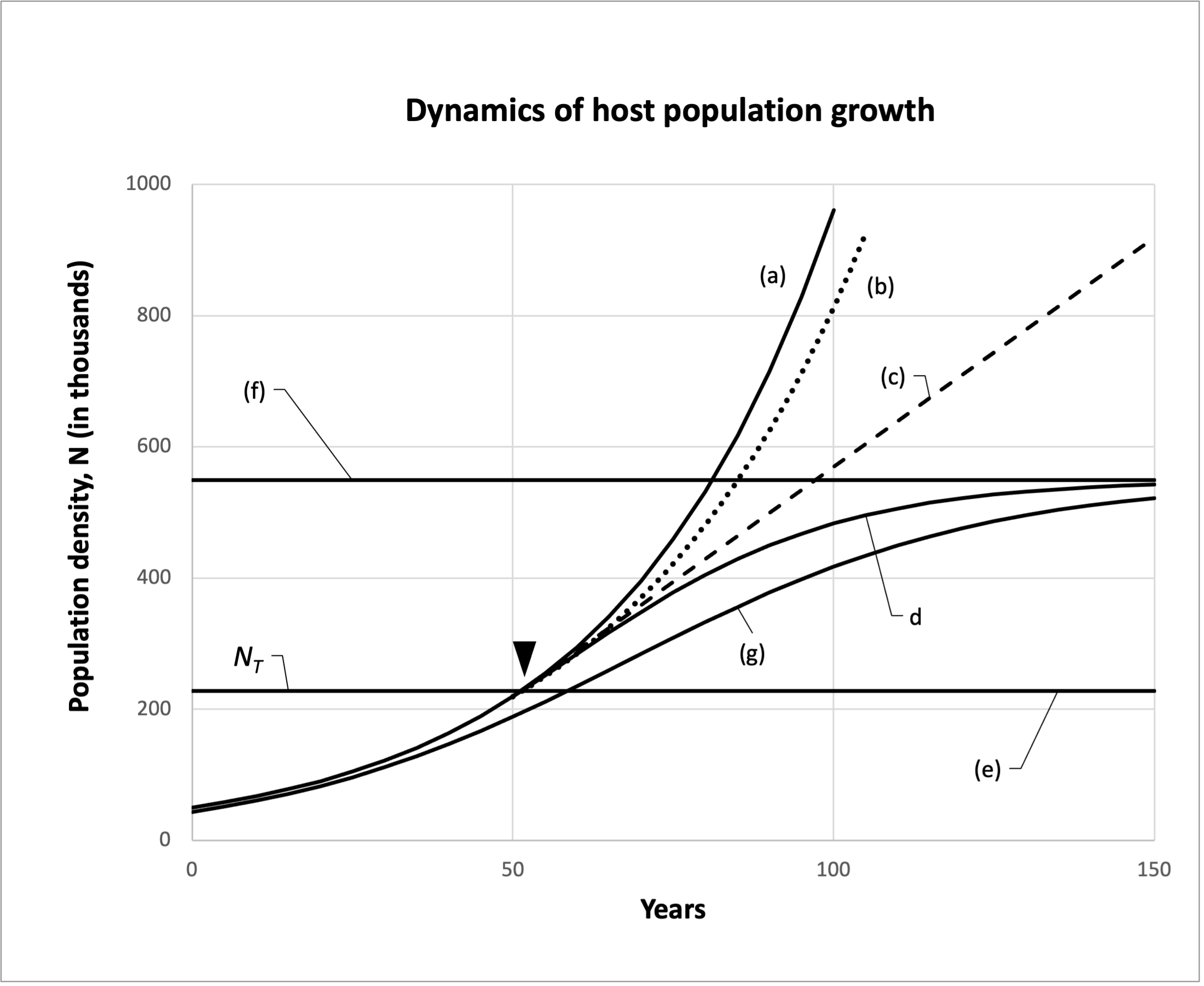}\caption{Adapted from a similar figure in \citep{Anderson1979-jn}. The timeline
in Years and Population density are both imported from the original.
Line (g) is a plot of the logistic equation whose asymptote, $N^{*}$,
is the same as for (d). The arrow head indicates the time point at
which the population density exceeds the threshold for disease transmission
effects to emerge.}
\end{figure}

Our Fig.1 also contains a logistic equation plot (g) whose asymptote,
$N^{*}$, is the same as a population brought to steady-state by disease
(Eq.14 in \citep{Anderson1979-jn}). This has a direct bearing
on the model outlined in the next subsection (2.4) since, as in the
previous subsection (2.2), the population is divided into distinct
molecular species. It is assumed the Anderson-May model can be applied
independently to each major HLA genotype population. This assumption
is underpinned by a direct connection to the extensive review by Anderson
and May \citep{May1983-fk} in which the idea is put forward that
polymorphism of host defence alleles can be modeled by density-dependent
logistic functions. This, in turn, has direct relevance to the stability
of co-existing populations at steady-state \citep{May2001}.

\subsection{The Verhulst logistic equation and Loka-Volterra competition equations}

The idea that populations expand geometrically and indefinitely can
be traced back to Euler \citep{Euler1748,Euler1767,Klyve2014-kb}.
Malthus was apparently the first to be interested in limitations to
geometric growth imposed by limited resources, and his Essay on the
subject \citep{Malthus1798} was widely influential. It had a major
impact on Darwin, who read the Essay in 1838. Verhulst also read the
Essay and published a general equation for capped population growth

\begin{equation}
\frac{dN}{dt}=rN-\varphi(N)
\end{equation}
where \emph{r} is the net growth (birth \emph{minus} death) rate constant,
$N$ is the population density, and $\varphi(N)$ is an additional
mortality rate that is some function of $N$ \citep{Verhulst1838,Verhulst1845}.
(Verhulst's symbols have been changed from those in his paper to their
common modern usage.) Verhulst chose $\varphi(N)=\alpha N^{2}$ from
a number of possibilities, since it best fitted the census data on
the rate of growth of populations in the first decades of the 19th
century, of which the French census data were the most important because
of their size and thoroughness. The Verhulst equation is, therefore,

\begin{equation}
\frac{dN}{dt}=rN-\alpha N^{2}
\end{equation}
or, expressed as the per capita change, is

\begin{equation}
\frac{1}{N}.\frac{dN}{dt}=r-\alpha N
\end{equation}
where\emph{ r} is Fisher's intrinsic fitness. The integrated form
of Eq. 2 is Verhulst's logistic equation.

Verhulst's work fell into desuetude in the latter part of the 19th
century. His logistic equation was developed independently by McKendrick
and Kesava Pai \citep{McKendrick1912}, and rediscovered by Pearl
and Reed in 1920 \citep{Pearl1920}. Pearl and Reed gave it the widely-used
form

\begin{equation}
\frac{dN}{dt}=mN=r\biggl(1-\frac{N}{K}\biggr)N
\end{equation}
where

\begin{equation}
\frac{1}{N}.\frac{dN}{dt}=m=r\biggl(1-\frac{N}{K}\biggr)
\end{equation}
\emph{m} is the effective fitness and Fisher's Malthusian parameter.
Verhulst obviously knew that the value of \emph{K} would be the upper
limit for the population density when \emph{t=}\ensuremath{\infty}
and $dN/dt=0$\emph{ }since he called it \emph{\textquoteleft la limite
supérieure de la population}\textquoteright{} \citep{Verhulst1838}.
It subsequently came to be regarded as a carrying capacity, acquiring
the widespread status of an independent variable. In reality, it is
set by the balance between \emph{r} and $\alpha$, since $r/\alpha=K$
at steady-state. Mallet has written an extensive review of $r-\alpha$
and $r-K$ formulations, arguing, correctly in our view, for the merits
of the $r-\alpha$ form \citep{Mallet2012}.

This is an important distinction in the context of our paper since
the resource layer has been set aside as a source of population regulation.
We have argued instead that, to a first approximation, we are dealing
with a contest between expansion of a focal host population of haplotypes
and an upper trophic layer that responds negatively to emerging haplotype
densities in the host guild.

\subsection{Fitness and natural selection in a population at steady-state}

Steady states in the current context exist when $dN/dt=m=0$. At steady
state, $r=\alpha N^{*}$, where $N^{*}$ is the steady-state density
($N^{*}$ being the formulation used in \citep{Anderson1979-jn}).
Since \emph{r} is a measure of fitness, so must $\alpha N^{*}$ be.
The values of $\alpha$ and \emph{r} cannot be measured directly,
but the values of\emph{ $N^{*}$ }emerge as relative abundances or
frequencies. For molecular species within a single Linnean species,
one can make the assumption that \emph{r}, or at least the average,
$\tilde{r}$, is the same for each genotype. 

If $\tilde{r_{i}}=\alpha_{i}N_{i}^{*}$ for the \emph{i}th population
at steady-state, and $\tilde{r_{1}}=\tilde{r_{2}}=\cdots\tilde{r_{i}}\cdots=\tilde{r_{n}}$,
then $\alpha_{i}N_{i}^{*}$ is the same for each genotype and

\begin{equation}
\alpha_{1}N_{1}^{*}=\alpha_{2}N_{2}^{*}=\cdots\alpha_{i}N_{i}^{*}\cdots=\alpha_{n}N_{n}^{*}
\end{equation}
However, the values of $\alpha_{i}$ and $N_{i}^{*}$ need not be
the same, since only their product is equal to $\tilde{r}$. Instead,
their values pivot; the larger the value of$N^{*}$, the smaller the
value of $\alpha$, and vice-versa.

A simple example will illustrate how useful this can be in the case
of infectious disease. We take an immunologically naïve population
and immunise a proportion against a disease such as smallpox. We assume
the immunisation itself is harmless and produces no morbidity or mortality.
The total population is then exposed to smallpox. The immunised population
shows a fitness represented by curves (a) or (b) in Fig. 1. By contrast,
the unimmunised population is capped at a much lower density. If we
assume both populations reach steady-state, the density of the immunised
population is much higher than that of the unimmunised population.
Its value of $N^{*}$ is higher, and its $\alpha$ is smaller although
the \emph{per capita} rates of growth are zero in both cases. We regard
the immunised steady-state population as fitter because it has a higher
density and higher frequency. A rank order of genotypic frequencies
is a rank order of fitnesses, and the two can be plotted against each
other, as in Figs. 2-7 below.

One can also re-write Eq. 6 as a special case where the logistic equation
is applied to the full range of population expansion:

\begin{equation}
\frac{m}{r}+\frac{N}{N^{*}}=1
\end{equation}
Stated verbally, the ratio of the Malthusian parameter to Fisher's
fitness, plus the ratio of actual population density to the ultimate
steady-state value, is equal to unity. This equation demonstrates
a conservation of fitness.

This is also a convenient opportunity to introduce conditions around
stability of steady-states involving the co-existence of multiple
species. An important, widely quoted, contribution was made by Levin
\citep{Levin1970}:
\begin{quote}
No stable equlibrium can be attained in an an ecological community
in which some \emph{r} components are limited by less than \emph{r}
limiting factors. In particular, no stable equilibrium is possible
if some \emph{r }species are limited by less than \emph{r} factors.
\end{quote}
Like us, Levin eliminated the restriction that all species are resource
limited and wrote the following in his Summary:
\begin{quote}
.., if two species feed on distinct but superabundant food sources,
but are limited by the same single predator, they cannot continue
to coexist indefinitely. Thus these two species, although apparently
filling distinct ecological niches, cannot survive together. In general,
each species will increase if the predator becomes scarce, will decrease
where it is abundant, and will have a characteristic threshold predator
level at which it stabilizes. That species with the higher threshold
level will be on the increase when the other is not, and will tend
to replace the other in the community. If the two have comparable
threshold values, which is certainly possible, any equilibrium reached
between the two will be highly variable, and no stable equilibrium
will result.
\end{quote}
(Levin's threshold level is the steady-state.) The case of smallpox
infection is of a single limiting factor. It follows that only inoculated
individuals will prosper long-term and, other things being equal,
immunologically-vulnerable individuals will either exist at low densities
or drift to extinction.

\subsection{Lotka-Volterra competition equations}

The parallel use of equations such as Eq.4 led to a large literature
on the co-existence, stable or otherwise, of Linnean species. The
simplest case is of two capped species that live independently and
make no claims on resources used by the other. This simple arrangement
is likely to be rare in practice, because some competition for resources
of one sort or another is almost inevitable in a resource-based model,
and claims on each other are likely to arise. This is the territory
of the Lotka-Volterra competion (LVC) equations where two or more
species compete in the same trophic layer. The equations for the two
populations are

\begin{equation}
\frac{dN_{1}}{dt}=\bigl(r_{1}-\alpha_{1}N_{1}-\alpha_{12}N_{2}\bigr)N_{1}
\end{equation}

and

\begin{equation}
\frac{dN_{2}}{dt}=\bigl(r_{2}-\alpha_{2}N_{2}-\alpha_{21}N_{1}\bigr)N_{2}
\end{equation}
where the third term in the bracket of Eq.8 represents the claims
by population \{2\} on the resources that would otherwise be available
entirely to population \{1\}, and Eq.9 represents the reverse case.
When $dN_{1}/dt=dN_{2}/dt=0$, one solution is that $(r_{1}-\alpha_{11}N_{1}-\alpha_{12}N_{2}\bigr)$
and $\bigl(r_{2}-\alpha_{22}N_{2}-\alpha_{21}N_{1}\bigr)$ are both
equal to zero. The equations can be drawn graphically, with linear
isoclines, $(r_{1}-\alpha_{11}N_{1}-\alpha_{12}N_{2})=0$ and $(r_{2}-\alpha_{22}N_{2}-\alpha_{21}N_{1})=0$.
Stable co-existence occurs if $r_{1}/\alpha_{12}>r_{2}/\alpha_{22}$
for $N_{1}=0$ and $r_{2}/\alpha_{21}>r_{1}/\alpha_{11}$ for $N_{2}=0$.
Put verbally, the two species may coexist at some steady-state or
equilibrium, or one may always exclude the other. It is sometimes
said that one species can win depending on initial conditions, implying
that a win is not always inevitable if initial conditions are different.
The simple co-existence case arises if $\alpha_{12}$ and $\alpha_{21}$
both equal zero.

The use of LVC equations has been extended to much larger numbers
of distinct populations. More generalised forms of the equations were
introduced some 40 years ago \citep{MacArthur1969,MacArthur1970}
for a community of Linnean species that take the general form

\begin{equation}
\frac{dN_{i}}{dt}=N_{i}\left(r_{i}-\alpha_{i}N_{i}-f\left(\sum_{\textnormal{\emph{j}=1}}^{S}\alpha_{ij}N_{j}\right)\right),j\text{\ensuremath{\neq}i}
\end{equation}
where \emph{i} runs over a pool of \emph{S} species' populations,
with $N_{i}$, \emph{$r_{i}$}, and $\alpha_{i}$ as their population
density, intrinsic growth rate constant (= intrinsic fitness), and
self-regulation (density-dependent mortality) constant, respectively:
\emph{f} is the functional response, and $\alpha_{i,j}$ are the interaction
coefficients of other species making claims on resources that would
otherwise be available entirely to species \emph{i}. Versions of Eq.10 have been used to model resource competition in ecological communities
\citep{MacArthur1969}, community food webs \citep{Cohen1985}, plant
communities \citep{Law2000}, and plant-animal mutualistic networks
\citep{Bascompte2007}.

\subsection{The Neher-Shraiman (NS) model (2009)}

Interpretation of data in the current paper has been informed by a
theoretical approach developed by Neher and Shraiman \citep{Neher2009-ou,Neher2013}.
Their paper used computer simulation to examine the balance between
selection on recombination and epistasis respectively in determining
the outcome for a set of polymorphic loci on a hypothetical haploid
chromosome that undergoes sexual reproduction. A key finding is that
``clonal condensation'' occurs when the recombination rate falls
enough for epistatic alleles derived from a group of loci to behave
as a single \emph{super}-allele or haplotype and undergo selection
accordingly. If the positive epistatic effects are large enough, a
haplotype rises to fixation through competitive exclusion and is the
fittest haplotype, measured in terms of Fisher's fitness.

An important theoretical component of \citep{Neher2009-ou,Neher2013}
is the separation of haplotype fitness, $F$, into two components,
epistatic, \emph{E}, and additive, \emph{A}, such that:

\begin{equation}
F=E+A
\end{equation}
The fitness of the additive component is heritable in the sense that
haplotype fitness is determined by the independent fitnesses of the contributory
alleles, with no positive epistasis. If haplotypes differ in additive
fitness, it is due to differences in the intrinsic fitness of the
alleles. By contrast, the epistatic component reflects the ability
of some allelic combinations to perform more effectively than others,
with heightened fitness and selection for expansion to high frequencies.
The highly-fit haplotype passes unaltered to progeny but is occasionally
broken up by recombination, thereby imposing a recombinational load.
Underlying the NS model is an environment where all haplotype clones
that come under selection are subject to competitive exclusion and
only one emerges as the fittest. Polymorphism would be inherently
unstable if haplotypes had markedly different fitnesses. This outcome
is obviated if the high-frequency haplotypes are independently regulated,
irrespective of the source of regulation. We nominate disease because
it is manifestly connected to HLA function and the theoretical epidemiological
underpinng has been known and accepted for decades.

A major focus of the NS model is the ability of clonal populations
with one or a few major clones to adapt in a timely way to a shifting
environment that requires adoption of new mutants if population fitness
is to be maintained. The possibility that HLA haplotypes are at risk
of slow response to a pathogen environment that can change rapidly
to avoid peptide presentation is obviously a matter of concern, not
least because so much of the analysis in this paper collapses if medium-
to long-term steady-states are unattainable.

A brief interpolation is given here to help interpretation of the
NMDP data. Concerns, inasmuch as they affect the HLA presentation
of peptide, are two-fold. First, standing variation is greatly increased
if many different haplotypes are stably retained in parallel, rather
than eliminated. Multiple morphs can be stabilised by the limiting
similarity that follows haplotype specialisation in complex environments.
The second concern is an assumption that a rapidly changing pathogen
environment requires HLA haplotypes to change rapidly. We argue this
assumption is fundamentally wrong. The evolved solution to the challenge
of pathogen escape is the stable existence of numerous, moderately
low-affinity HLA haplotypes, not rapid turnover of HLA specificities.

\subsection{Belevitch}

The following is a brief summary of a linguistic analysis developed
by Belevitch \citep{Belevitch1959}. It is included because it provides
an insight into outcomes obtained from log rank-log frequency plots,
which are used in this paper to analyse the frequency distributions
available from the US NMDP.

Any plain language text has elements (letters, words, sentences, etc.)
that can be counted. Dividing these counts (say, of specific words)
by the total number of words in the text gives relative frequencies
or \emph{a priori} probabilities for each word. One can also construct
catalogues for each linguistic element (dictionaries for words, for
example) where each distinct element (word) is listed with its probability
of occurrence in the text. These words can be ranked in order of non-increasing
probabilities. Some elements (words) only occur once and these define
the unit probability or frequency of occurrence for a particular text.
All other elements have frequencies in the text that are whole-number
multiples of the singleton frequency. Since the number of ranks is
equal to the number of dictionary entries, each rank is associated
with a whole-number multiple of the singleton frequency.

There are two additional properties of statistical linguistics that
are absent in general statistics. One is the closure condition

\begin{equation}
\sum_{i=1}^{i=N}\mbox{N}_{i}p_{i}=1
\end{equation}
The other is that, in information theory, there is an entropy, $x$,
that is related to probability by

\begin{equation}
x=-\mbox{log}p
\end{equation}
The entropy \emph{x} can be plotted against rank. A property of dictionaries
with large numbers of low-frequency elements is that the behaviour
of the bulk of the population is represented by a relatively small
number of entropies in a small part of the rank order range. For that
reason, it is more useful to plot log rank against entropy, which
is rank \emph{versus} probability (or frequency) on logarithmic co-ordinates.

There can be no \emph{a priori }expectation for the shape of the resulting
distribution, but the local value of the slope of a tangent at a particular
point in the distribution potentially has value. Belevitch established
the gradient of the log rank-log frequency distribution for an arbitrary
distribution function $\varphi(x)$ in the neighbourhood of a point
$x_{0}$ using a Taylor expansion and arrived at the expression

\begin{equation}
\mbox{log}\frac{p_{i}}{p_{0}}=-A\:\mbox{log}\frac{i}{i_{0}}
\end{equation}
where an element with entropy $x=x_{i}$ has a probability $p_{i}=e^{-x_{i}}$
and rank $i$. Similar notations $p_{0}$ and $i_{0}$ apply to a
nearby reference point $x_{0}$, which may be the frequency of the
highest-ranked element. Eq.14 is independent of any assumption
about the distribution law, with $A$ merely measuring the slope of
the tangent at $x_{0}$ to the rank-frequency characteristic on logarithmic
scales. There is an obvious connection to a community of elements,
such as haplotypes, at steady-state.

\subsection{Our model}

Considerations covered in subsections 2.1 to 2.8 lead to a simple
model, stated here for a single HLA haplotype. Haplotype \{1\} presents
peptides from pathogen A but not B. \{1\} is immune to A by virtue
of its successful peptide presentation and a population of \{1\} would
expand, in the absence of B, following curve (b) in Fig.1 (this paper).
In the presence of B, \{1\} will be regulated and follow a curve such
as (d) to a density-regulated steady-state population size. At worst,
B may be sufficiently pathogenic that \{1\} cannot grow above the
Kermack-McKendrick threshold. Haplotype \{1\} is therefore regulated
by B but safe from extinction. The same applies to haplotype \{2\},
which presents peptides from pathogen B but not A. 

This model is scaled up to account for extensive HLA polymorphism,
which can be regarded as a large-scale community of genotypes. Large-scale
communities have raised issues about stability for more than 50 years
\citep{Gardner1970,May1972,May2001}. Random matrix theory (RMT) has
played an important rôle over that time (see introduction to \citep{Krumbeck2021}
for a brief up-to-date overview). A key element in the approach to
stability has been the asymptotic linear stability of an equilibrium
point. Large-scale Lotka-Volterra models of the form in Eq.10 are
increasingly commonly used. For example, a recent two-trophic-level
predator-prey model has 200 predators and 800 prey species \citep{Krumbeck2021}.

We argue in the current paper that the number of HLA haplotype species
under selection in the Caucasian dataset is of the order of 350-2000.
Bearing in mind Levin's restrictions on the number of species that
can co-exist in a stable equilibrium \citep{Levin1970}, there must
be a similar or greater number of limiting factors. In this paper,
the limiting factors are pathogens. A set of 350-2000 haplotypes is
consistent with an estimated 1400 or so human pathogens \citep{Anon2011},
some of which will exist as multiple major strains \citep{Penman2013-eu,Gupta2024}.

\noindent \medskip{}

\section{\textrm{Results: Data analysis}}

\subsection{Log-log rank-frequency plots: HLA 5-locus haplotypes}

A distinguishing feature of HLA polymorphism is its scale. Even now,
after decades of work, the rate of discovery of new HLA class I and
II alleles shows no sign of saturation (see, for example, Fig.2 in
\citep{Barker2023}). The term \emph{hyperpolymorphism} has entered
the language \citep{Barker2023}. The largest datasets that give direct
access to the scale of HLA polymorphism are those assembled by the
US National Marrow Donor Program (NMDP). These are accessible through
https://www.nmdp.org. A fuller description of the development of the
NMDP database is given in \citep{Gragert2013}. The NMDP files have
the bottom \ensuremath{\sim}1\% of the census population removed.
We also use a set of frequencies derived from the NMDP data using
an expectation-maximisation (EM) algorithm \citep{Alter2017-df}.
This set is complete. Assuming the minimum frequency is the unit frequency,
the sample has a census value of 14,947,683 haplotypes, covering 101,230
different haplotype species. There are no accompanying identifiers
in the EM set. A log-log plot of the two sets against each other shows
no obvious anomalies, but extends only to haplotype species of rank
37645. This accounts for 99.08\% of the census population.

NMDP haplotype frequencies are listed under 13 self-identified ethnicities
(Caucasian, Hispanic, American Indian, etc.). We have examined data
for all 13. However, we have mostly used data for the Caucasian population,
which is substantially larger than the others, particularly the five-locus
Caucasian haplotype HLA-A\ensuremath{\sim}B\ensuremath{\sim}C\ensuremath{\sim}DRB1\ensuremath{\sim}DQB1.
Other similar sets of data are available from the Anthony Nolan register,
but these are much smaller. For example, the British/Irish/North-West
European (BINWE) set in the Nolan registry, which is likely to have
a similar ethnicity to the NMDP Caucasian set, has only 4\% of the
NMDP coverage. The distribution of Caucasian 5-locus HLA haplotype
frequencies from the NMDP set is shown in Fig.2 (red markers). It
demonstrates close approximation to a power law distribution for the
first 300 or so haplotypes by rank (black line). The blue line represents
the expected values for the same 5-locus HLA haplotypes calculated
from the equilibrium frequencies of individual contributing alleles.

\begin{figure}
\hspace{4cm}\includegraphics[scale=0.4]{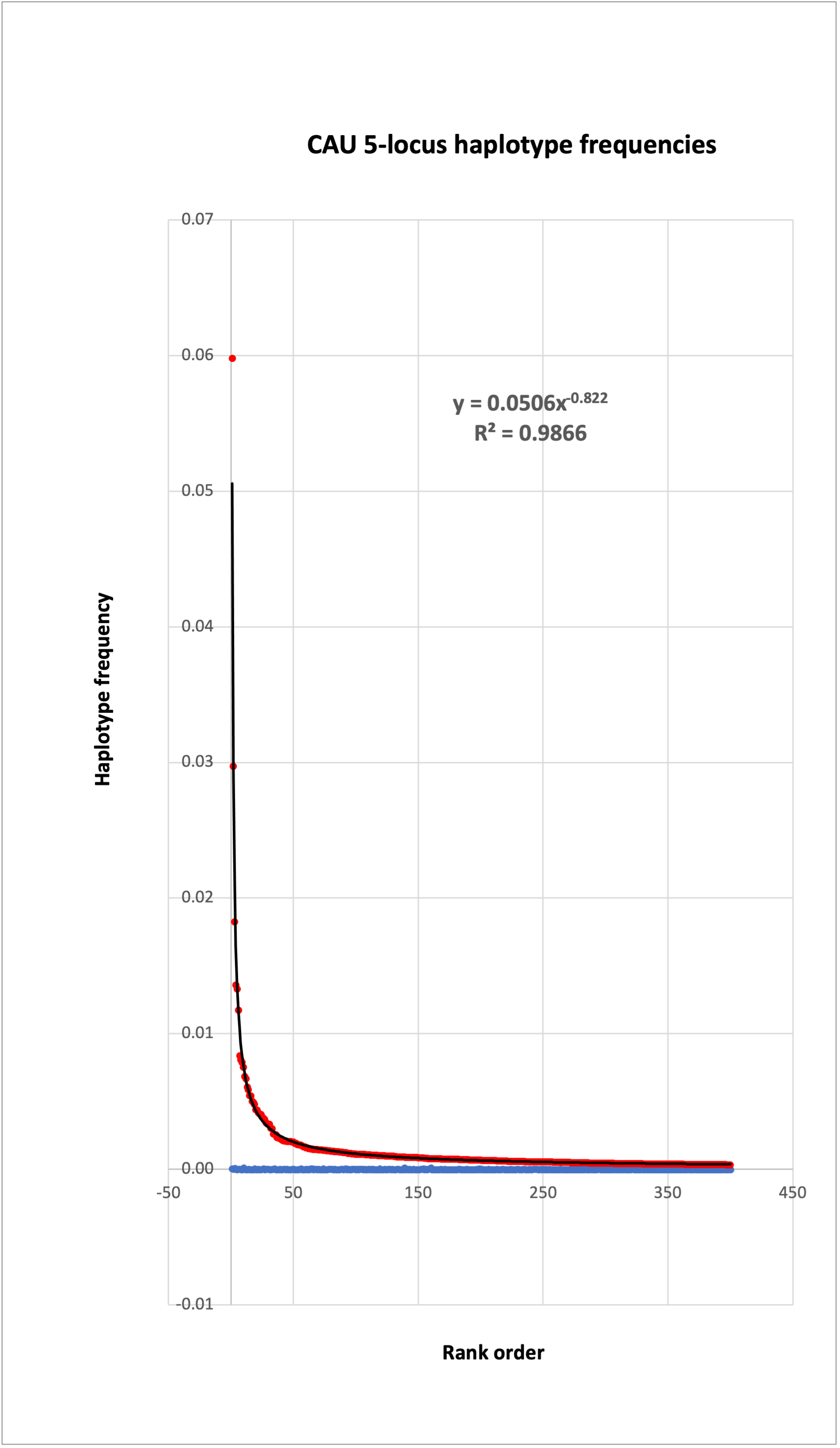}

\caption{The US NMDP Caucasian HLA 5-locus data set plotted on linear co-ordinates
for the 400 haplotypes of highest frequency. Red markers are actual
frequency values. Blue markers reflect calculated haplotype frequencies
based on multiplication of allele frequencies; that is, equilibrium
values. The black line is the least-squares fit to the red data. The
curve is closely fitted by the simple power relationship, $y=0.0506x^{-0.822}$.}
\end{figure}

The data from frequency data in Fig. 2 can be plotted on logarithmic
co-ordinates (Fig. 3). Following Belevitch \citep{Belevitch1959},
the ordinate can also represent entropy values. The red markers represent
Shannon entropies. The first \ensuremath{\approx}350 haplotypes (\ensuremath{\approx}58\%
of the census population) lie on a straight line whose gradient of
-0.813 is the constant -A in Eq.14. The extensive linear portion
of the distribution indicates underlying niche apportionment.
\begin{figure}
\hspace{2.5cm}\includegraphics[scale=0.3]{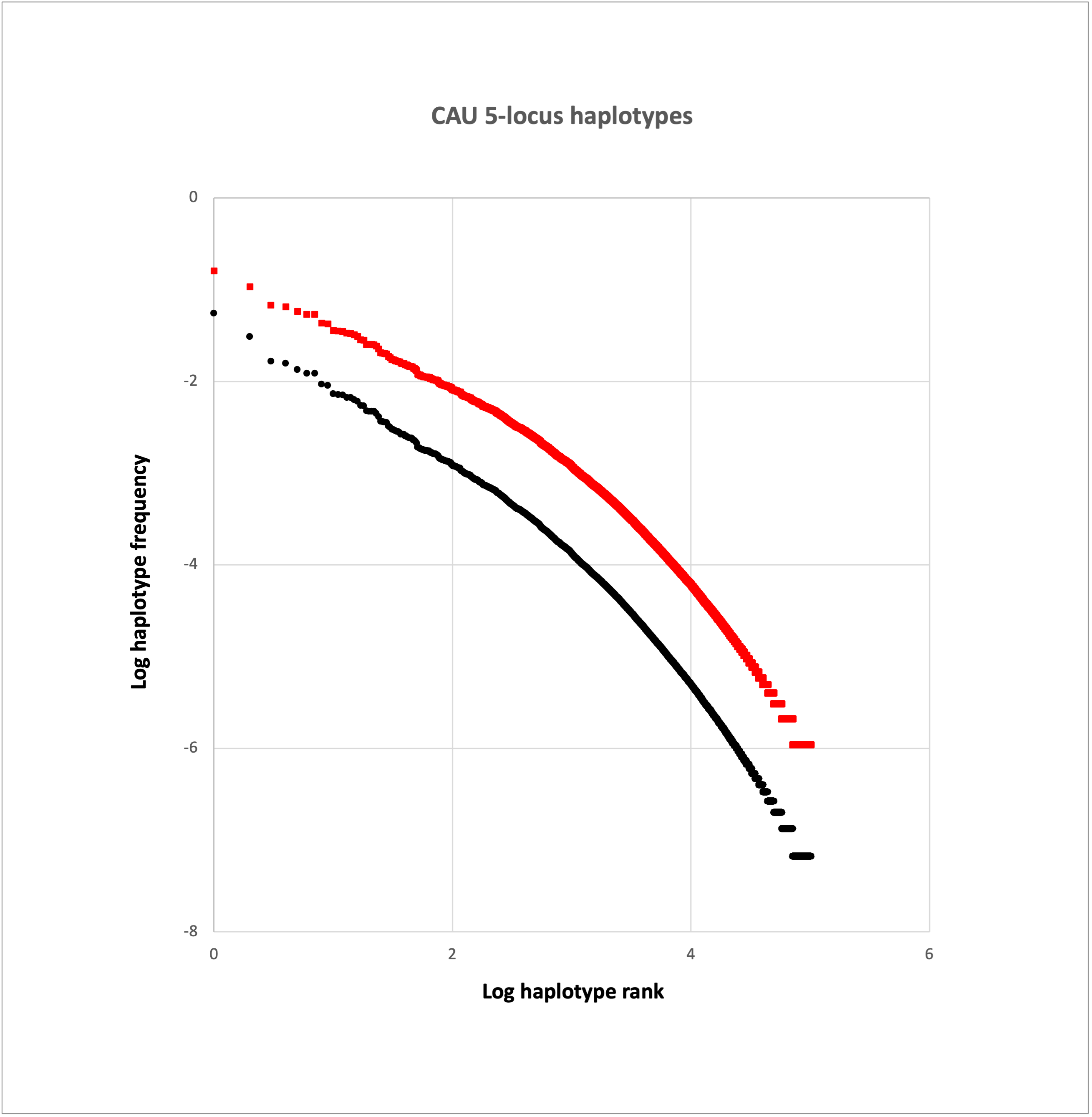}\caption{The NMDP Caucasian HLA 5-locus data set after use of an EM (Expectation-Maximisation)
algorithm \citep{Alter2017-df} (black markers). The frequencies in
the linear portion are consistent with positive selection on haplotypes,
with an underlying niche apportionment possibly acting on a haplotypic
HLA population behaving as a quasi-species. The curved portion reflects
progressive loss of positive selection and its replacement by drift
and possibly negative selection, consistent with changing sign for
$D_{ij}^{'}$ \citep{Alter2017-df}. The red markers represent the
distribution of Shannon entropies of individual haplotypes ($S_{i}=-p_{i}lnp_{i}$)
on the same axes.}
\end{figure}

The data in \citep{Alter2017-df} can be plotted in other ways. For
example, each haplotype species is associated with a census frequency
that indicates how many copies of that haplotype exist in the sample.
These frequencies are integral multiples of the unit frequency shown
by a haplotype species that occurs only once. One can construct histograms
from these data in which the frequency multiples, \emph{n}, including
those for which there are no actual haplotypes, are listed in order
on the abscissae of Figs.4a,b. The ordinate represents the number
of times, \emph{h}, the sample includes different haplotype species
with a given frequency multiple (described as bin size, $n$). By
way of illustration, there are $h=30002$ haplotype species in the
Alter et al. sample \citep{Alter2017-df} that occur only once $(n=1)$,
$h=13604$ haplotype species occur twice $(n=2)$, $h=8163$ three
times $(n=3)$, and so on. At the upper end of the frequency spectrum,
there are separate haplotype species that occur $n=826429,462799,250191$
times, etc. but these frequency multiples are each associated with
a single haplotype that is different in each case $(h=1)$. Most high-frequency
multiples are absent. The outcome of a plot in which $hn$ is plotted
against $n$ is highly informative (Figs.4a,b). The ordinate is
linear and the abscissa is logarithmic. The sum of the $hn$ values
for all actual $n$ is the CAU haplotype sample size (= 14,947,683).

\begin{figure}
\hspace{0.5cm}\subfloat[Abscissa: bin size, \emph{n}, is the number of times a haplotype appears
in the catalogue. Bin sizes are integers and run continuously from
1 to 10 million on logarithmic scale, irrespective of whether bins
are filled or empty. Ordinate: the total number of haplotypes in a
bin, \emph{h} x \emph{n}, where \emph{h} is the number of haplotypes
whose catalogue frequency matches a particular bin size. The scale
is linear. The plot shows two distinct populations with limited overlap.
That on the left of the minimum contains haplotypes that are subject
either to drift or mild negative selection. That on the right of the
minimum contains haplotypes under increasing positive selection as
one moves to the right.]{\hspace{7.5mm}\includegraphics[scale=0.3]{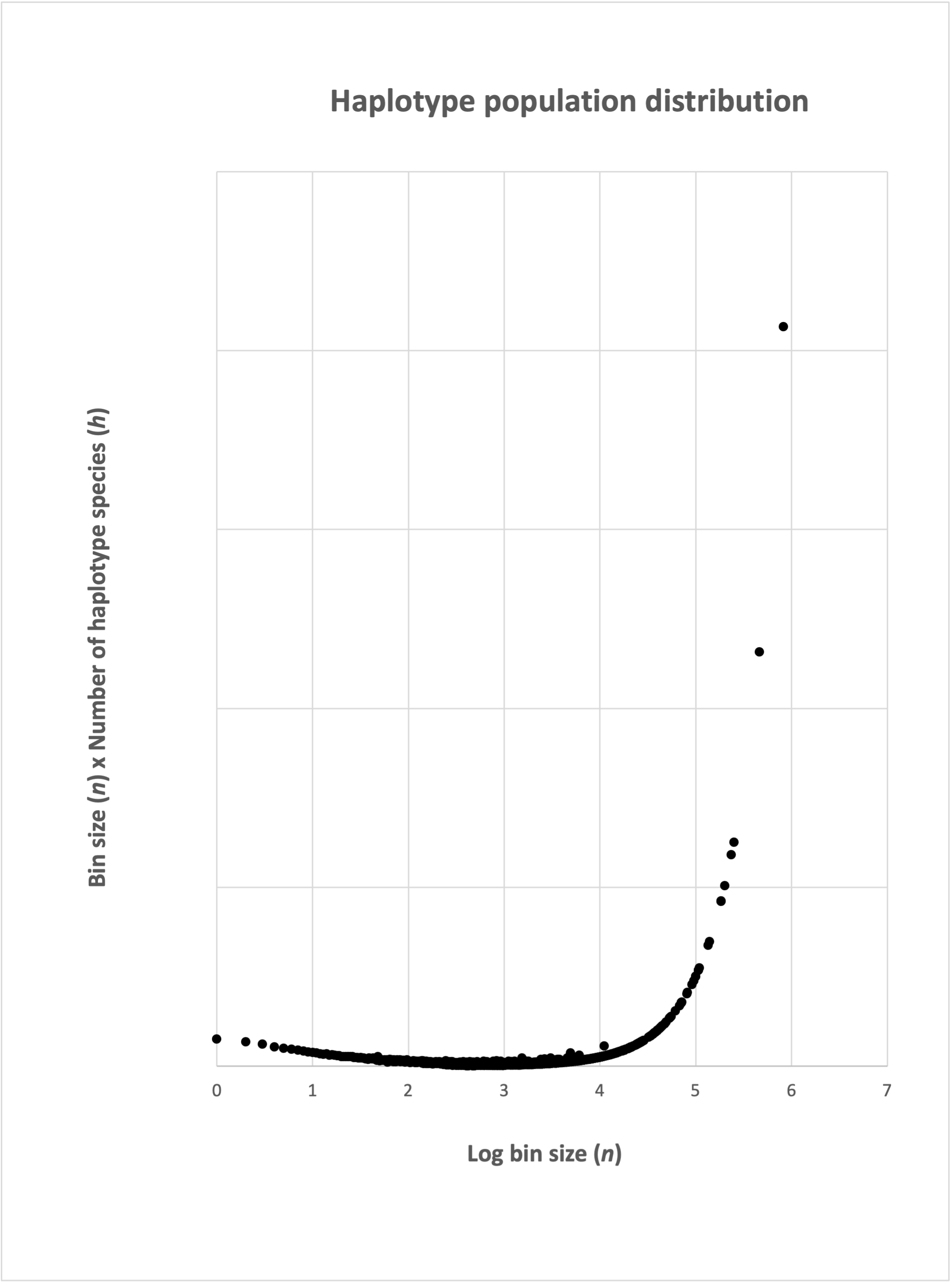}}\hspace{1cm}\subfloat[Data used in Fig.4a showing fitting of two exponential curves. That
on the right is for bins in which $h=1$, with one exception where
$h=2$. Since \emph{n} is plotted against \emph{n}, one on linear,
the other on logarithmic co-ordinates, the curve is exponential ($y=1.0378e^{2.2947x};R^{2}=0.9999$).
The best fit of an exponential for the left-side population is slightly
poorer ($39074e^{-0.993x};R^{2}=0.9514$), reflecting the increasing
stochasticity of $h$ values associated with small haplotype numbers
in small bins. Blue markers show data used for left-side population
exponential, red markers for right-side.]{\includegraphics[scale=0.3]{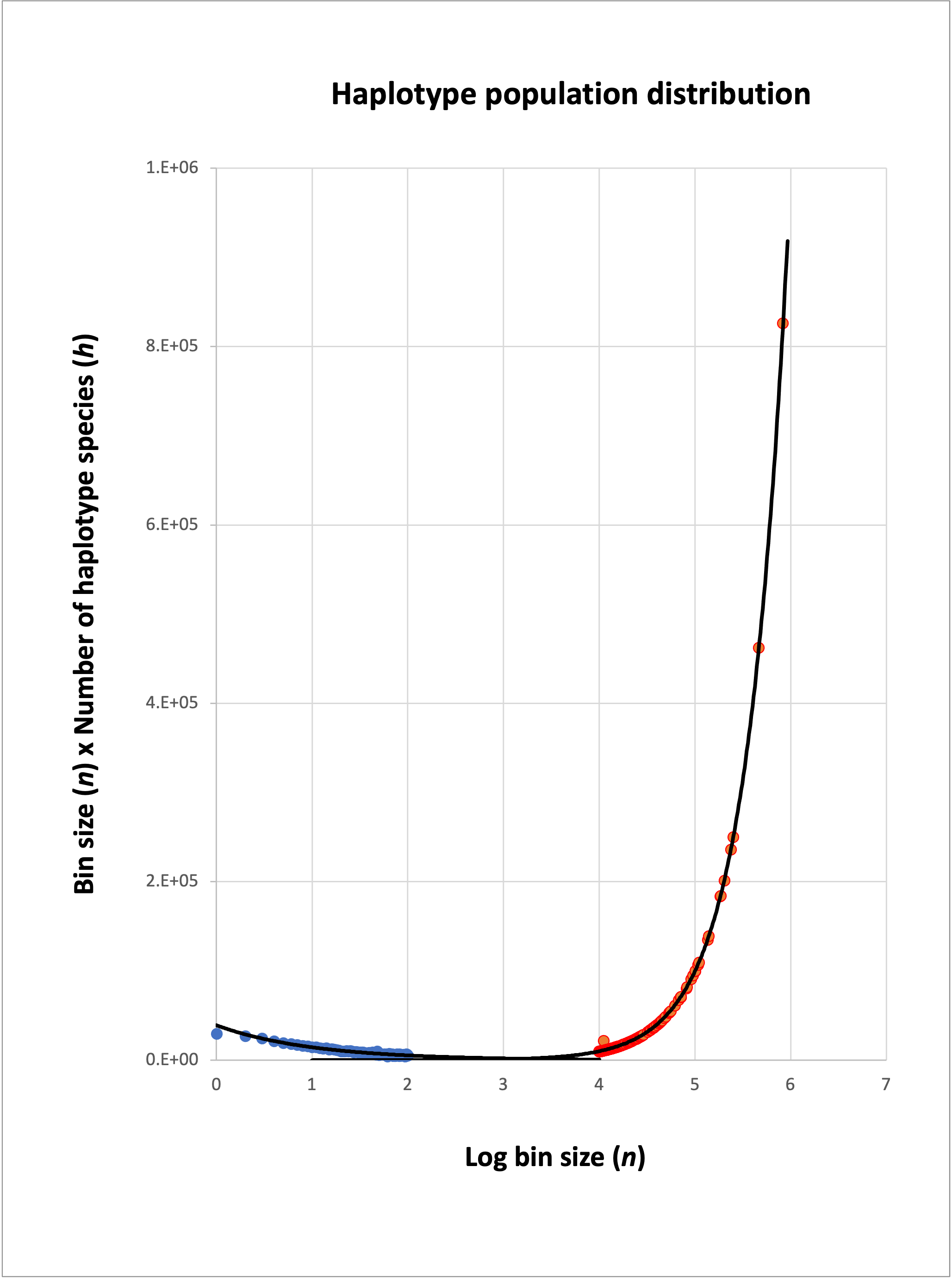}

}\caption{Distribution of census haplotype numbers by product of haplotype species
number and bin size, $h\mbox{ x }n$, against bin size, $n$. }
\end{figure}
Figs.4a,b provide clear evidence of two haplotype populations that
have limited overlap. The results support the proposition that the
population on the right-side of the minimum represents haplotypes
under positive selection whereas that on the left-side represents
those haplotypes subject to drift or negative selection. They are
consistent with the conclusions in \citep{Alter2017-df} that linkage
disequilibrium is positive $(D_{ij}^{'}>0)$ in high-frequency haplotypes
and negative $(D_{ij}^{'}<0)$ among rare haplotypes, and the evidence
that the Ewens-Watterson test for homozygosity shows excess homozygosity
for common 5-locus haplotypes.

The value of $hn$ in the minimum is about 1020, or rank 1730.
Although the two exponentials extend on either side of this minimum,
it is useful to denote rank 1730 as an approximate transition point
\emph{t }in the rank order. This point produces an \ensuremath{\approx}80:20
split in the census population, with those haplotypes under selection
(\ensuremath{\approx}80\%) being the majority by census.

\subsection{Log-log rank-frequency plots: HLA class I and II alleles}

Data for allele frequencies are derived by the US NMDP directly from
their haplotype dataset and shown in the 5 panels of Fig. 5.

\begin{figure}
\hfill{}\includegraphics[scale=0.3]{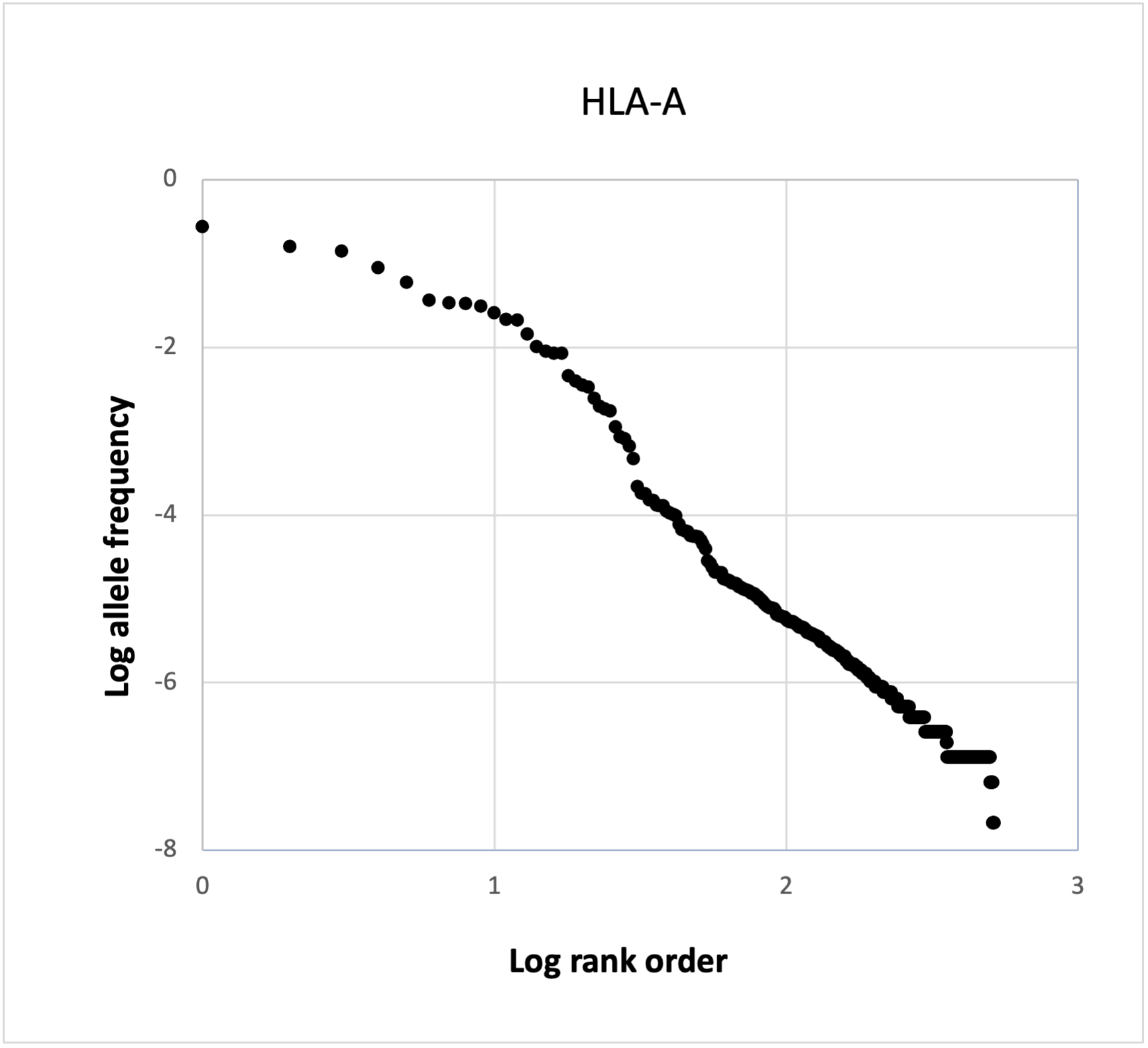}\hfill{}\includegraphics[scale=0.3]{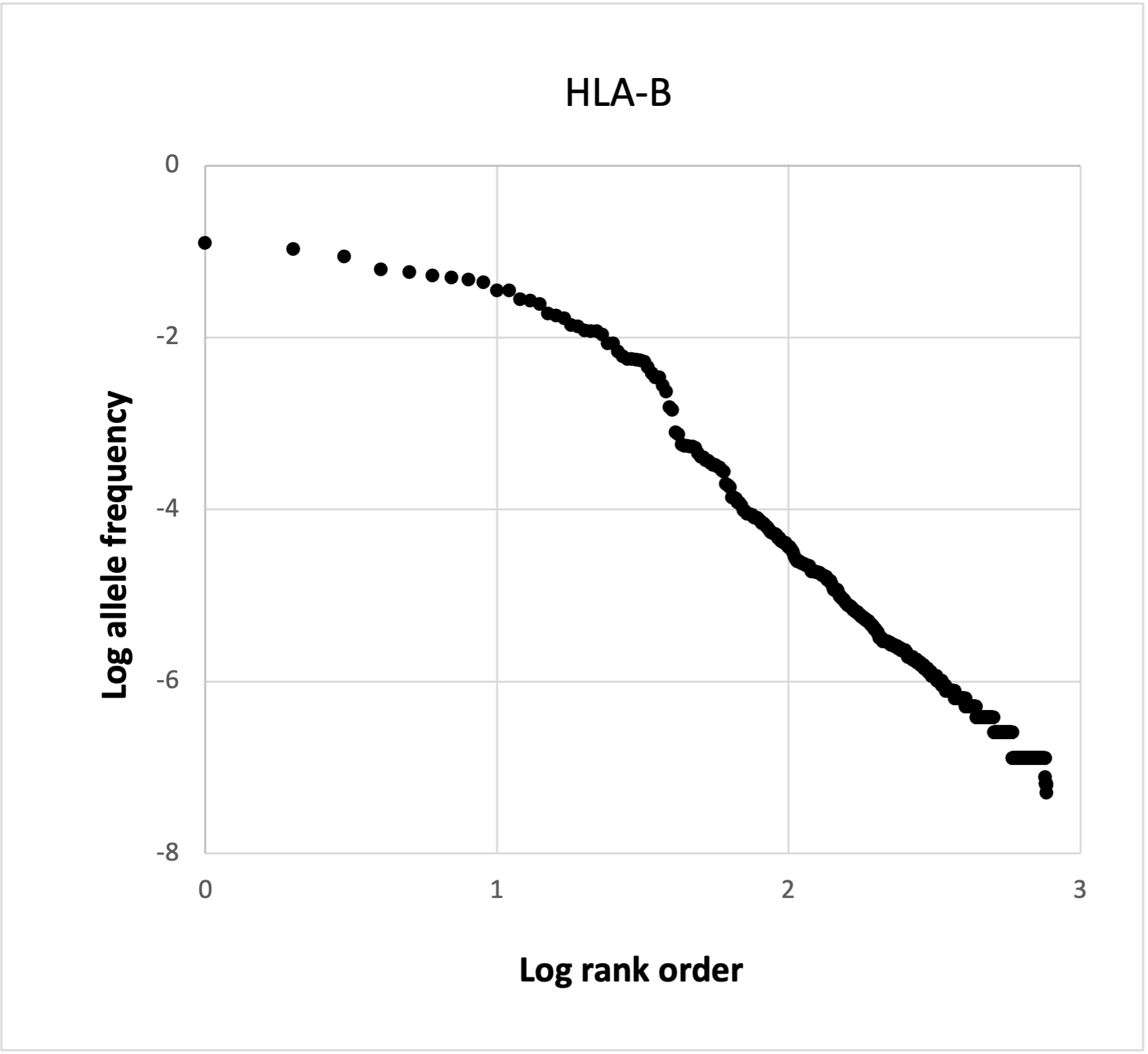}\hfill{}\includegraphics[scale=0.3]{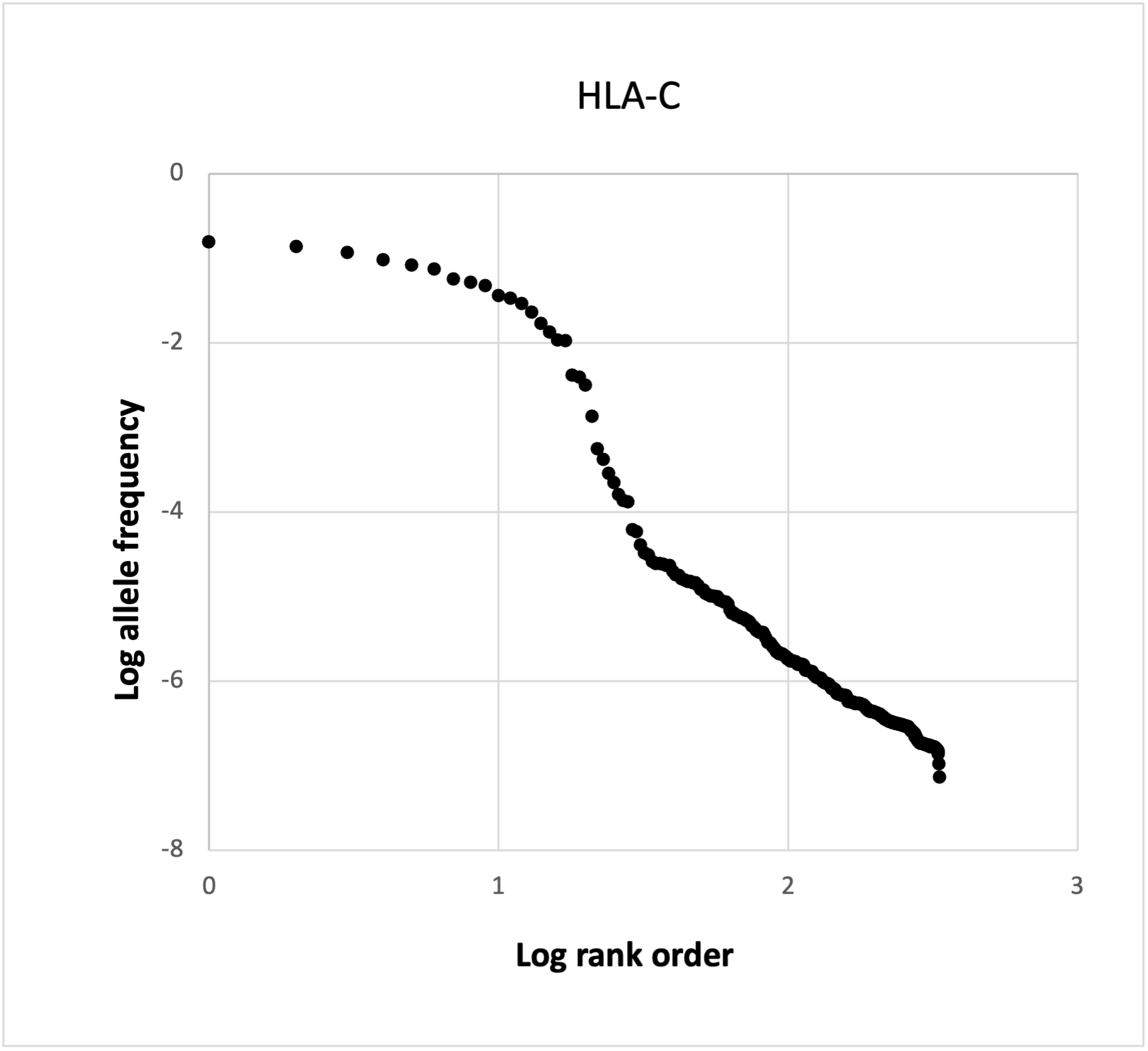}\hfill{}

\bigskip{}

\hspace*{3cm}\includegraphics[scale=0.3]{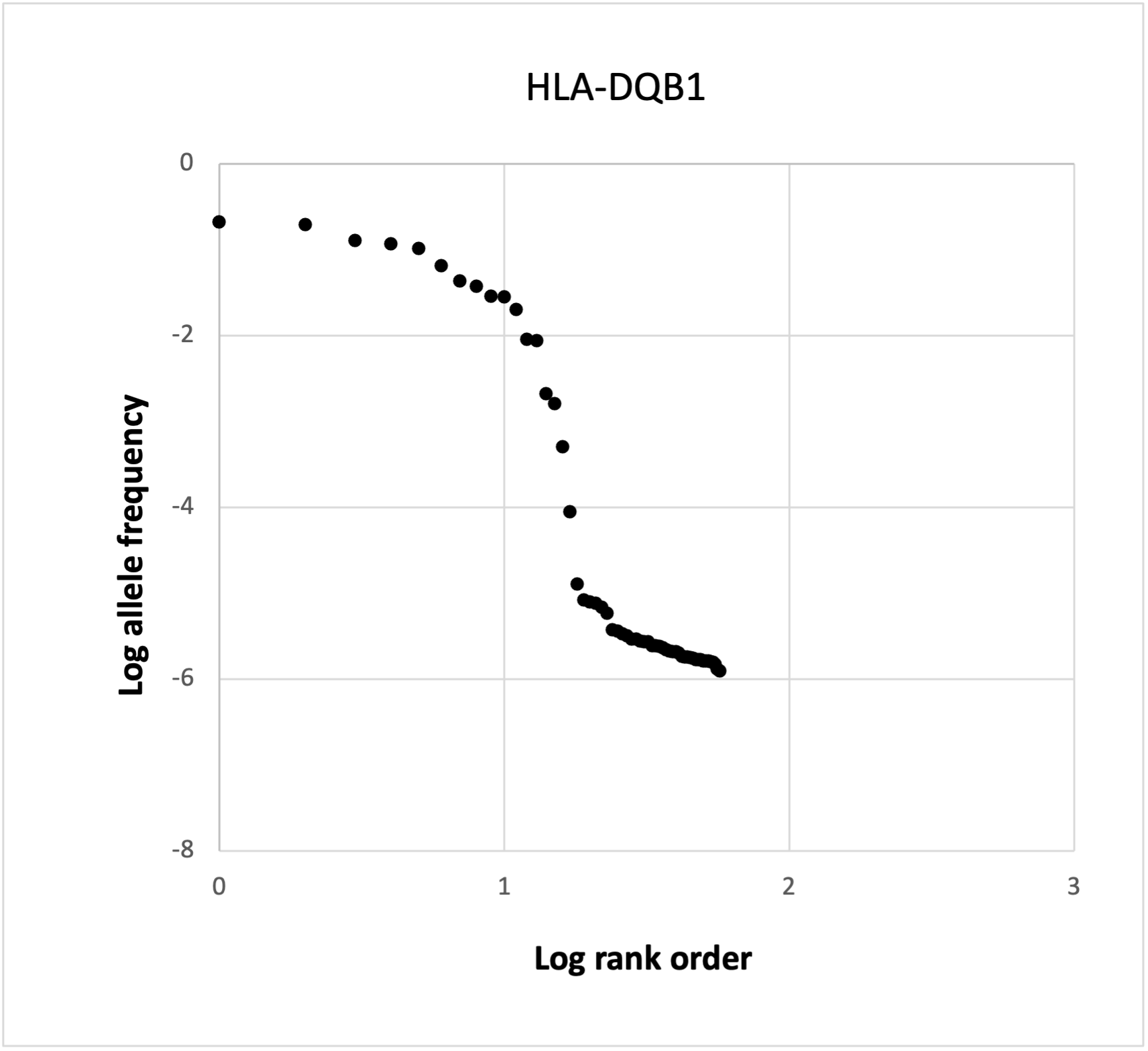}
\hspace*{1cm}\includegraphics[scale=0.3]{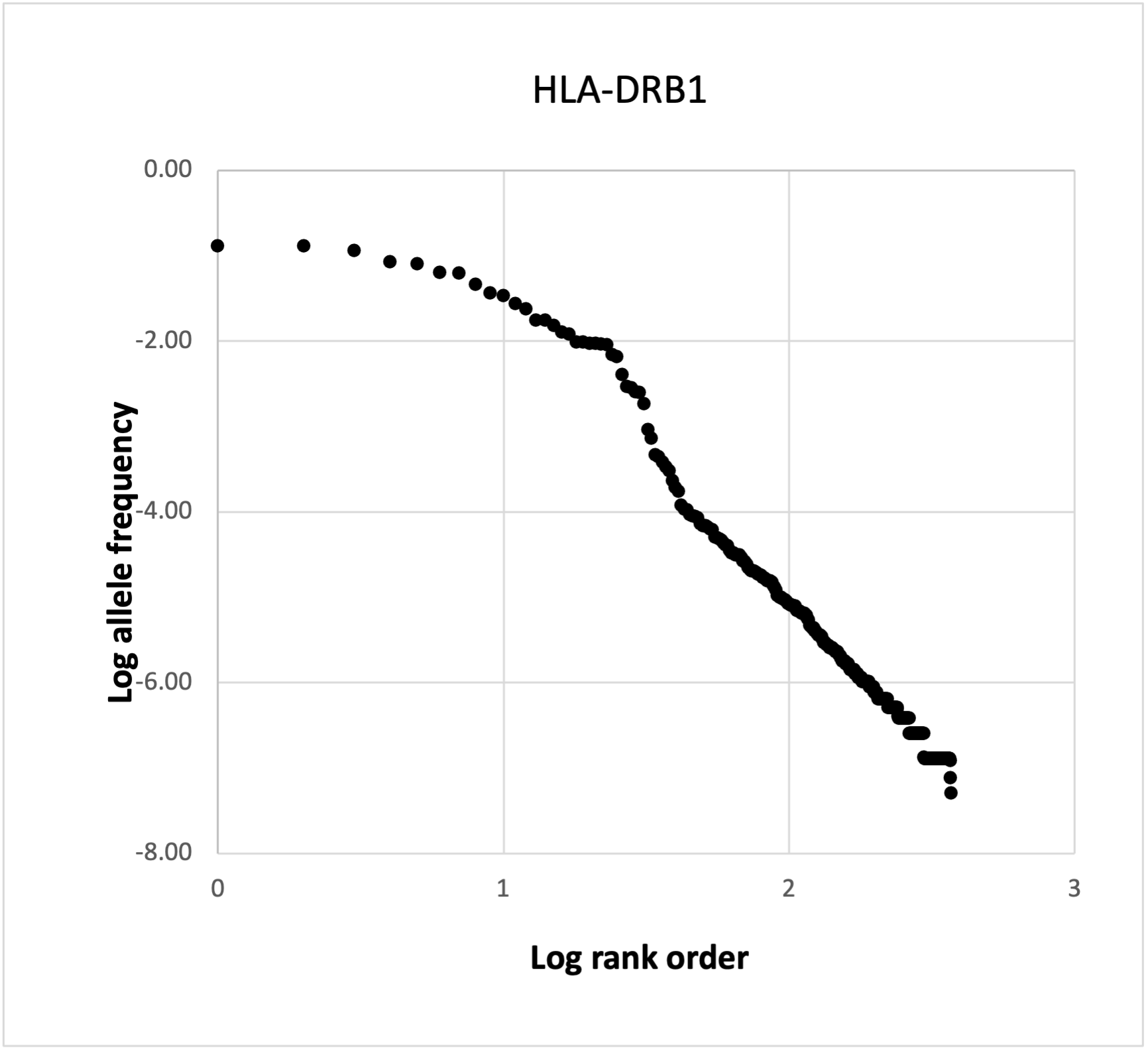}\hfill{}

\caption{HLA allele frequencies plotted against rank on logarithmic co-ordinates
for the five loci that make up the HLA-A\ensuremath{\sim}B\ensuremath{\sim}C\ensuremath{\sim}DRB1\ensuremath{\sim}DQB1
Caucasian 5-locus haplotype dataset. All data from the Caucasian allele
files were assembled by the US NMDP and are available as separate
files from their website.}
\end{figure}
HLA alleles do not follow a power law distribution. Rather, their
distributions show a relatively small number of alleles that break
away from a trend-line for the frequencies of low-frequency alleles.
The discontinuity is obvious for HLA-C and HLA-DQB1 but is present
at all loci. Reasons why the discontinuities occur is discussed shortly.
High-frequency alleles show logarithmic decay, as shown by plotting
the data in Fig.5 on log-normal co-ordinates (Fig.6). The plots
in Fig.6 indicate a geometric decline for high-frequency alleles
in which each allele in ascending rank occupies a fixed fraction of
the available population after deducting the allelic frequencies of
lower rank.

This was an unexpected result, not least because the frequency values
for the lowest ranked alleles are themselves composites of many hundreds
of separate contributions from each haplotype species to which the
allele in question contributes (see evidence below, subsection 3.3).
The underlying orderliness of the logarithmic distribution suggests
a form of niche allocation, potentially produced by density-dependent
capping of alleles. The numbers of alleles identified for each locus
in the high-frequency category, together with their cumulative frequencies,
are: A/30/0.997; B/40/0.986; C/20/0.996; DQB1/15/0.999; DRB1/31/0.993,
making 137 alleles in all. Permutations restricted solely to this
group of 137 alleles account for \ensuremath{\approx}97.1 \% of all
haplotypes in the census sample. 

\begin{figure}
\hfill{}\includegraphics[scale=0.3]{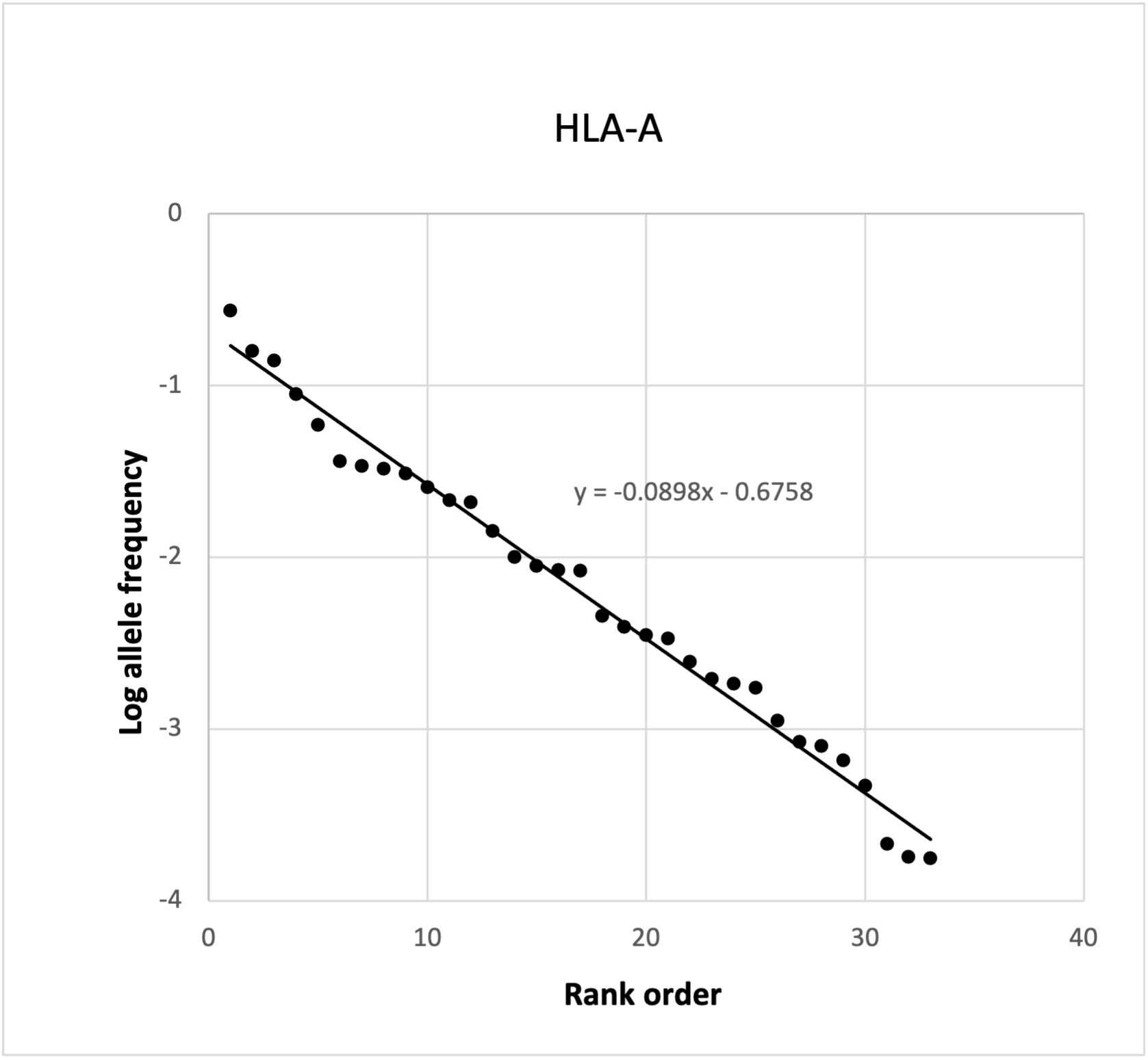}\hfill{}\includegraphics[scale=0.3]{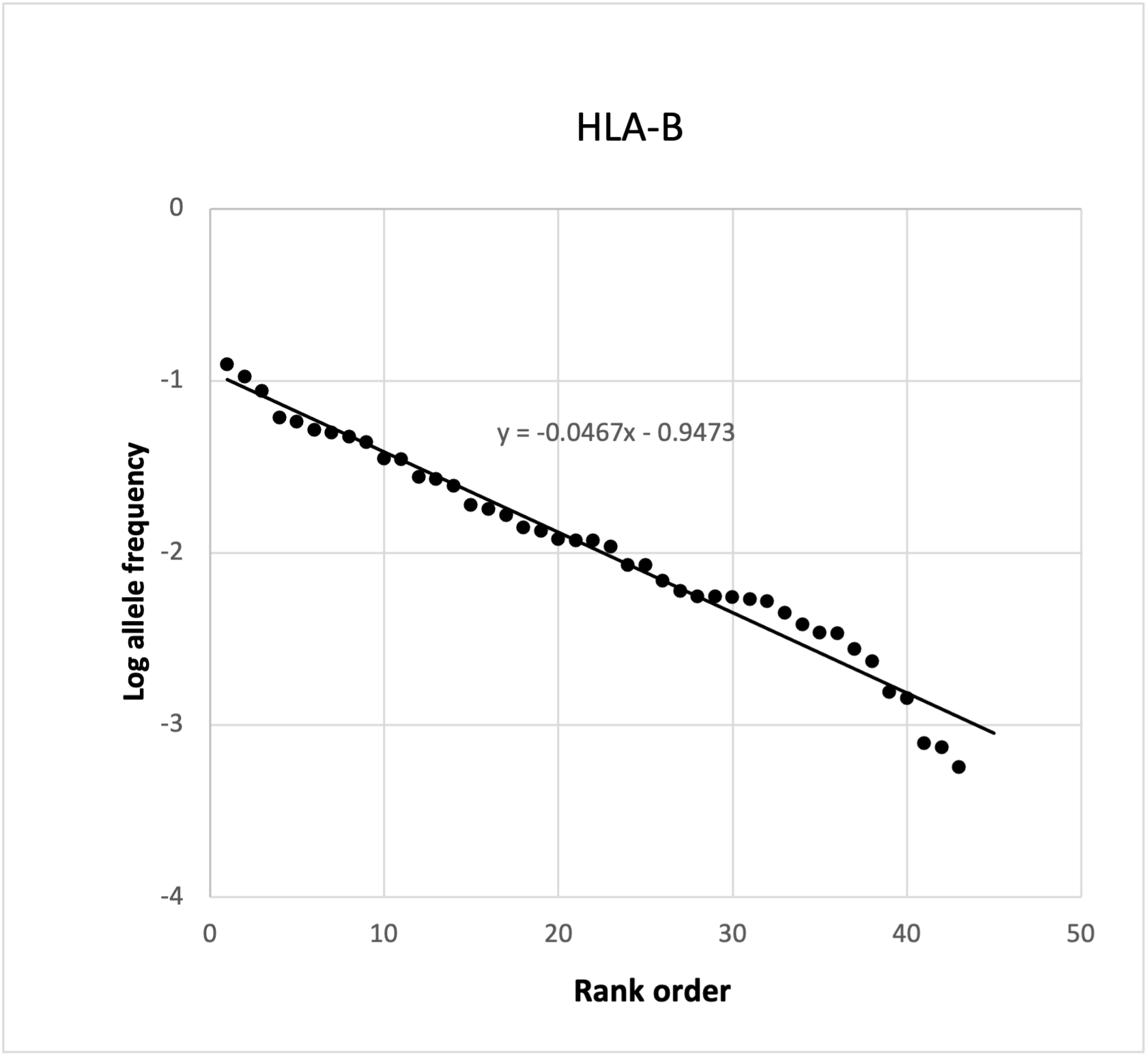}\hfill{}\includegraphics[scale=0.3]{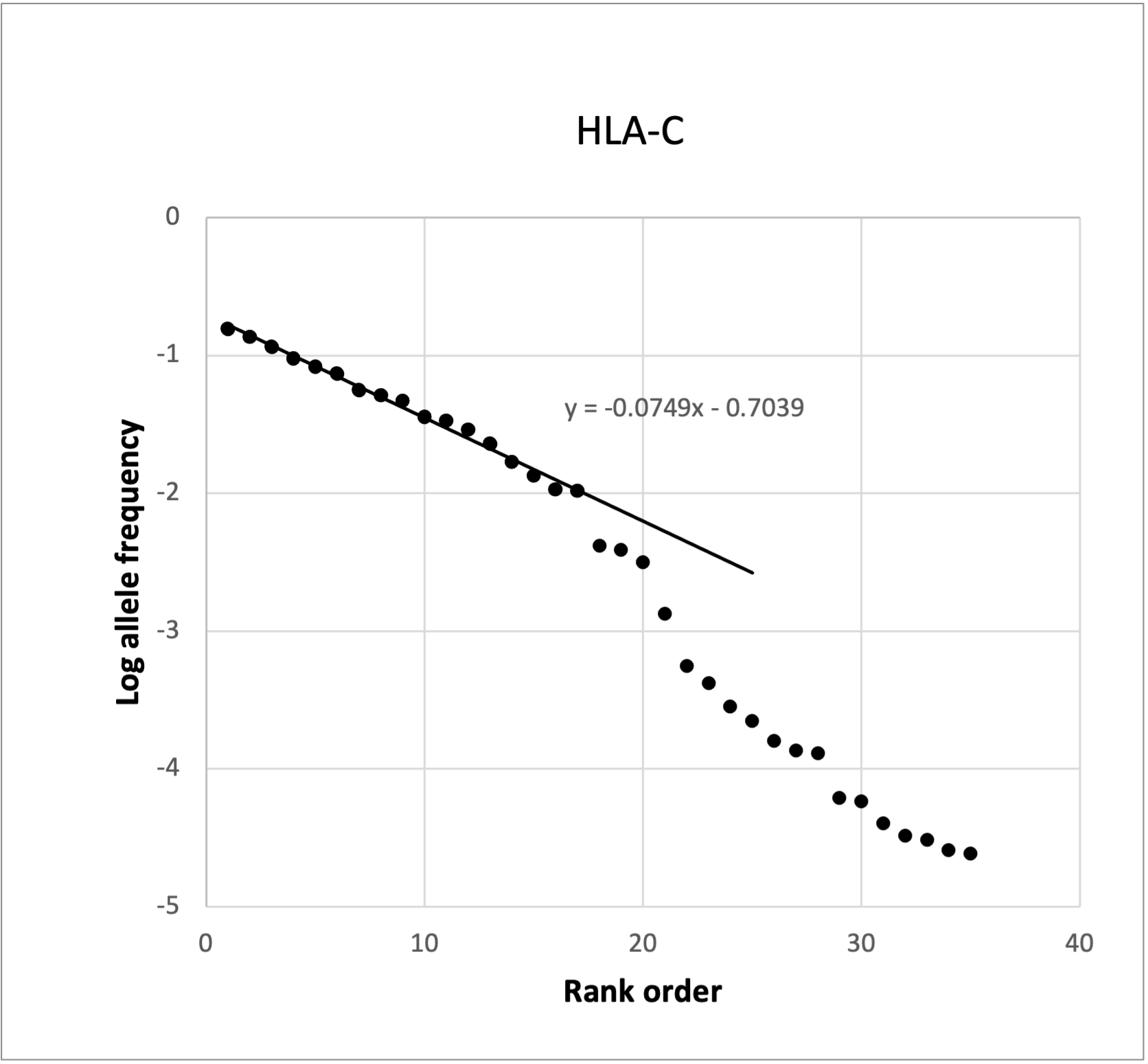}\hfill{}

\bigskip{}

\hspace*{3cm}\includegraphics[scale=0.3]{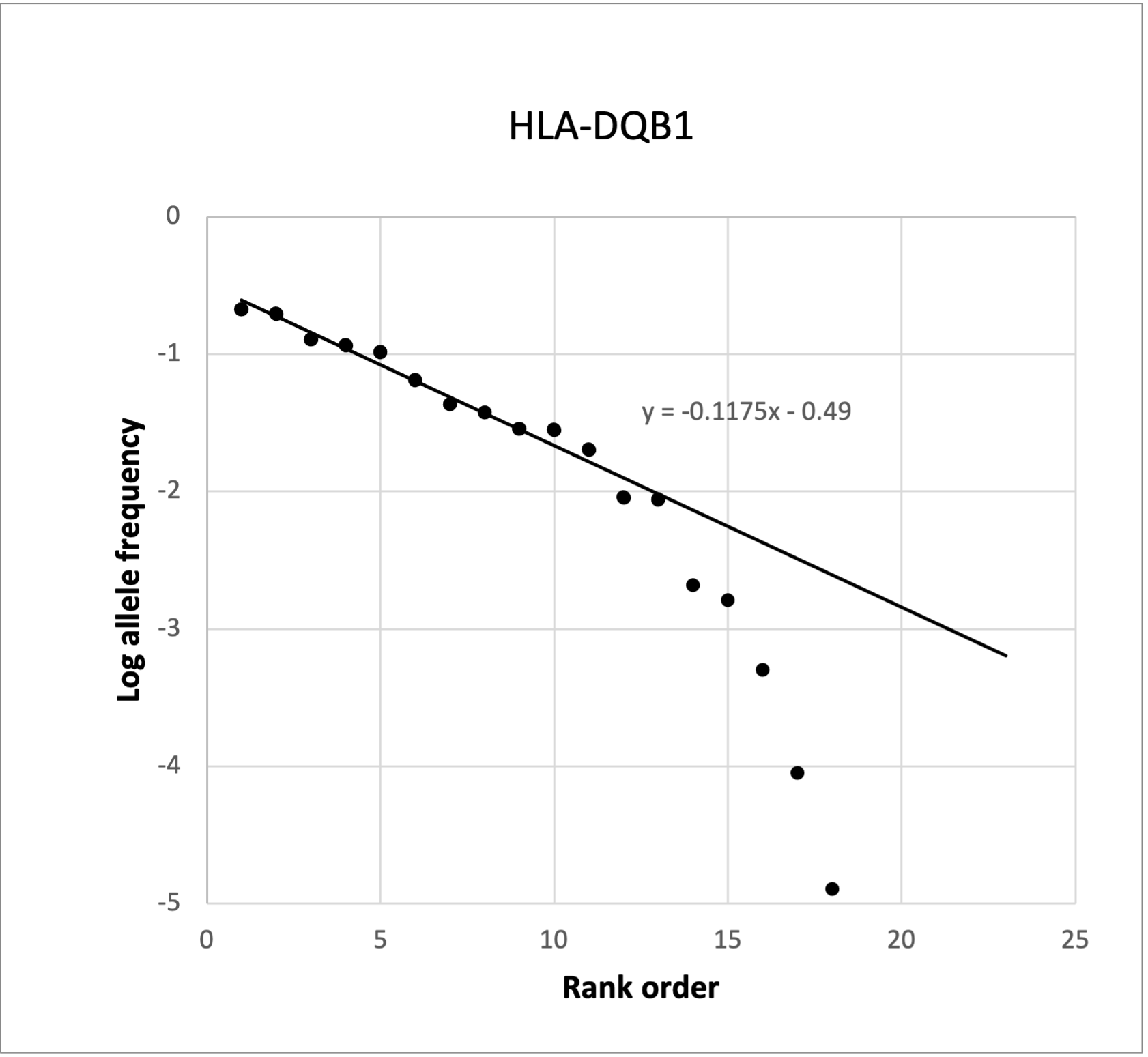}\hspace*{1cm}\includegraphics[scale=0.3]{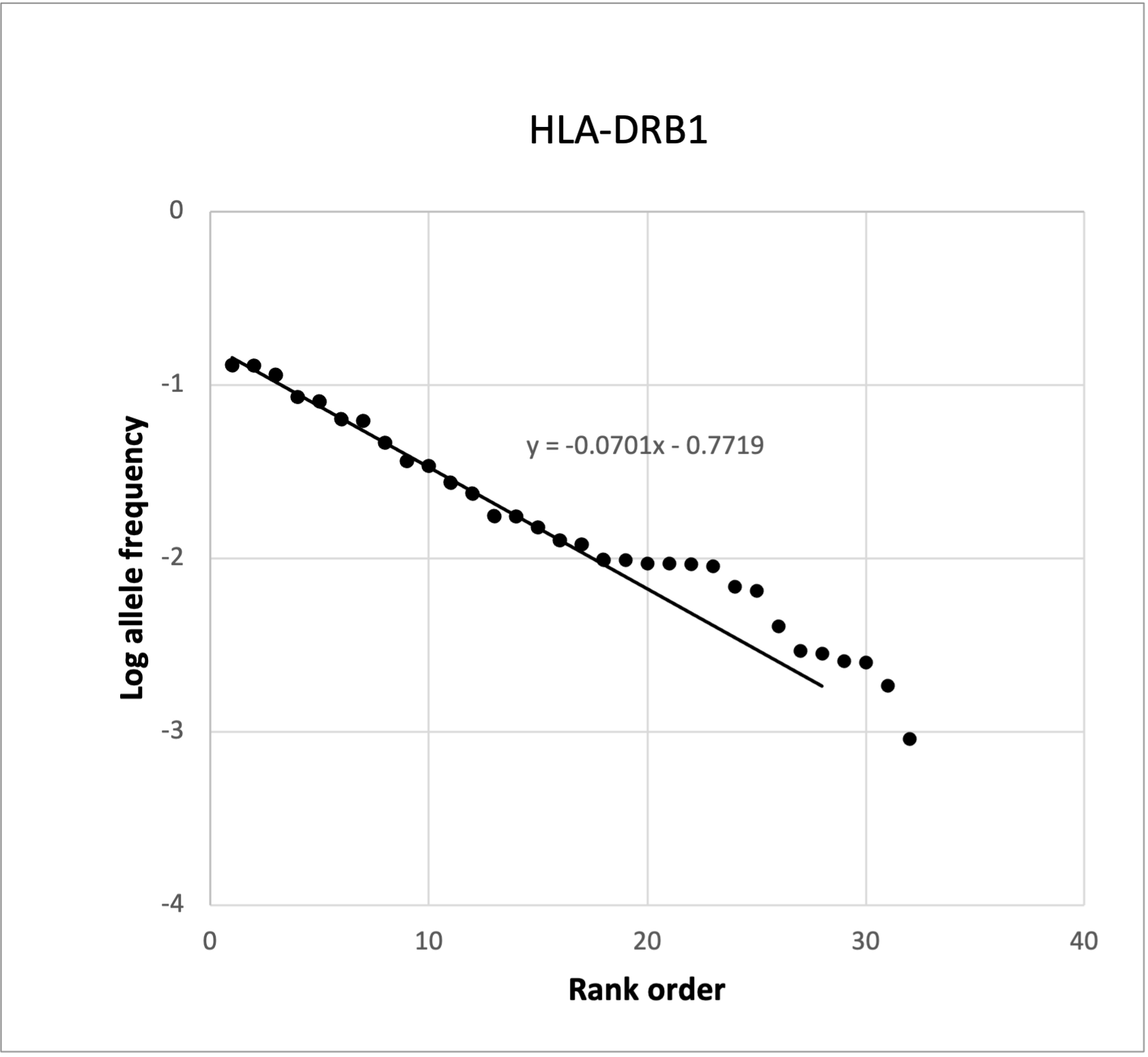}\hfill{}$ $

\caption{The data used for Fig.5 but plotted here on a linear scale for rank.
The data show remarkable linearity for frequencies of numerically
lowest rank.}
\end{figure}

The shapes of all five distributions in Fig.5 are similar to that
in Figure 2B in the Neher-Shraiman paper \citep{Neher2009-ou} (discussed
in 2.7 above). However, the two graphs plot different measurements.
The ordinate in the NS paper measures estimates of linkage disequilibrium
(LD) whereas it is allele frequency in Fig.5. The abscissa in the
NS model is crossovers per chromosome but is rank order in Fig.5.
Despite these differences, the two plots share commonality.

The starting point for the NS analysis is a group of polymorphic loci
\citep{Neher2009-ou}. Under Quasi-Linkage Equilibrium (QLE), the
potential for emergence of positive epistasis between allelic combinations
is frustrated by high rates of recombination. As the recombination
rate is lowered, there is a recombination rate below which allelic
combinations remain intact long enough for selection to switch to
the haplotype. Competition between the various haplotypes then follows
and one haplotype emerges by competitive exclusion.

We assume, as a starting point, that the abscissæ in Fig.5 do not
represent a range of recombination rates; rather there is a single
recombination rate that is below the critical value identified in
\citep{Neher2009-ou}. Instead, we regard those alleles to the right
of the respective discontinuities as incapable of generating positive
epistatic interactions. By contrast, the 137 alleles to the left of
the discontinuities do show positive epistatic interactions. We show
below that high-frequency alleles contribute extensively to numerous
haplotypes in both populations identified in Figs.4a,b. Their frequencies
are pulled up by selection on high-frequency haplotypes that contain
them. Recalling Eq.11, the fitness of these alleles has both epistatic
and additive components and the overall frequency of a particular
allele is the sum of the haplotype frequencies to which it contributes
is

\begin{equation}
f(A_{x})=\sum_{1}^{i_{t}}f(H_{A_{x}})_{i}+\sum_{i_{t}+1}^{i_{max}}f(H_{A_{x}})_{i}
\end{equation}
The frequency of allele \emph{x} is the sum of the frequencies of
haplotypes that carry the allele. These haplotypes are partitioned
between those of high frequency (that is, low rank) that show the
outcome of positive epistasis (first term on RHS of Eq.16) and those
of low frequency (high rank) that are neutral or deleterious (second
term on RHS of Eq.16). The composite allele frequencies are those
for haplotype rank $i$, and the transition point occurs at rank $i_{t}$.
We propose the data points in the five panels in Fig. 5 to the left
of the discontinuities are those whose overall frequency has both
components of Eq.11, whereas those to the right have no epistatic
component. Viewed in this light, the ordinates in Fig.5 are log frequencies, $\mbox{log}$$f(A_{x})$,
that include additive components and, in some cases, an additional
epistatic component.

We have built on this approach to gain an approximate measure of the
positive epistatic effect that provides the basis for additional selection
in high-frequency haplotypes. The results are shown in Fig.7. By
splitting the frequency of individual alleles into a component that
only arises in high-frequency haplotypes, we gain access to $A+E$
for a particular allele. The frequency gap between the two lines represents
an approximate measure of the positive epistasis, and this appears
as an increase in linkage disequilibrium. Convergence of the two lines
gives an assessment of the point at which positive epistasis ceases
to emerge, allowing us to identify the 137 alleles around which an
HLA network or cloud emerges.

\begin{figure}
\hfill{}\includegraphics[scale=0.3]{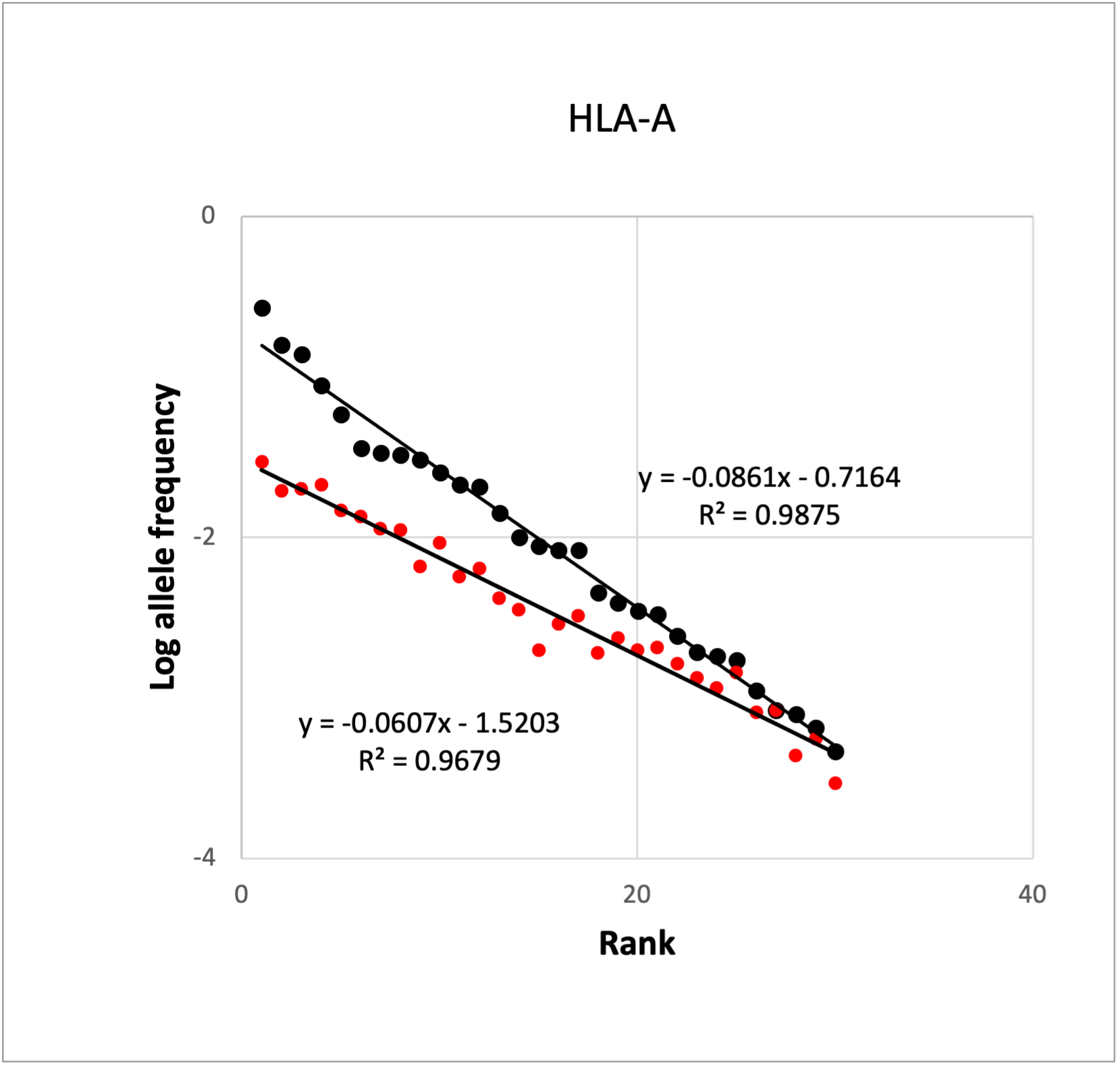}\hfill{}\includegraphics[scale=0.3]{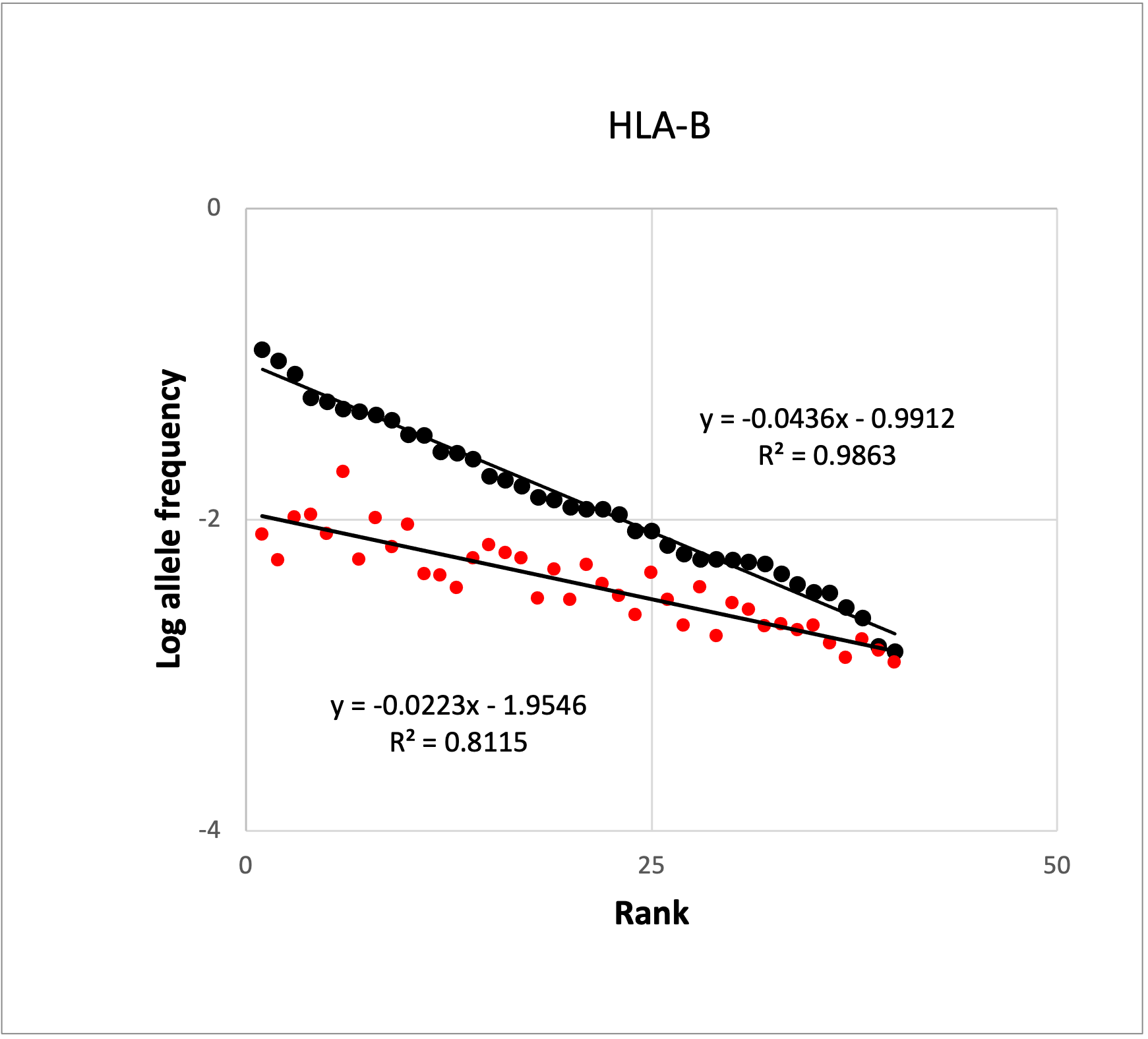}\hfill{}\includegraphics[scale=0.3]{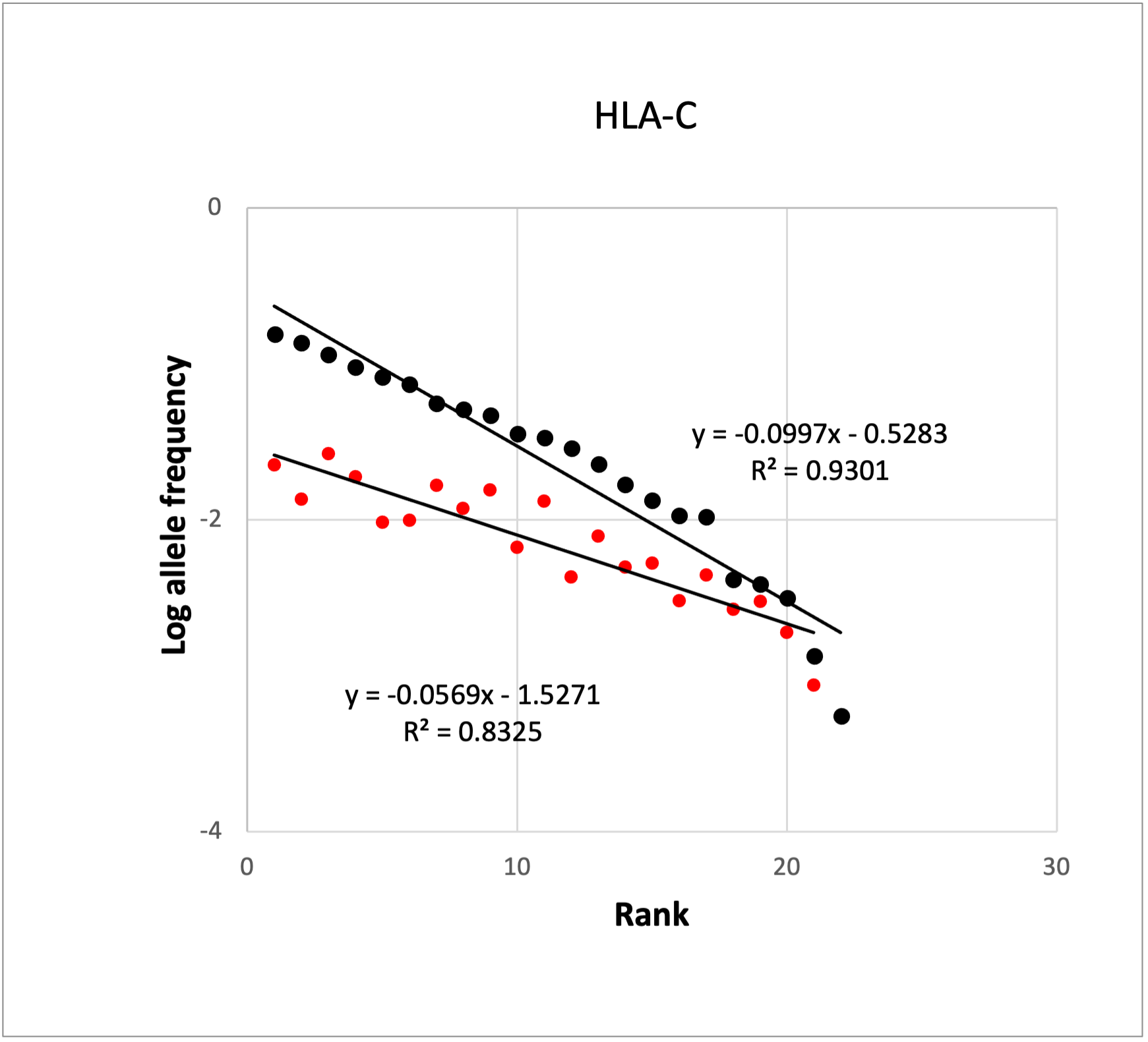}\hfill{}

\bigskip{}

\hspace*{3cm}\includegraphics[scale=0.3]{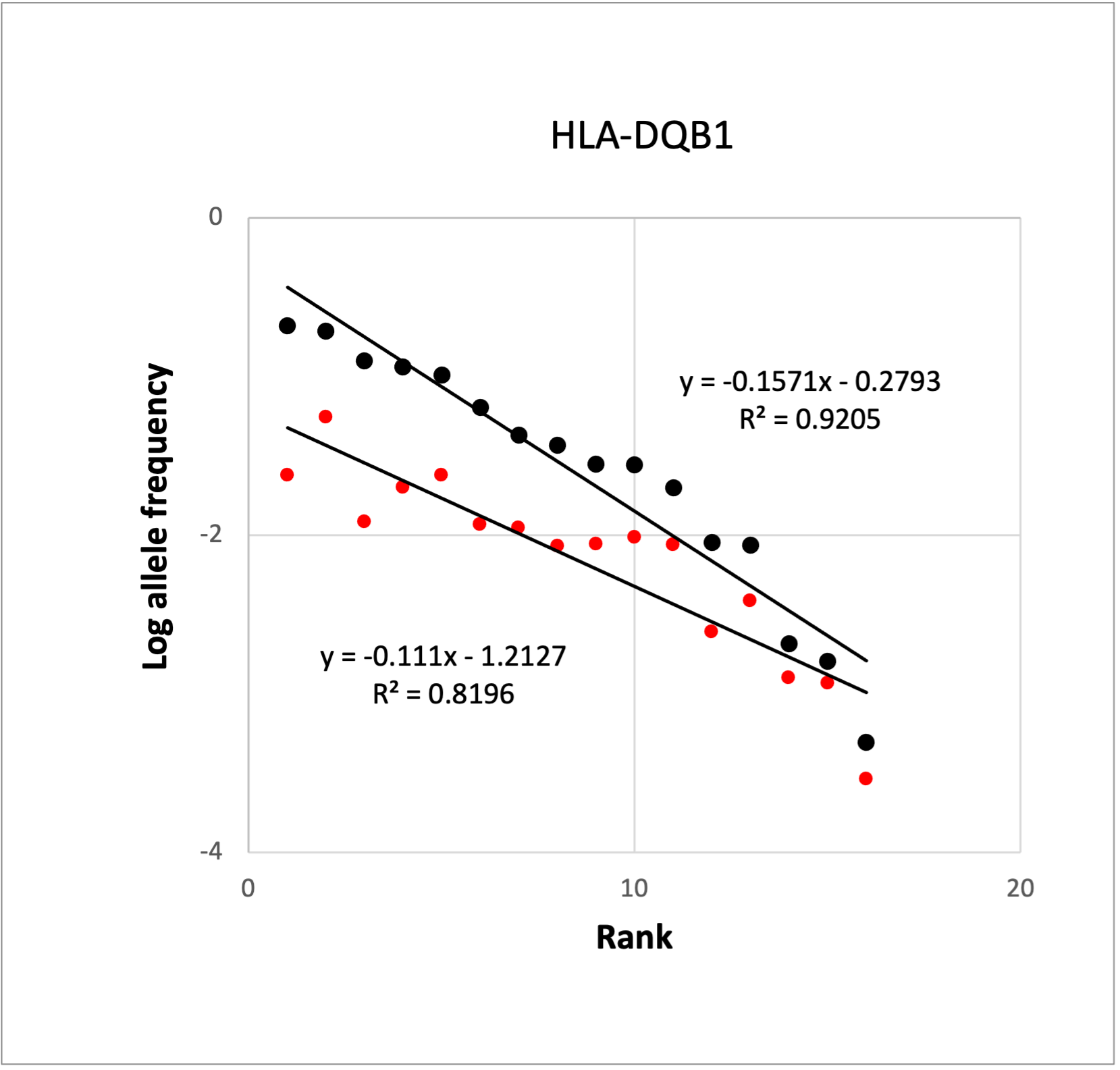}\hspace*{1cm}\includegraphics[scale=0.3]{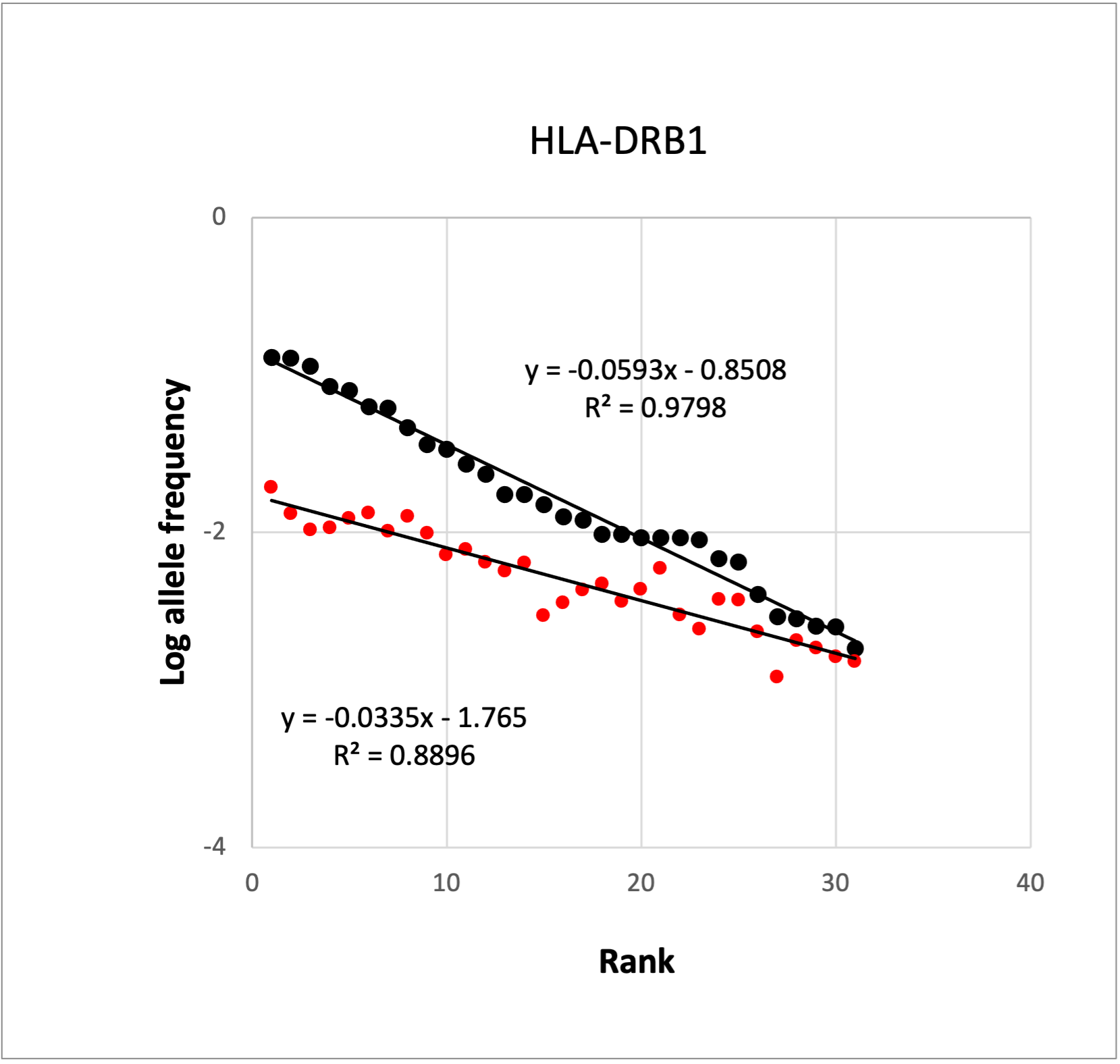}\hfill{}$ $

\caption{Data as in Fig.6 (black markers), limited to allele frequencies above
the discontinuities shown in Fig.5. Red markers indicate residual
allele frequencies after deduction of a frequency component attributable
to positive epistasis, \emph{E}, as defined in \citep{Neher2013}.
Somewhere between 9 and 12\% of the highest frequencies at each locus
is attributable to the heritable component, \emph{A}.}
\end{figure}

We are now in a better position to see why there should be a similarity
between the plots in Fig.5 and that in Figure 2B of Neher and Shraiman
\citep{Neher2009-ou}. The ordinate in the NS paper is log-linkage
disequilibrium (LD) and is log-frequency in Fig.5. The two are connected
because the highest allele frequencies reflect positive selection
on the haplotypes that contain them, and positive epistasis and low
recombination rates are the basis for the LD shown by major haplotypes.
The discontinuities in Fig.5 represent the point at which there is
no epistatic contribution to lower allele frequencies.

\subsection{Interim assessment I}

The evidence, so far, is consistent with two distinct haplotype populations
(Figs.4a,b). One population accounts for at least 80\% of the haplotype
census population, and we suggest this population is under different
degrees of positive selection. The second population consists of low-frequency
haplotypes that are either neutral or mildly deleterious. The allele
frequency data are consistent with the elevation of a cohort of 137
alleles to substantial frequencies due to positive epistasis at haplotype
level, leading these haplotypes to come under purifying selection.
We attribute the orderly spread, both of haplotype and allele frequencies,
as indicative of an HLA system whose components are held at intermediate
frequencies due to density-dependent effects on growth. The obvious
external agency for density-dependent host population regulation is
disease transmission.

We can also begin discussion of stability of haplotype distributions
as complex systems. The two layers, haplotype and allele, have some
properties of trophic layers but can be recognised as having different
general characteristics. The haplotype layer is characterised by a
large number of individual haplotypes (somewhere between 350 and 1850)
whose combination in diploids is continually broken up by segregation.
The number of effective alleles is much smaller per locus, with mutualism
restricted to specific allelic combinations. It is worth quoting May
\citep{May1972} at this point:
\begin{quote}
``Roughly speaking, this suggests that within a web species which
interact with many others (large \emph{C}) should do so weakly (small
$\alpha$), and conversely those which interact strongly should do
so with but a few species. This is indeed a tendency in many natural
ecosystems, as noted, for example, by Margalef$^{7}$: ``From empirical
evidence it seems that species that interact feebly with others do
so with a great number of other species. Conversely, species with
strong interactions are often part of a system with a small number
of species ...''{}''
\end{quote}
The reference to Margalef is page 7 in \citep{margalef1968}. The
sense of May's statement is improved by the insertion of a comma between
\emph{web} and \emph{species} in the first sentence. May's paper ``Will
a large complex system be stable?'' has been profoundly influential
(see introduction to \citep{Krumbeck2021}, for example). ``The notion
of stability referred to by May and these later works is that of asymptotic
linear stability of an equilibrium point'' \citep{Krumbeck2021}.
That notion is core to the approach in this paper, although it should
be noted that the haplotype cloud that emerges from the data will
not be that of a random matrix. At this point, we would note that
in the two trophic layers mentioned in the first paragraph of this
section, that containing the haplotypes have large connectedness (large
\emph{C}) and weak interactions (small $\alpha$), whilst the allele
layer has the opposite: low connectedness and strong interactions.

The remainder of this Results section is taken up with detailed evidence
to support the conclusions reached so far.

\subsection{Allele spreads across 5-locus haplotype distributions}

This subsection looks specifically at the spread of class I and class
II alleles across the NMDP Caucasian 5-locus haplotypes. The purpose
of the analysis is the better to understand allele distribution across
the haplotype board for single alleles identified from Fig.5 as
potentially under positive epistatic selection. Alleles at all five
loci were examined. These frequencies are taken from the slightly
truncated lists available from the NMDP website.

The data presented here are for four of the 137 alleles identified
earlier. The difference in rank, in all four panels, between two adjacent
appearances of an identified allele in the ranked list of HLA haplotypes,
is plotted linearly on the \emph{y}-axis against the rank position
of the first allele of the pair. Every data point represents a different
haplotype. The results are shown in Fig.8 for (a) A{*}02:01g, (b)
DRB1{*}07:01, (c) A{*}01:03g, and (d) DRB1{*}04:13 as representative
distributions. Panel (a) shows the most frequent class I HLA-A allele,
A{*}02:01g and (b) the most frequent class II HLA-DRB1 allele, DRB1{*}07:01.
These two alleles are components of many hundred haplotypes each.
That is, they are highly pleiotropic (for fuller discussion of pleiotropy,
see \citep{Zhang2023}). Those alleles under selection constitute
no more than the first 2000. Panels (a) and (b) therefore demonstrate
the extensive involvement of both alleles in haplotypes that are subject
to drift. The average rank difference is 5.3 for the top 5000 haplotypes
containing A{*}02:01g and 9.3 for DRB1{*}07:01. By contrast, the average
for haplotypes of ranks from 32,000 to 37,000 is 8.4 and 14.4 respectively.
This indicates a degree of interpolation at low frequencies by mutant
alleles that are individually rare but numerous in aggregate.

Panels (c) and (d) have an extended linear scale on the ordinate (x40
that in the first two bands). Allele A{*}01:03g (c) has rank 31 in
the HLA-A catalogue, which places it at the discontinuity in the HLA-A
panel of Fig.5. Allele DRB1{*}04:13 (d) has a rank of 75 and places
it in the portion of the HLA-DRB1 panel in Fig.5 whose alleles
show no epistatic component of fitness. These two lower panels are
characteristic of less frequent alleles, showing little or no presence
in the first 5000 alleles by frequency.

\begin{figure}
\subfloat[A{*}02:01g]{\hspace{1cm}\includegraphics[width=0.9\textwidth]{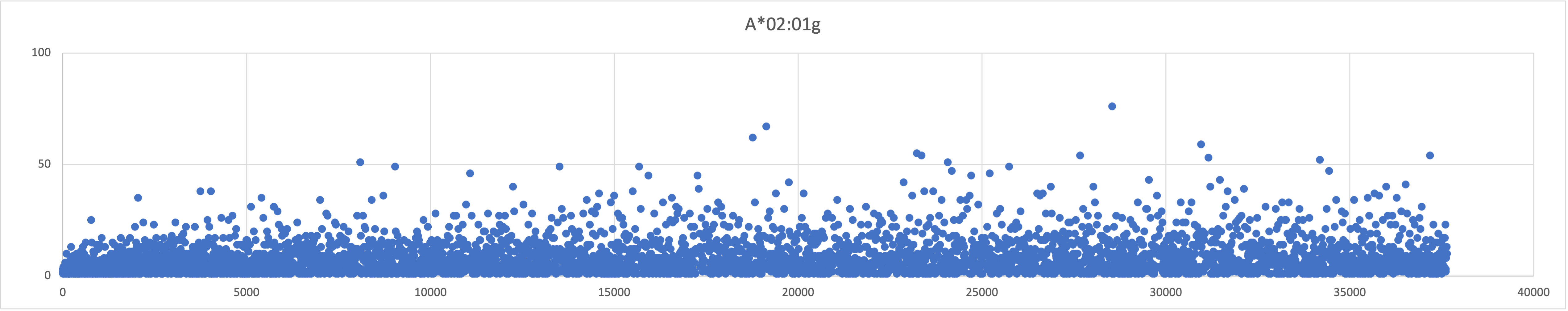}\hfill{}

}

\subfloat[DRB{*}07:01]{\hspace{1cm}\includegraphics[width=0.9\textwidth]{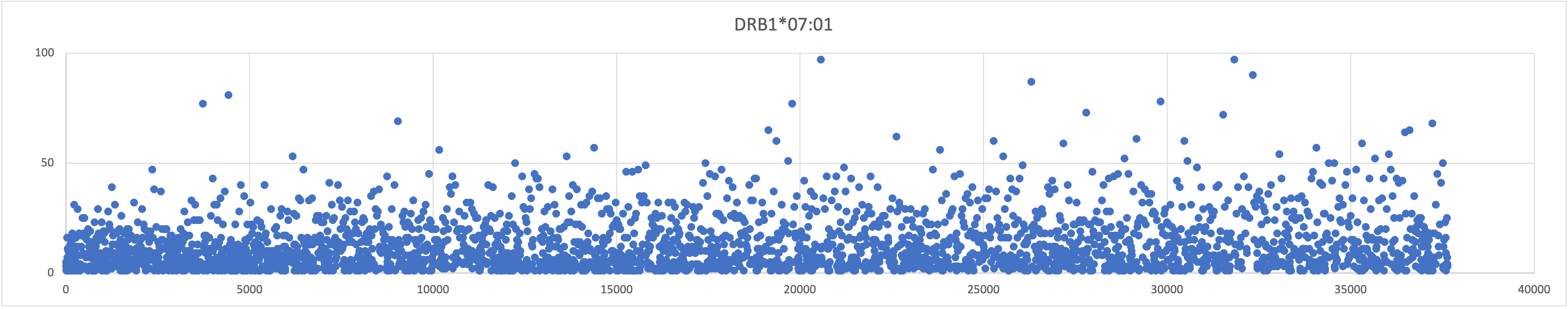}\hfill{}

}

\subfloat[A{*}01:03g]{\hspace{1cm}\includegraphics[width=0.9\textwidth]{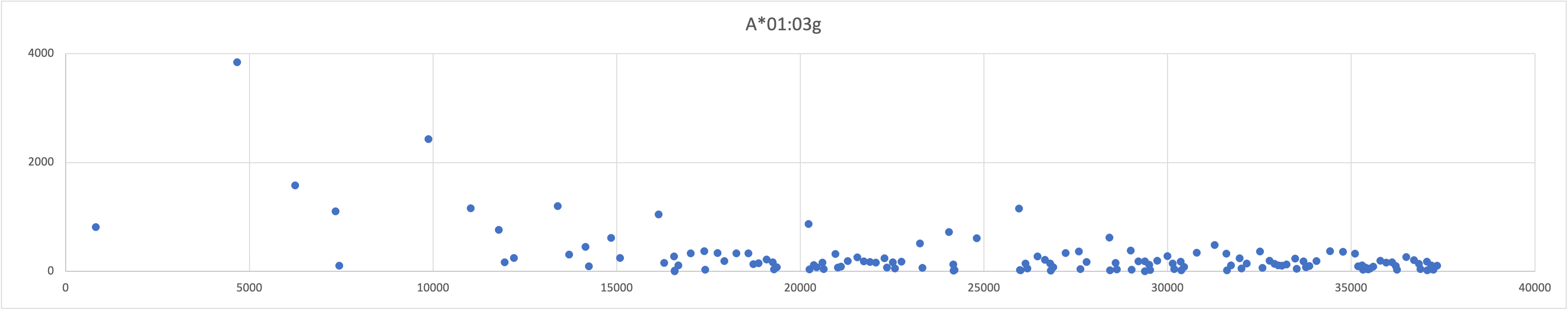}\hfill{}

}

\subfloat[DRB1{*}04:13]{\hspace{1cm}\includegraphics[width=0.9\textwidth]{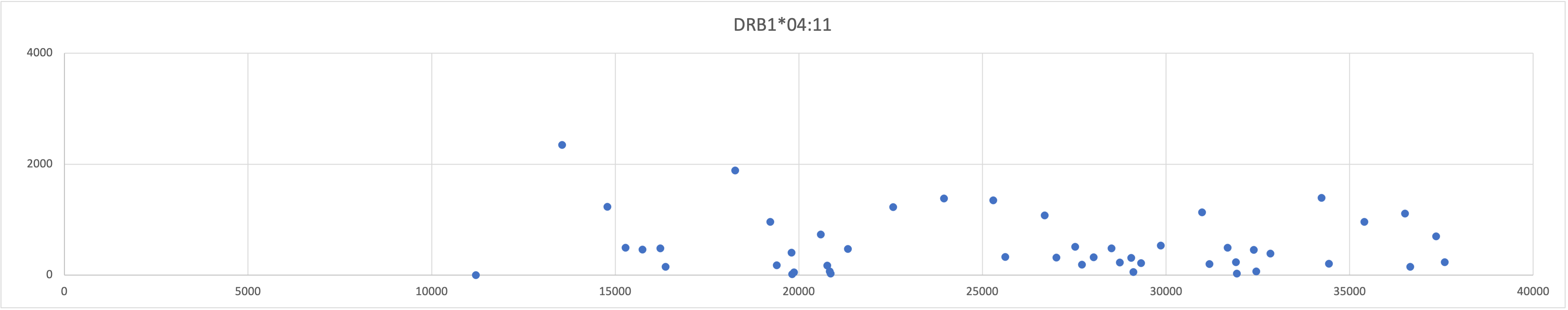}\hfill{}

}

\caption{Allele spreads across ranked haplotypes. Ranks run from 1 to 37,000
in all four panels.}

\end{figure}

The two major alleles contribute extensively to minor haplotypes,
certainly down to rank 37,000. We should recall that the minimum in
Fig.5(a) is located between ranks 1000 to 2000, emphasising that
the spectra in the panels in Fig.8 are dominated by the low end of
the frequency spectrum. Panel (c) for A{*}01:03g, which has the lowest
frequency of the HLA-A alleles above the discontinuity in Fig.5,
shows that the distance between successive allele appearances is initially
much greater than for a high-frequency allele such as A{*}02:01g,
before settling into a more regular pattern of rank separation.

The overall number of haplotypes by census population associated with
mutant alleles is <3\% of the total census population size, where
a mutant is an allele that is not one of the 137 alleles identified
earlier. Since the census haplotype population to the left of the
minimum in Fig.4a is \ensuremath{\approx}20\% of the total, most of
the haplotypes by census population in this 20\% do not contain mutants;
rather, they are permutations of the 137 alleles but without generation
of positive epistasis. This suggests, in turn, that the additive component
of fitness in most haplotypes is intrinsically low.

\subsection{Distribution of epistatic effects}

Fitness is higher in some haplotypes but not others due to the uneven
distribution of positive epistasis. This can be demonstrated by looking
at the frequencies of haplotypes derived from the top 5 class I alleles
(-A, -B, -C) by frequency. There are 125 possible permutations and
all 125 are found in the NMDP CAU catalogue. The results are shown
in Fig.9a. The red markers indicate the actual frequencies of the
125 class I haplotypes on a logarithmic scale. The black markers indicate
the calculated frequencies of the 125 possible class I haplotypes
obtained as the product of the actual frequencies of the individual
alleles. The calculated frequency values sit in a band that is approximately
one order of magnitude wide. By contrast, the actual frequencies span
approximately 5 orders of magnitude. One interpretation is that the
right-hand cluster of theoretical frequencies is characterised by
HLA class I haplotypes that are of low frequency and not subject to
selection. The haplotypes of interest are the 21 haplotypes on the
left-hand side. These are under positive selection if assumptions
about steady-state or quasi-steady-state nature of haplotype frequencies
are accepted. These 21 haplotypes are still spread over two orders
of magnitude and reflect steady-state fitnesses accordingly.

The conclusion of interest is that randomising the alleles in the
class I haplotypes of highest frequencies uncouples the positive epistasis
seen in a minority of allelic combinations. It demonstrates at first
hand the cost of the recombinational load.

\begin{figure}
\hspace{1.5cm}\subfloat[Results for randomisation of the highest frequency class I alleles
(top 5 alleles for -A, -B, and -C loci), generating 125 possible class
I haplotypes. All haplotypes are represented in the NMDP sample. Calculated
haplotype frequencies (black); actual haplotype frequencies (red).
The plot shows that randomisation of alleles has limited effect on
calculated haplotype frequencies. Actual haplotype frequencies, by
contrast, are widely spread, indicating the existence of a recombinational
load.]{\includegraphics[scale=0.3]{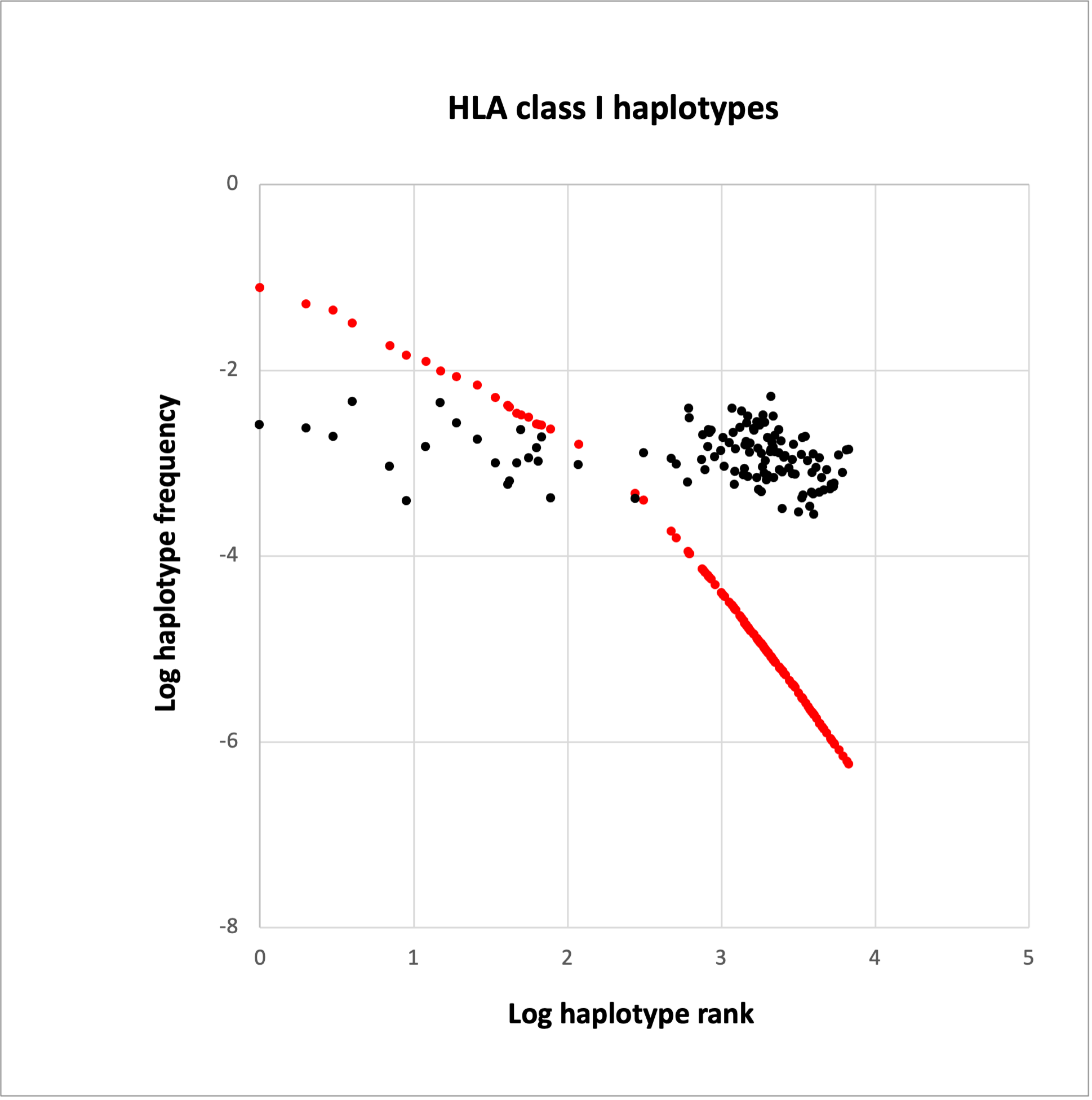}}\subfloat[Results for randomisation of the highest-frequency class I and class
II haplotypes (top 10 each), generating 100 possible 5-locus haplotypes.
Calculated 5-locus haplotype frequencies (black); actual 5-locus haplotype
frequencies (red). The plot shows that randomisation of class I and
II haplotypes has limited effect on calculated 5-locus haplotype frequencies,
but a marked effect on actual haplotype frequencies. The plot shows
that randomisation of alleles has limited effect on calculated haplotype
frequencies. Actual haplotype frequencies, by contrast, are widely
spread. The effect is less marked than in (a).]{\hspace{1cm}\includegraphics[scale=0.3]{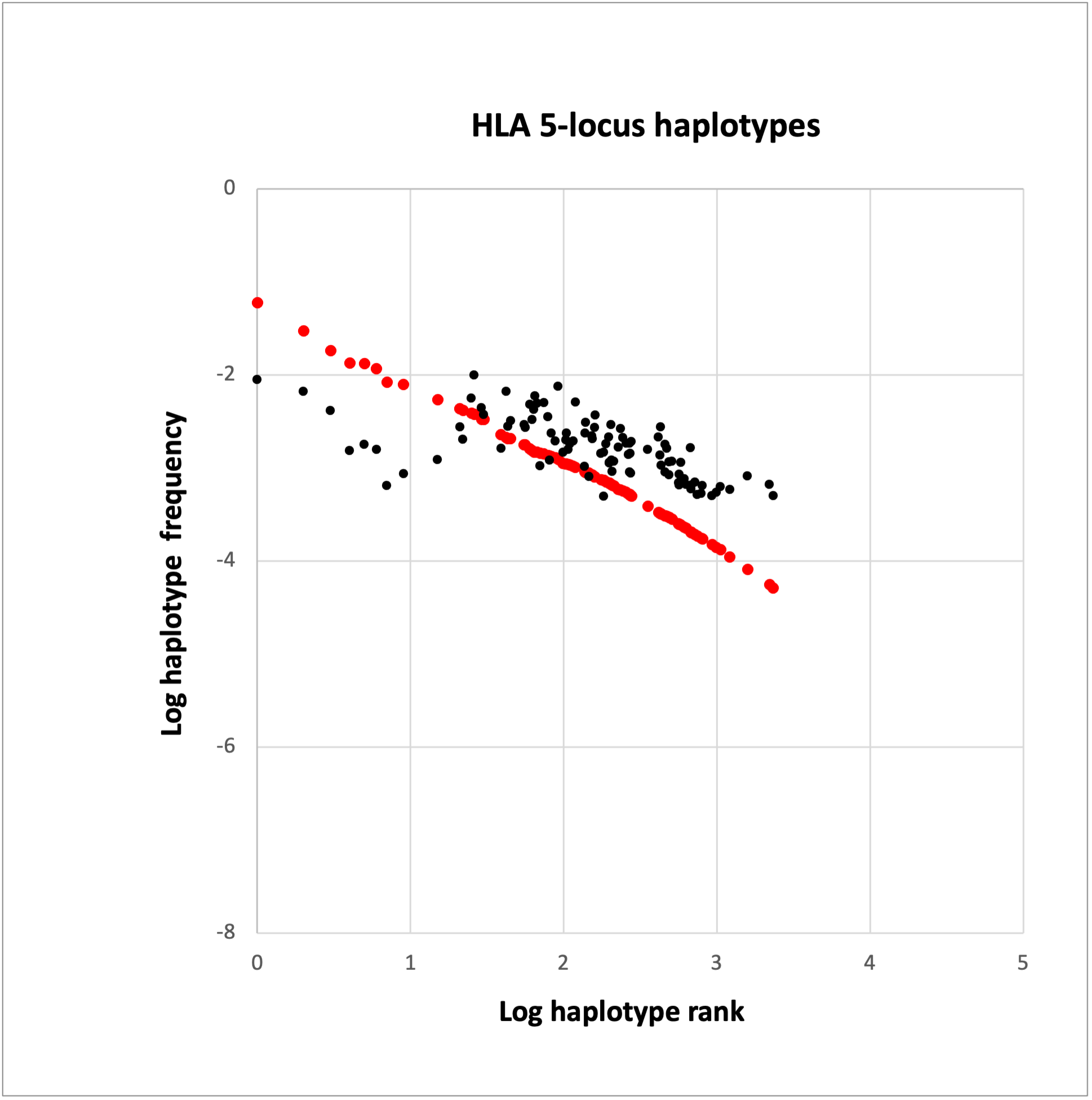}

}

\caption{Plots of log haplotype-frequency against log haplotype-rank. }
\end{figure}

A similar effect can be seen in combinations of class I and II haplotypes
(Fig.9b). This figure shows the outcome when the top 10 haplotypes
of each class are used. These give an overall set of 100 possible
5-locus haplotypes, again all of which are in the NMDP Caucasian datasets.
The spread of actual haplotype frequencies is noticeably smaller,
at three orders of magnitude. This plot of 5-locus haplotypes can
be mapped directly to Figs.4a and b, and covers the range of haplotypes
under selection. This should not be surprising, since the contributing
class I and II haplotypes are all under selection themselves. Despite
this preselection, the evidence suggests that randomisation of class
I and II haplotypes generates a range of steady-state fitnesses, subject
to qualifications that have already been discussed.

It should be noted that the calculated frequencies in both Figs.9a
and 9b reflect constituent allele frequencies that are elevated through
their presence in haplotypes showing positive epistasis. The calculated
frequencies for haplotypes using allele frequencies that are devoid
of any involvement in epistatically-selected haplotypes would be lower.

\subsection{Further evidence of linkage}

The data in the previous section provide evidence that the positive
epistasis evident in high-frequency HLA haplotypes is not simply due
to random combinations of high-frequency alleles. Rather, the epistasis
arises from specific combinations. This is no surprise. Evidence published
in 2013 \citep{Penman2013-eu} provided evidence of non-random association
of HLA class I and II alleles, citing work on two relatively isolated
populations, the Burusho population of Pakistan and the Hutterite
population of South Dakota in the US \citep{Penman2013-eu}. An important
part of that paper addressed the issue of recombination between class
I (beta block) and class II (delta block) alleles, since these are
separated by the gamma block, with a known possibility of recombination between the
blocks. Evidence of coupling between B and DRB1 alleles supported
a model in which selection on discrete allelic combinations was maintained
by pathogen interactions with hosts.

We tested the ability of the much larger NMDP Caucasian dataset for
evidence of long-range associations, focusing here on the results
for HLA-B\ensuremath{\sim}DRB1. Fig.10 shows the frequency of HLA-B\ensuremath{\sim}DRB1
pairs plotted against rank, both on logarithmic co-ordinates. It has
the same broad shape as the 5-locus haplotype distribution of Fig.3, but the linear portion of the distribution is limited to the first
\ensuremath{\approx}150 pairs by rank, out of a possible number of
1240 ($=40\times31$) pairs. Possibly 250 pairs in total, or 20\%,
could be considered as being under selection.

\begin{figure}
\hspace{5.1cm}\includegraphics[scale=0.3]{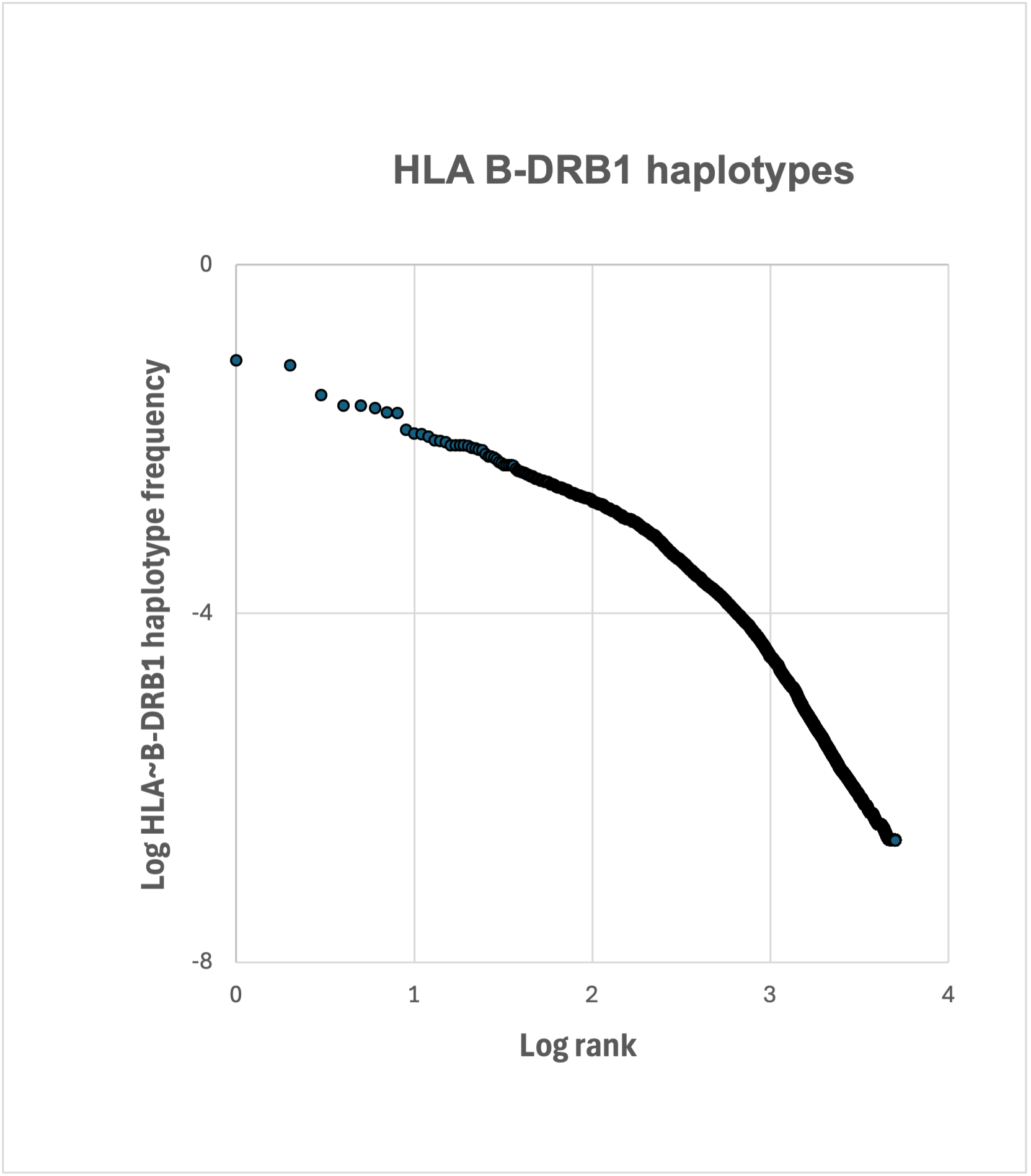}

\caption{Frequencies of Caucasian HLA B-DRB1 allele pairs plotted against rank
on logarithmic co-ordinates}
\end{figure}

A frequency plot of the top 18 HLA-B alleles by frequency against
the top 15 HLA-DRB1 alleles is shown in Fig.11, with coloured bars
indicating their frequencies. The frequency associated with each allele
pair is itself an aggregate, because the contributing alleles are
pleiotropic. The data indicate two major pairings, B{*}08:01g/DRB1{*}03:01g
and B{*}07:02/DRB1{*}15:01, and about 10 moderate pairings, out of
the \ensuremath{\approx}250 under selection.

\begin{figure}
\includegraphics[scale=0.6]{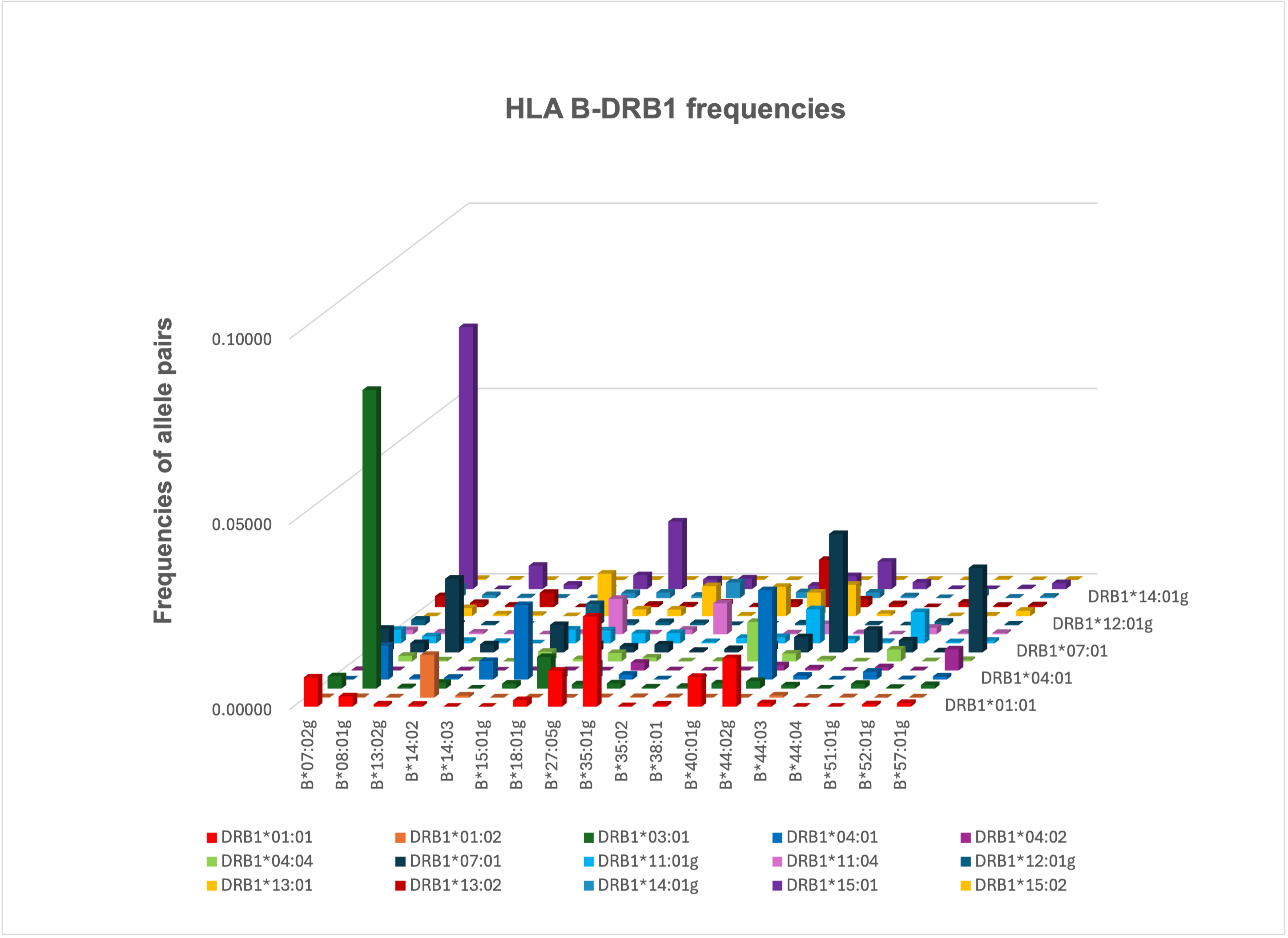}

\caption{Frequency distribution of HLA-B\ensuremath{\sim}DRB1 pairs; The two
loci are separated by a potential recombination region in the wider
HLA locus. The data suggest the region has the ability to set recombination
at a sufficiently low level for selection on positive epistasis
and linkage disequilibrium.}

\end{figure}

The two major allele pairs account for \ensuremath{\approx}8\% and
\ensuremath{\approx}6.9\% of the HLA-B\ensuremath{\sim}DRB1 population
respectively. Common HLA-B\ensuremath{\sim}DRB1 pairs are associated
with high-frequency 5-locus-haplotypes.

\subsection{Interim assessment II}

The data presented so far provide evidence of selection on haplotypes,
with major haplotypes benefiting from positive epistasis. This epistasis
arises from specific interactions between the 137 alleles from the
five transplantation loci that show demonstrable effects on their
frequencies from an epistatic component. There are \ensuremath{\approx}11.1
million possible permutations of the 137 alleles of which 350-1850
are under selection. The precise boundary is difficult to determine
because the tail of those under selection overlaps with the population
behaving neutrally or under mild negative selection. Those under selection
account for \ensuremath{\approx}80\% of the census population, whilst
further permutations of the 137 core alleles account for a further
17\%. There is then the remaining 3\% of haplotypes by census that
represent haplotypes with one or more alleles that lie outside the
137. These alleles are mutations. Many of the haplotypes that are
not under selection may have functional value, not least in diploids,
by interposing further hurdles to avoidance of presentation \citep{Klitz2012}.

The picture that emerges is a core network or community of highly
related molecular species. The high level of relatedness directly
reflects the pleiotropic properties of the higher frequency alleles.

\subsection{Mapping core HLA class I alleles onto HLA supertypes}

A key consideration in considering community structure is its stability.
Of direct interest, therefore, is evidence that some major HLA alleles
show longevity that exceeds the generation times of most pathogens
by orders of magnitude.

A major computational analysis of 10,956 distinct HLA class I allele
DNA sequencess (HLA-A, 3489; -B, 4356; and -C, 3111) was published
in 2017 \citep{Robinson2017-ze}. The data pointed to two origins
for the variation between alleles at each of the three loci: single
nucleotide polymorphisms (SNPs) and recombinational events. Alleles
with SNPs differ by point mutation from older, more common alleles.
Removal of these SNP-carrying alleles from the set of 10,956 revealed
a residue of 1171 allele types related to each other by recombination
(-A, 236; -B, 775; -C, 160). Removal of these alleles, in turn, revealed
a set of 42 core allele types (-A, 11; -B, 17; -C, 14) that represent
all functionally significant variation in exons 2 and 3 that cannot
be derived by recombination events and point mutation. They are regarded
as older in their origins than either the SNP alleles or recombinant
alleles, and approximately half of them by census fraction are consistent
with Denisovan or Neanderthal origins.

An older classification of HLA class I alleles has been into supertypes
\citep{Sette1999-ry,Sidney2008-dk}. A recent update of supertype
classification uses an approach based on structural similarity \citep{Shen2023-bk}.

These two classifications can be combined with the class I allele
frequency distributions shown in Fig.5 to identify the relationship
between the antiquity, the lack of relatedness, and the frequency
of particular alleles. The results are shown in Figs.12-16. Fig.12 shows the principal supertypes identified in \citep{Shen2023-bk}
as headers of each column, and the ordinate shows the rank order of
each allele. Blue represents core alleles as identified in \citep{Robinson2017-ze}
of potentially Neanderthal/Denisovan origin, orange represents remaining
core alleles, and green represents alleles of more recent origin that
are related to the core alleles. Two additional entries in the data
for class I HLA-A alleles for ranks 32 and 33. Gaps at the lower levels
of class I HLA-C alleles reflect the absence of these alleles in the
supertype catalogue list. All three class I loci are represented at
the highest allele frequencies by core alleles as identified in \citep{Robinson2017-ze},
but there are subtle differences.

\begin{figure}
\hspace{3cm}\includegraphics[scale=0.3]{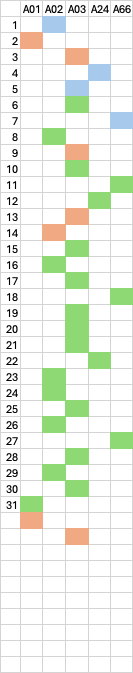}\hfill{}\includegraphics[scale=0.3]{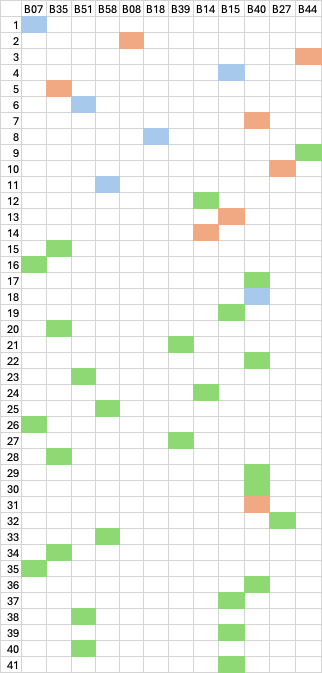}\hfill{}\includegraphics[scale=0.3]{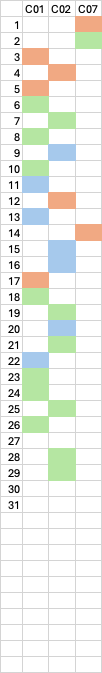}\hfill{}

\caption{}
\end{figure}

We have visualised frequency distributions for each supertype in Figures
14-16.

\begin{figure}
\hfill{}\includegraphics[height=7cm]{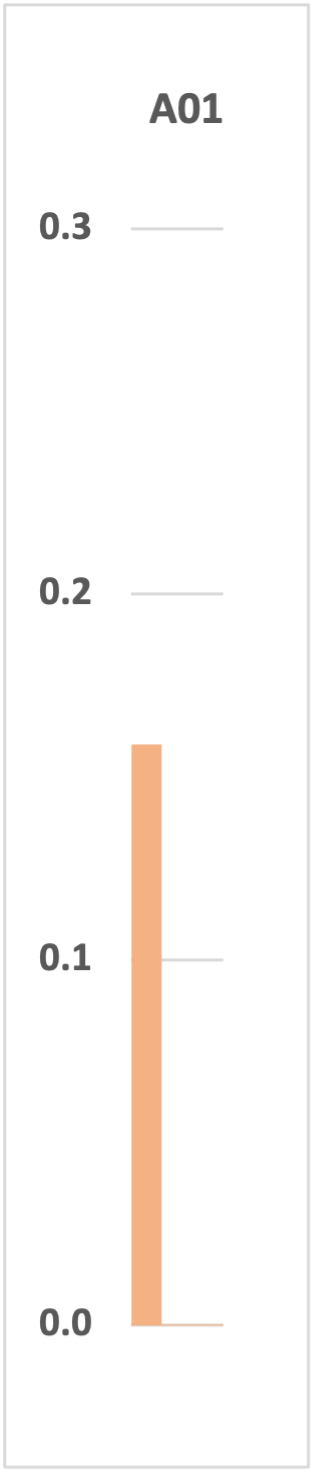}\hspace{0.5cm}\includegraphics[height=7cm]{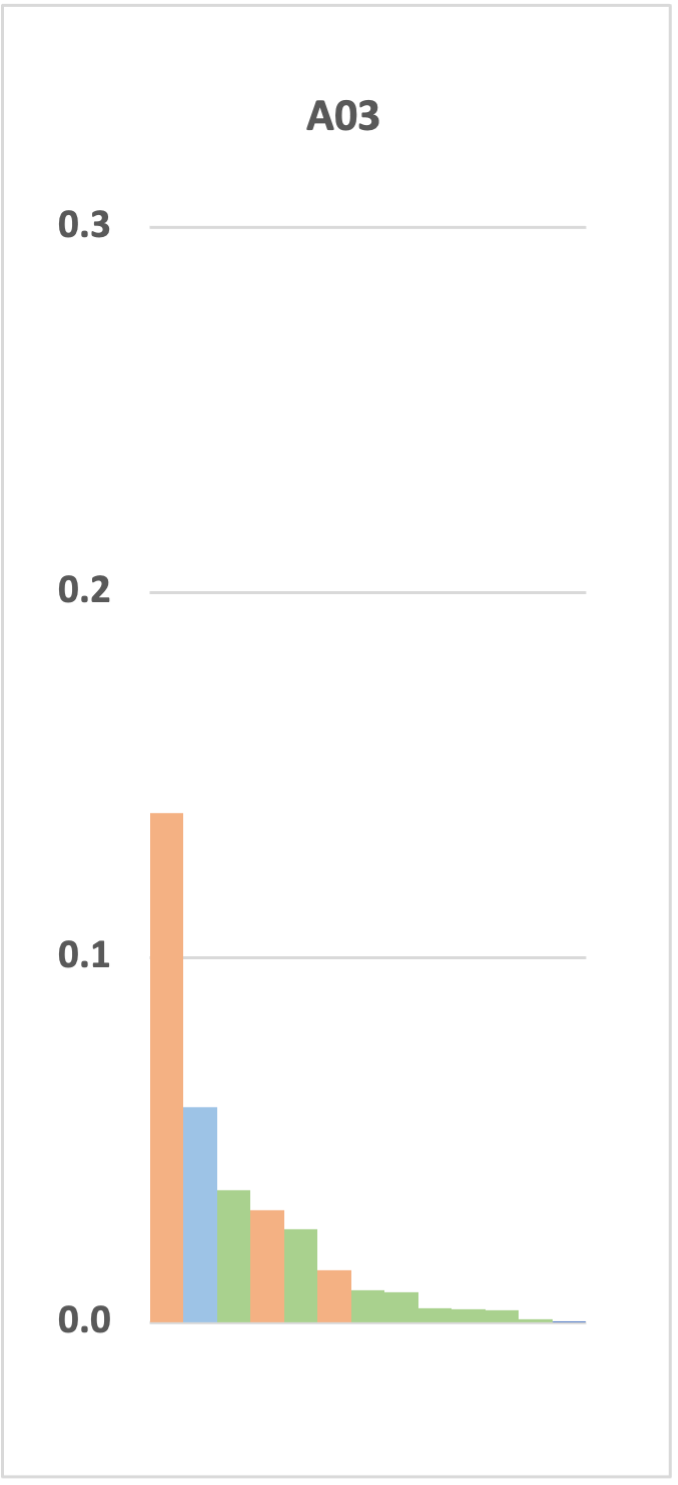}\hspace{0.5cm}\includegraphics[height=7cm]{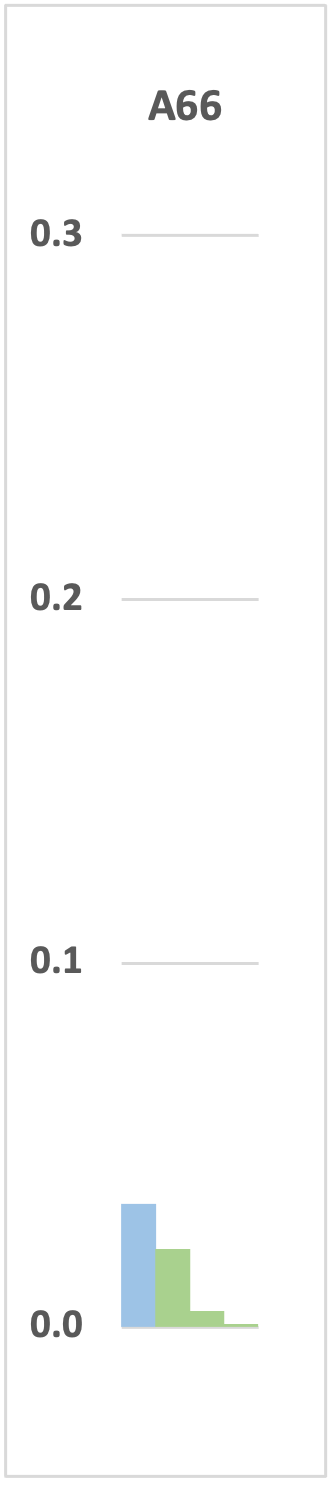}\hspace{0.5cm}\includegraphics[height=7cm]{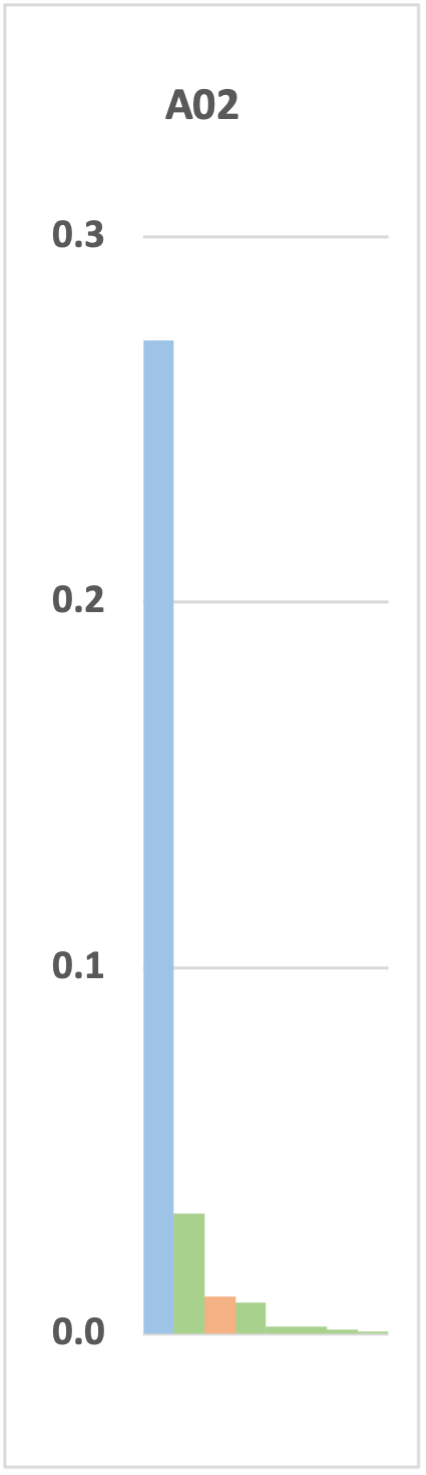}\hspace{0.5cm}\includegraphics[height=7cm]{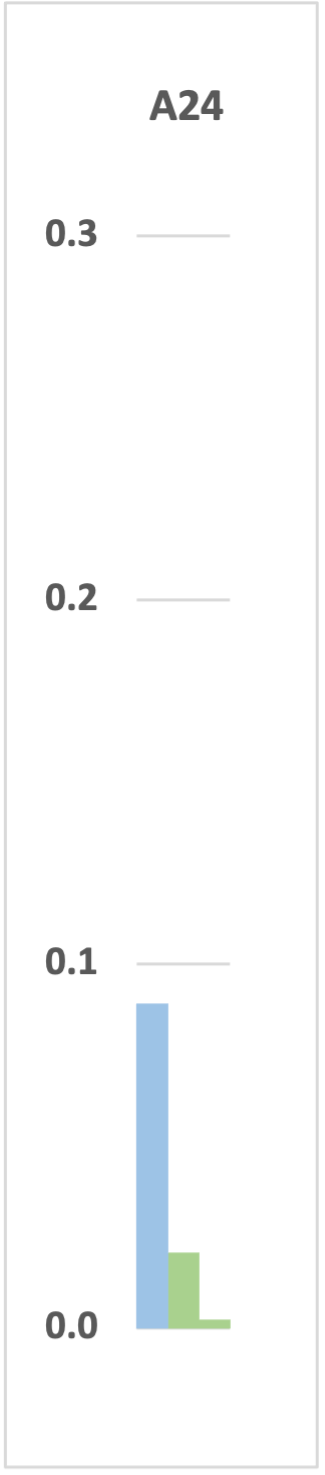}\hfill{}

\caption{Frequencies of major \emph{HLA-A} allele distributed into their respective
supertypes (as defined in \citep{Shen2023-bk}). Blue represents core
alleles as identified in \citep{Robinson2017-ze} of potentially Neanderthal/Denisovan
origin, orange represents remaining core alleles, and green represents
alleles of more recent origin that are related to the core alleles. }
\end{figure}

\begin{figure}
\hfill{}\includegraphics[height=7cm]{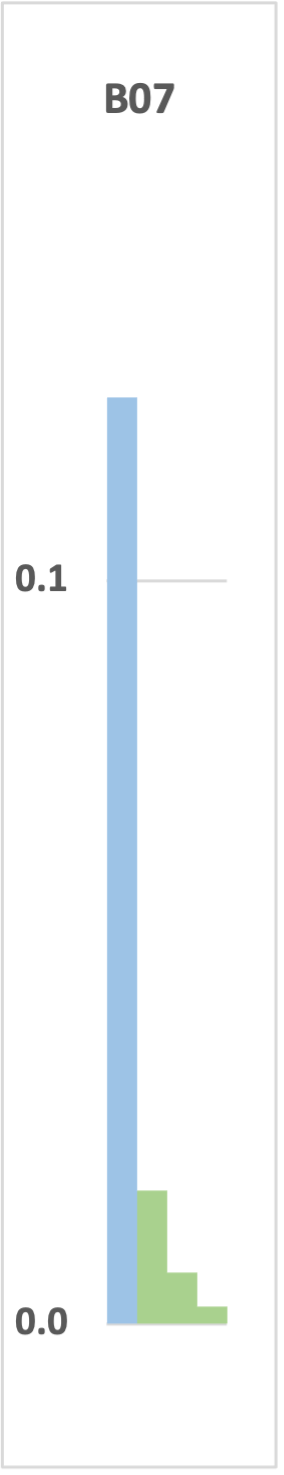}\hspace{0.5cm}\includegraphics[height=7cm]{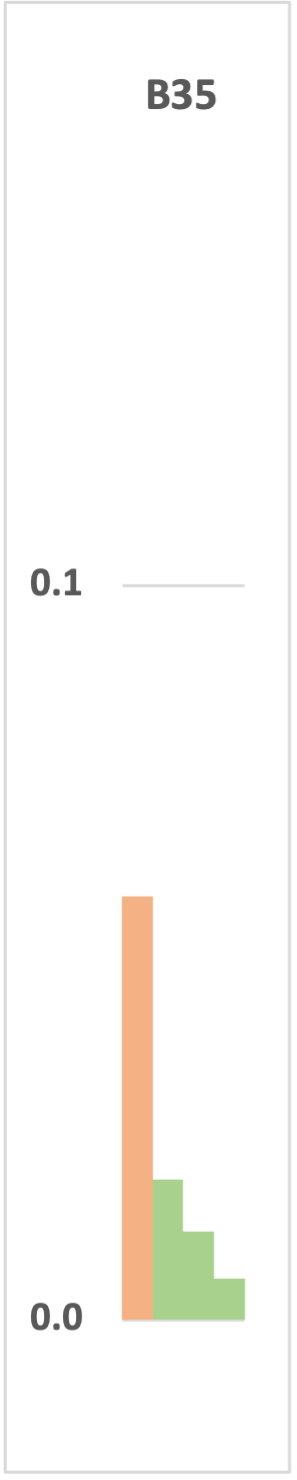}\hspace{0.5cm}\includegraphics[height=7cm]{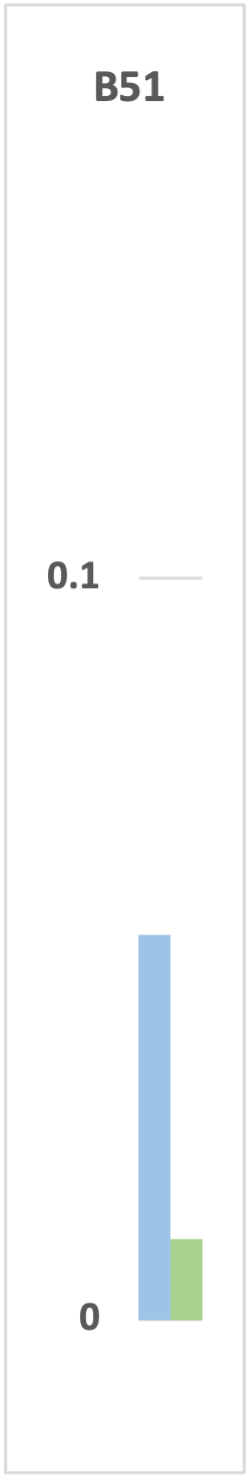}\hspace{0.5cm}\includegraphics[height=7cm]{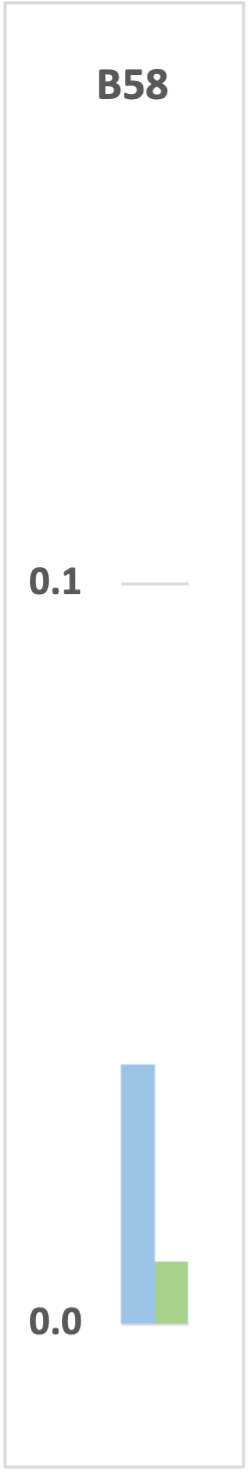}\hspace{0.5cm}\includegraphics[height=7cm]{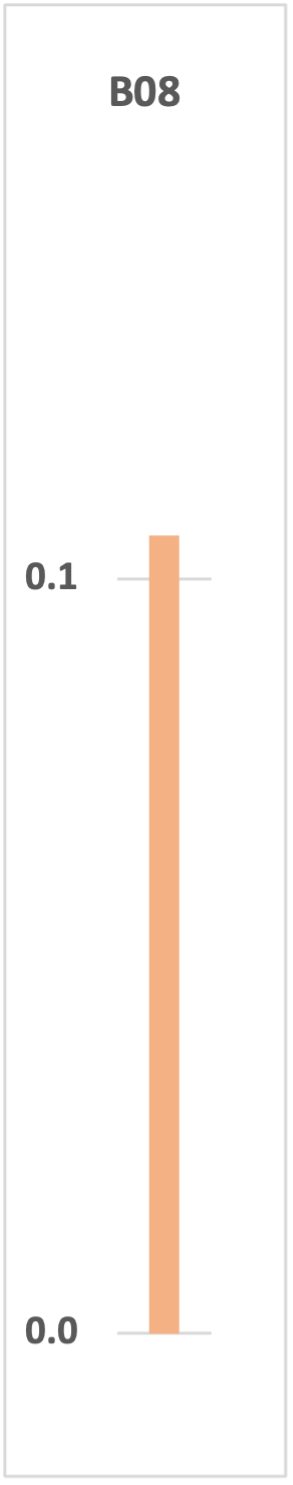}\hspace{0.5cm}\includegraphics[height=7cm]{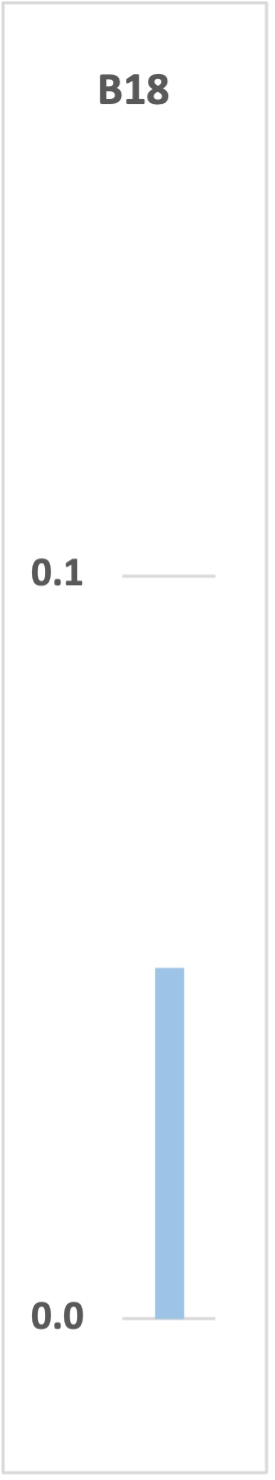}\hspace{0.5cm}\includegraphics[height=7cm]{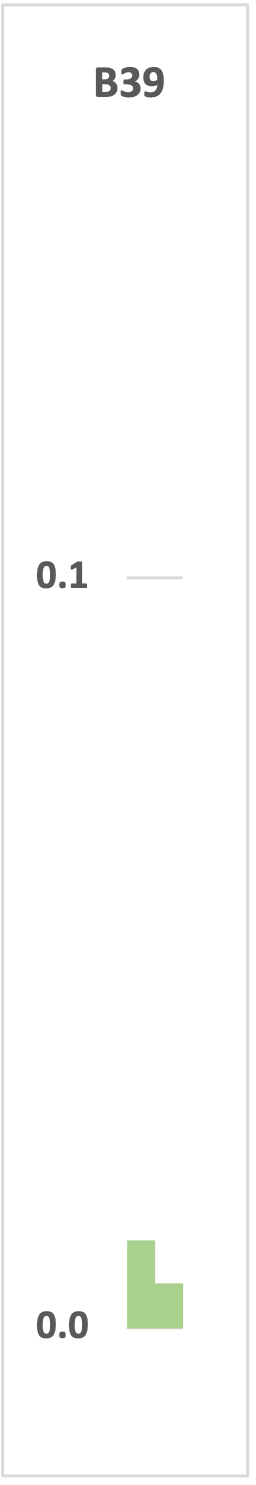}\hfill{}

\bigskip{}

\hfill{}\includegraphics[height=7cm]{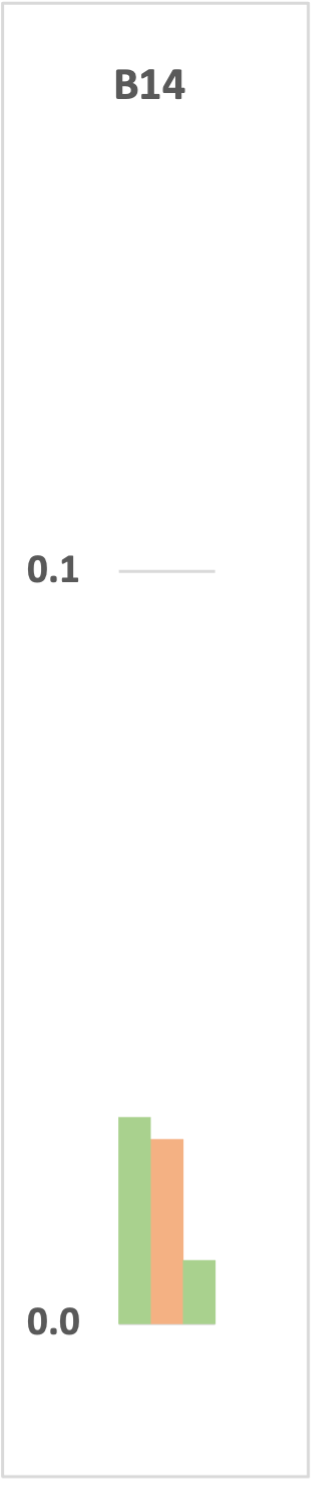}\hspace{0.5cm}\includegraphics[height=7cm]{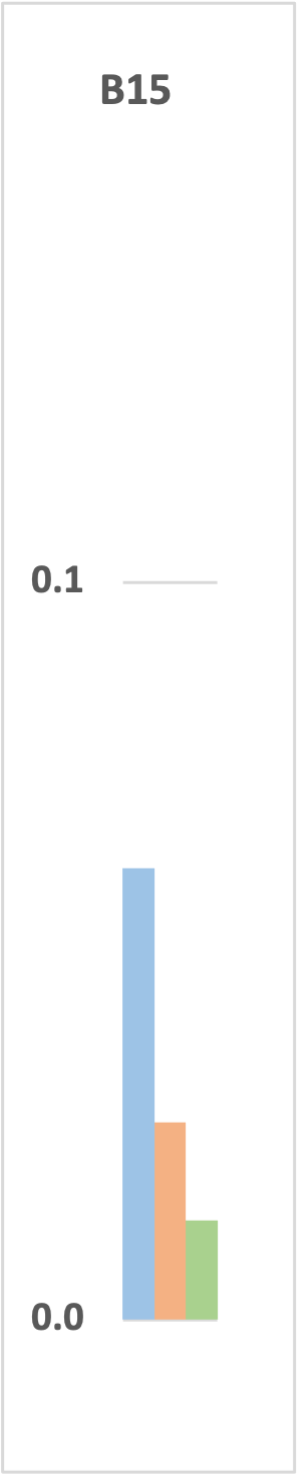}\hspace{0.5cm}\includegraphics[height=7cm]{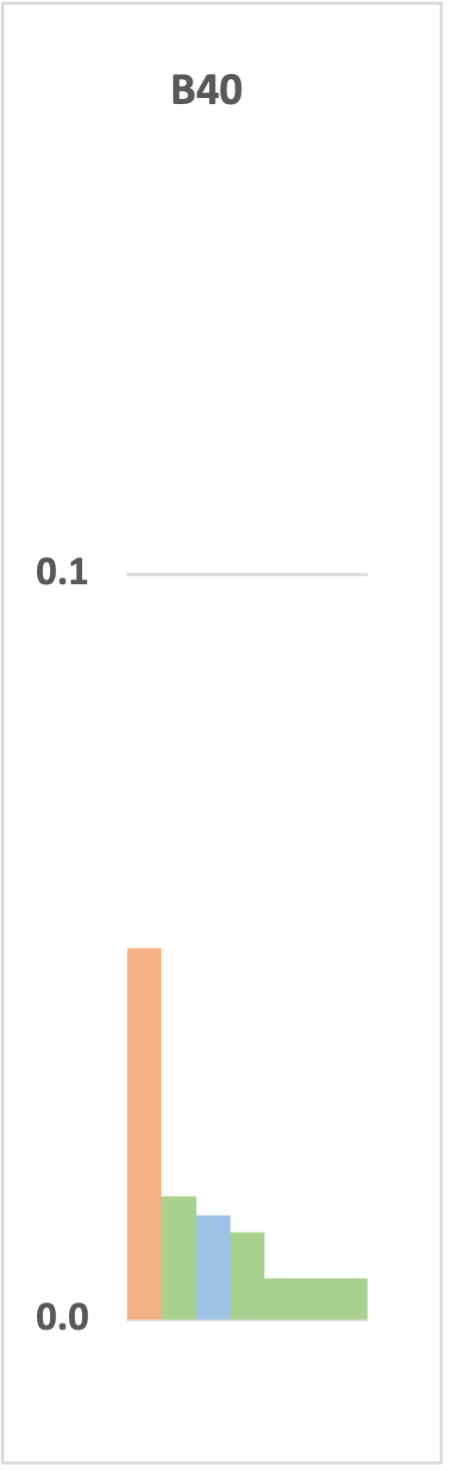}\hspace{0.5cm}\includegraphics[height=7cm]{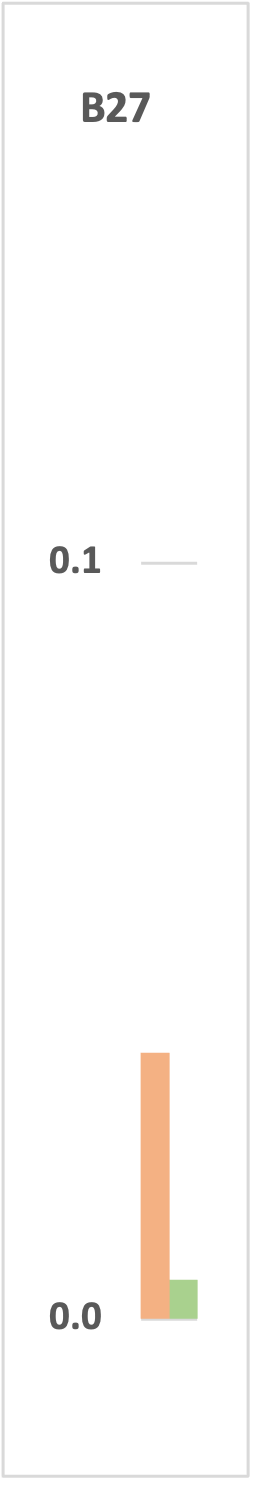}\hspace{0.5cm}\includegraphics[height=7cm]{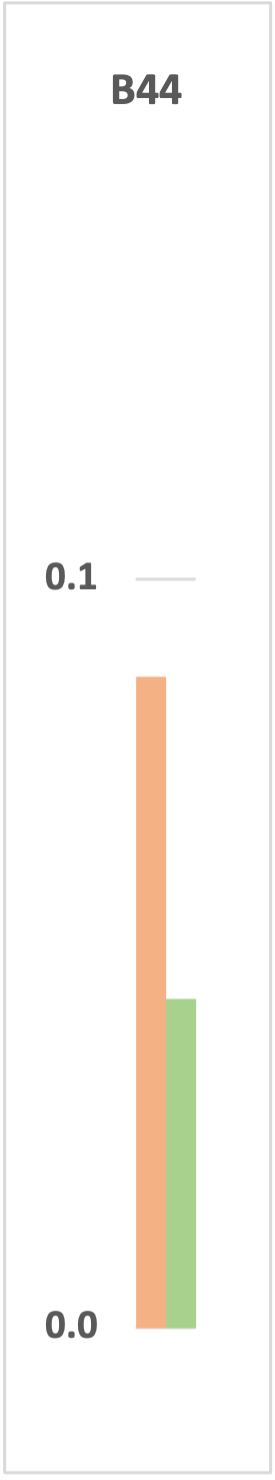}\hfill{}

\caption{Frequencies of major \emph{HLA-B} allele distributed into their respective
supertypes (as defined in \citep{Shen2023-bk}). Blue represents core
alleles as identified in \citep{Robinson2017-ze} of potentially Neanderthal/Denisovan
origin, orange represents remaining core alleles, and green represents
alleles of more recent origin that are related to the core alleles. }
\end{figure}
\begin{figure}
\hfill{}\includegraphics[height=7cm]{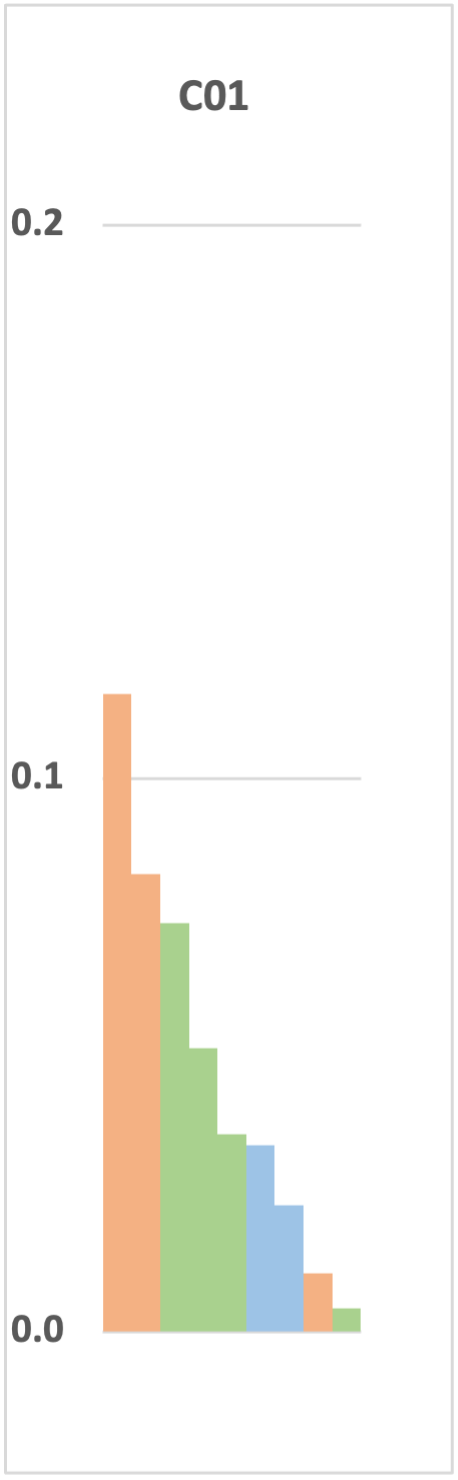}\hspace{0.5cm}\includegraphics[height=7cm]{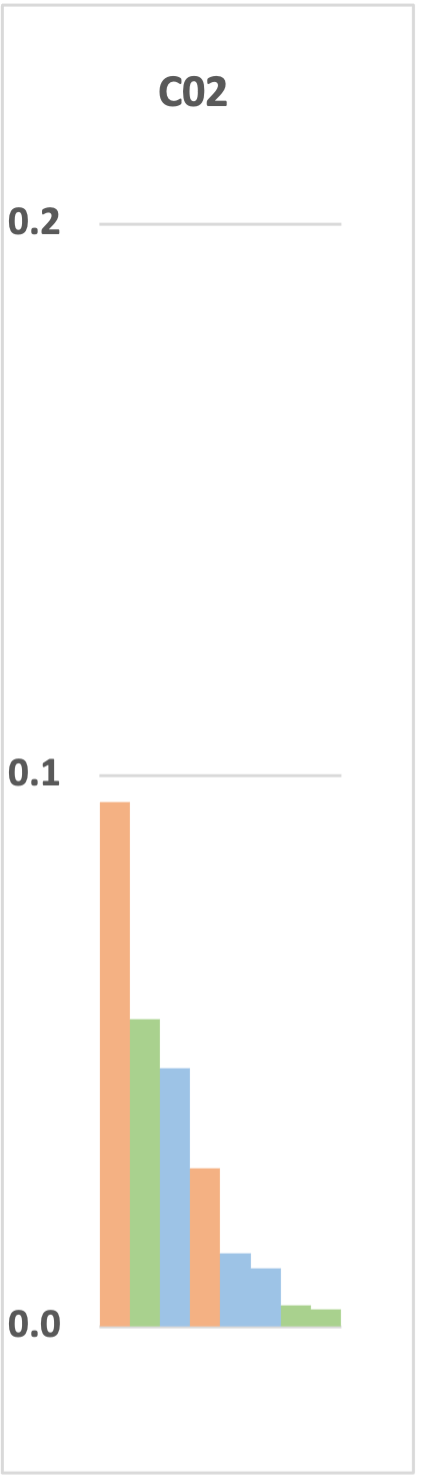}\hspace{0.5cm}\includegraphics[height=7cm]{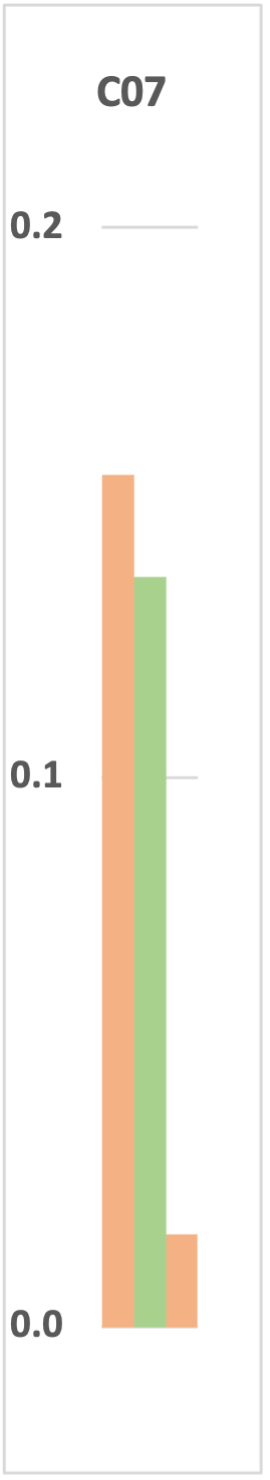}\hfill{}

\caption{Frequencies of major \emph{HLA-C} allele distributed into their respective
supertypes (as defined in \citep{Shen2023-bk}). Blue represents core
alleles as identified in \citep{Robinson2017-ze} of potentially Neanderthal/Denisovan
origin, orange represents remaining core alleles, and green represents
alleles of more recent origin that are related to the core alleles. }
\end{figure}

The current position on the temporal origin of these alleles appears
currently to be the following. The Neanderthal split from the common
ancestor of modern humans occurred about 530 kya and from Denisovans
about 400 kya. Modern humans arrived in Europe in an out-of-Africa
migration, passing through a bottleneck that is thought to have eliminated
Neanderthal/Denisovan ancestry as it migrated. The Neanderthal DNA
that exists in modern Caucasian populations is considered to have
arisen in a subsequent period of reproductive contact that occurred
about 45-50 kya. Neanderthals went extinct about 40 kya, preventing
more recent introgression. The level of Neanderthal introgression
from 50 kya is thought to have been \ensuremath{\approx}10\% of the
Caucasian genome, abating to its current level of an average of \ensuremath{\approx}2\%
through conventional purifying selection. However, modern Caucasians
have a frequency of Neanderthal/Denisovan HLA alleles that is higher
than 10\%. These frequencies are thought to represent positive selection
for alleles that are better adapted to local pathogens. There appears
to have been sustained survival of alleles of Neanderthal/Denisovan
allele candidates at A and B loci.

An important modification to this view is that positive selection
acts on haplotypes. High-frequency haplotypes acquire fitness through
positive epistatic effects between alleles. Allele frequencies are
high where epistatic effects across multiple haplotypes leads to frequency
aggregation at the allelic level. Alleles are themselves regulated
by the same negative-dependent disease transmission factors that also
regulate haplotype frequencies. Adaptation to local pathogens is likely
to be reflected in nuanced shifts in haplotype frequencies that are
then reflected in changes in allele frequencies.

The existence of Class I supertype silos is strong evidence of niche
apportionment through limiting similarity, an expected outcome from
the model we have proposed. The existence of a major allele in many
HLA-A and -B silos is consistent with competitive exclusion within
the boundaries of those silos. It may also be that the classification
into supertypes requires further refinement, as multiple alleles within
a single allelic silo may represent different epistatic properties.

There is also longstanding evidence of even longer survival of HLA
class II alleles, as evidenced by\emph{ trans}-species polymorphisms.
The most recent separation of humans from apes was that with chimpanzees
between 7 and 9 million years ago \citep{langergraber2012}, or about
250,000-300,000 generations ago.

One obvious explanation for the longevity of major class I and II
alleles is that they represent those with high long-term fitness.
This fitness is generated by their ability to enter numerous
positive epistatic relationships with alleles from other HLA loci.
Although it is difficult to date core alleles that are not of Neanderthal/Denisovan
origin, there is no reason why they should not be a legacy of the
out-of-Africa migration population that met the incumbent Neanderthal
population in Western Europe some 45,000 years ago. However, this remains to
be demonstrated. Second, it might be argued that supertype silos,
such as A03, B40, B44, C01, C02, and C03, where more recent alleles
are present at considerable frequencies are evidence of changes in
frequency that are under way. That possibility has to be set against
two other factors: (i) the survival of ancient alleles in the face
of ongoing attempts by pathogens to avoid HLA peptide presentation,
particularly by viruses, and (ii) the evidence that allele frequencies
can be used to track human migrations, suggesting that a degree of
frequency stability is possible even in the face of disease. If frequencies
generally are stable, then wild fluctuations that fully dislodge alleles
would seem less plausible. Indeed, the only major selective sweep
leading to elimination of a class I supertype for which there is good
evidence is the loss of A*02 in west-African chimpanzees \citep{deGroot2002}.
This has been attributed to the emergence of a devastating Simian
Immunodeficiency Virus (SIV).

\section{Discussion}

\subsection{HLA haplotypes form a large inter-connected network }

The extent of the polymorphism associated with the HLA transplantation
loci has emerged progressively over the past 8 decades or so. It is
now recognised as very extensive. Two of the principal explanations
for the polymorphism are also longstanding: negative frequency-dependent
selection (NFDS) and heterozygote advantage (HA). They focus on alleles,
not haplotypes, and they assume open-ended population growth as envisaged
in Fisher's Fundamental Theorem of Natural Selection. Fisher's association
of fitness with the rate constant for exponential population growth
is an in-built assumption of these models, with non-existent consideration
of Fisher's Malthusian parameter and its attenuation to zero with
increases in population density. Both NFDS and HA mechanisms require
similar fitnesses to sustain multiple alleles or heterozygotes and
to avoid competitive exclusion. This is at odds with evidence of widespread
differences in frequency that are at least quasi-stable, itself supported
by evidence of substantial linkage disequilibrium in major HLA haplotypes.

We propose a substantial departure from these assumptions. At the
evidential level, there is support for the view that it is HLA haplotypes
that carry a major part of the selective burden, not the allele. In
addition, the problem of achieving zero effective fitness of haplotypes
is attributed directly to the density-dependence of pathogen transmission.
There are sufficient human pathogens to allow approximately 1500 agencies
to determine the number of HLA haplotypes. Collectively, the haplotype
network drives selection of the 137 HLA alleles whose pleiotropy drives
the epistatic effects that underpin the network. The orderly behaviour
of both haplotypic and allelic frequencies suggests the network itself
is under selective pressure. In this respect, it would have some of
the properties of a quasi-species.

This model has a number of implications. First, there are the evolutionary
ones. The focus in heterozygote advantage was a beneficial expansion
of the range of peptide recognition in diploid individuals through
possession of two functionally-distinct alleles in \emph{trans} at
each locus, a direct consequence of diploidicity. This pairing is
random, and associations are broken up in meiosis. Haplotypes, by
contrast, survive numerous rounds of segregation and meiosis, and
demonstrate substantial positive epistasis and linkage disequilibrium.
Complementation is undoubtedly important for overall HLA function,
but it manifests itself in \emph{cis}, not \emph{trans}. This looks
to be an evolutionary adaptation.

If so, we should look at the way that peptide binding operates across
the haplotype as a whole. This binding can be represented as $pHLA_{haplo}$
rather than $pHLA_{allele}$. As far as we know, HLA molecules are
not coupled physically on the peptide-presenting cell, but co-presentation
on cell surfaces potentially provides a form of of coincidence detection
in the target T-cell response for the 12 (=2x6) distinct HLA alleles
used by a diploid host. In pursuing this goal, we should keep a watchful
eye on the downstream consequences of presenting numerous peptides
to individual T-cells from 12 different HLA molecules. There is, for
example, an interesting case in which there is linkage between two
alleles in the DR region, the DR2 haplotype, where one allele modifies
the response of the other through activation-induced cell death \citep{Gregersen2006}.
The allelic pairing offers partial protection against multiple-sclerosis-like
disease. Given the evidence that numerous diseases are associated
with the broader HLA region of chromosome 6, distinctive interplay
between alleles of the HLA transplantation loci can be expected on
a significant scale.

Focus on the behaviour of major alleles and haplotypes does not mean
that rare HLA haplotypes are without utility. A useful account of
their potential can be found in \citep{Klitz2012}. The evidence that
most of the alleles by census in rare haplotypes are derived from
the 137 identified earlier (17\% out of 20\%) indicates that rare
haplotypes will be functional, even if there are too few to demonstrate
epistatic effects. In this respect, rare but functional haplotypes
will continue to complicate the challenges presented to pathogens
by individual hosts. This will be particularly important during reproduction
since it produces distinctively different HLA genotypes within members
of the same family. These have historically lived in close proximity
in small tribal communities. Diploidicity and sexual reproduction
would provide an acute protective benefit at the HLA locus, coupled
with additional effects through independent assortment of other polymorphic
sites involved in in the adaptive immune response, such as the KIR
locus on chromosome 19.

The model we propose conforms in broad outline to that predicted by
Margalef \citep{margalef1968} and May \citep{May1972} for stable
communities. In the case here, the allelic interactions in HLA haplotypes
are strong and the number of alleles involved is small (large $\alpha$,
small C), whereas interactions between haplotypes are weak and haplotypes
are numerous (large C, small $\alpha$).

\subsection{Stable networks, Red Queens, and Enigma machines}

The previous section cited some of the reasons for network stability.
The major alleles show evidence of longevity: putative evidence of
Neanderthal/Denisovan origins for class I, and trans-species polymorphism
(TSP) for class II. Additionally, the evidence of strong linkage disequilibrium
implicitly requires stability for the positive selection to manifest
itself. Selective positive epistasis generates a recombinational load
from breaking up successful haplotypes. The physical compactness of
frozen haplotypes limits the sites for recombination and recombinational
rates are lower in many cases than that for the genome as a whole
\Citep{Carrington1999,Dawkins1999}, although there is considerable
variation between haplotypes \citep{Ahmad2003}. The emergence of
positive epistasis and sharp increases in linkage disequilibrium are
contingent, on the NS model \citep{Neher2009-ou}, on recombination
rates that fall below a critical value. 

It greatly assists the theoretical analysis if the network is treated
as stable over the short term, since frequencies become measures of
fitness (section 2.4 of this paper). Many of the graphs become readily
interpretable on that basis.

However, the existence of haplotypes with medium-to-long-term stability
appears to sit uncomfortably with the rapid reproduction rates of
many pathogens and their ability to sustain high mutational loads.
Rapid mutation of pathogens provides an opportunity for them to probe
for weaknesses in slowly-adapting host defences, setting up an apparently
one-sided chase between rapidly-reproducing predator and slowly-reproducing
prey. This scenario differs markedly from conventional predator-prey
relationships, characterised by the Lotka-Volterra predator-prey equations,
where the predator reproduces more slowly than the prey.

The successful resistance to pathogens offered by human hosts through
the adaptive immune system has two components: presentation of pathogen-derived
peptides by HLA molecules followed by rapid amplification of the presentation
response through T-cell activation.

The parallel between the reproductive properties of pathogens and
that of the amplified response in humans can readily be seen in antibody
production, where responsive antibody-producing cells reproduce and
diverge rapidly through clonal expansion and hypersomatic mutation
of antibody candidates. High-affinity antibodies are selected by affinity
maturation. Pathogens such as \emph{Plasmodium falciparum} (malaria)
and \emph{Trypanosoma brucei }(trypanosomiasis), for example, provide
text-book examples of bait-and-switch played on the human antibody
response that are pure Red Queen. They show that human hosts do have
defensive responses that match rapidly-reproducing pathogens in speed.

By contrast, the challenge faced by the host presentation step is
that it remains unchanged for the lifetime of the human host, with
a generation interval that is 3-4 orders of magnitude slower than
many pathogens. That stability allows removal of self-reactive T-cells
during fœtal development, with the survivors sensitive to exogenous
peptide. But it makes pathogen avoidance of peptide presentation an
eminently achievable goal; or so it appears. The evolutionary response
that frustrates pathogen breakthroughs through this static defence
is to balkanise host presentation of pathogen-peptide by generating
a large network of presenting haplotypes.

Conceptually, it follows closely the use of Enigma machines in WWII.
The starting rotor positions for the three slots in the machine were
set each morning according to a cypher manual printed month's in advance.
The rotors carried 26 'letters' and the ensuing encryption was considered
to be unbreakable. Alan Turing realised that the encyphered messages
could be broken if all possible daily settings could be scrutinised
quickly. The search would be successful if a setting was discovered
that could produce readable German text. Turing developed an electro-mechanical
device (improved by Welchman) that greatly reduced the time needed
to break a daily setting.

The construction of an Enigma machine with slots for three rotors
to operate together has an obvious parallel in multi-locus HLA haplotypes,
where each rotor contains 26 letters that parallel HLA alleles. By
analogy, the electro-mechanical device was the pathogen, canvassing
rotor permutations in the hope that one would open up the rotor settings
for the day. Since distribution of Enigma messages was effectively
clonal, successful decoding resulted in a sharp increase in decoded
information.

Enigma security was progressively improved during the course of WWII
through introduction of a fourth slot and additional rotors. This
is paralleled by increasing haplotype size. The size of the information
gain for each breakthrough was reduced by partitioning of daily settings
into different groups. This reduction in decryption success is directly
paralleled by polymorphic HLA genotypes.

The design of the Enigma machines, and the history of their cracking
has clear messages for a biological peptide-presentation device such
as HLA haplotypes that pathogens wish to avoid. One strategy is to
increase the range of problems that pathogens face when they are degraded
and presented. A coupled set of HLA loci in close physical proximity,
as in a haplotype, allows for selection on allelic complementarity
between different alleles, making simultaneous avoidance challenging
for pathogens. The challenge of avoidance is increased by making HLA
molecules act as broad-spectrum, low affinity binders of peptides.
Thus the dissociation constants of HLA molecules are typically in
the $\mu$M-nM range. By contrast, antibodies are high-affinity, narrow-spectrum
binders with dissociation constants typically in the nM-pM range.
They are much easier to avoid using bait-and-switch. Finally, doubling
the number of HLA species in every host through diploidicity, with
each peptide-presenting cell carrying 12 different HLA alleles.

A second strategy is to limit the damage caused by pathogens by ensuring
that hosts they might infect are different from ones that have previously
shown susceptibility, an example of the benefits of genetic rarity,
both in parallel and in series.

There are also important differences between pathogens and Enigma
coding. For example, the Enigma system has no obvious benefit in selection
for some rotor settings over others, or positive epistasis.

Haldane canvassed the view that ``\emph{it is an advantage
to the individual to possess a rare biochemical phenotype. For just
because of its rarity it will be resistant to diseases which attack
the majority of its fellows.}'' Later in the same paper, he expressed
the view that: ``\emph{We have here, then, a mechanism which favours
polymorphism, because it gives selective value to a genotype so long
as it is rare. Such mechanisms are not very common.}'' \citep{Haldane1949}

Haldane was surely right. The limitation imposed by proponents of
NFDS was to assume that rarity has to rest in HLA alleles, whereas
it has to rest more widely in the defence genotype taken in the round.
The HLA haplotype is an interim stage. The proponents of HA failed
to foresee that complemention of HLA allelic function could be brought
under tighter genetic control if it occurred within the context of
haplotypes. Heterozygotes may indeed enjoy advantages because they
make life more difficult for pathogens but it is not the whole story.

\subsection{Balancing selection, Fisher's geometric model, and quasi-species}

Polymorphism of HLA loci is widely seen as an iconic example of balancing
selection \citep{Hedrick2019}. If we are to accept that view, we
had better understand what balancing selection means.

The first person to use the term \emph{balancing selection} is unclear.
Muller used the term\emph{ balancing lethal} in a 1917 paper to describe
the outcome of a breeding experiment in \emph{Drosophila melanogaster}
\citep{Muller1918}. Ford used the term \emph{balanced polymorphism}
in 1945 to describe the industrial polymorphism displayed in England
by the peppered moth, \emph{Biston betularia} \citep{Ford1945}. He
distinguished between \emph{transient} and \emph{balanced}, emphasising
a high degree of permanence in the ratios of the respective morphs
in balanced polymorphism.

We need to recognise two distinct types of occurrences. One is where
the two or morphs occupy the same space homogeneously. That is true
for human females and males, and the sex ratio is a balanced polymorphism.
It is not quite 50:50 for reasons explained by Fisher \citep{Fisher1930}.
The other type of occurrence is where two morphs occupy two distinct
regions of a heterogeneous space. This polymorphism is the one shown
by industrial melanism \citep{Saccheri2008-ft}. Wild-type peppered
moths occupied unpolluted regions to the west of Liverpool and Manchester.
The \emph{carbonaria} mutant, which is black and possibly more resistant
to air pollution, was prevalent in the two cities in the 19th century,
whose trees were covered in soot. The regulatory agent on both moth
populations was birds spotting resting moths on trees whose bark coloration
varied with the level of pollution. Establishing the density of moth
populations is likely to have been technically challenging, and observations
relied on catching nocturnal moth samples with light traps and measuring
frequency ratios. These inevitably add up to unity, running the risk
of giving a spurious semblance of balance. On Levin's principle \citep{Levin1970},
the two states, unpolluted and black, should be able to sustain two
morphs and, to a first approximation, that is what happens - the two
moth strains distribute according to their losses by avian predation.

There is a transitional zone to the west of Liverpool where unpolluted
trees give way to soot-covered trees. This zone is only 40 km at most
on the sampling isocline that runs from west Wales out to the North
Sea in the east \citep{Saccheri2008-ft}. The prevailing wind is from
the south-west, possibly compressing the zone. It is entirely possible
that the zone mid-point is almost entirely depleted of both morphs
because both are highly visible on partly-polluted trees. This would
potentially establish a third niche for moths that were genetically
part way between wild-type and black. It is no surprise, therefore,
that three genetic intermediates exist at low frequencies, but there
is no useable data on their distribution.

Transferring these ideas to HLA polymorphism, it is clear that some
HLA polymorphism might have arisen through spatial differences in
the pathogen environment. However, this is a special case. A more
general explanation requires an explanation for HLA polymorphism where
the environment is complex and the complexity is distributed homogeneously.

A recent summary of the current position of HA and NFDS is contained
in the introduction to \citep{Slatkin2022}:
\begin{quote}
``Two types of balancing selection have been commonly invoked, heterozygote
advantage, and rare allele advantage (also called negative frequency-dependent
selection). Both heterozygote advantage and rare allele advantage
are assumed to result from the role class I and II loci play in the
immune system. In general, alleles differ in their peptide-binding
regions and hence can present different antigens for inspection by
T cells and also regulate natural killer cell activity. Consequently,
an individual that can present more kinds of antigens efficiently
because it is more heterozygous will likely be able to mount an effective
immune response to a larger variety of pathogens. The difference between
the 2 hypotheses is, from a population genetics perspective, not very
important. In the heterozygote advantage model, heterozygosity at
each locus is itself favored by selection because it provides defense
against a pathogen pool that is regarded as unchanging. In the rare
allele advantage model, an allele in low frequency is favored because
pathogens are not yet well adapted to it. As an allele increases in
frequency, pathogens adapt and the allele\textquoteright s contribution
to fitness decreases. In the rare allele advantage model, HLA loci
and pathogens coevolve in a way that results in higher average fitness
of heterozygous individuals. In the heterozygote advantage model,
coevolution plays no role. The reason the difference between the models
is not important for population genetic analysis is that Takahata
and Nei (1990) showed the equations that govern the change in allele
frequency to be the same in the 2 models when the allelic fitness
in the rare allele advantage model is a linearly decreasing function
of frequency. Spurgin and Richardson (2010) provided further support
for Takahata and Nei\textquoteright s conclusion.''
\end{quote}
The references in this quotation are to Takahata and Nei \citep{Takahata1990},
and Spurgin and Richardson \citep{Spurgin2010}.

This quotation from 2022 shows how the debate around the causes of
balancing selection are still focussed on the allele, not the haplotype.
There is other material in the quotation that is contentious: for
example, the proposed timescales for the relationship between host
allele frequency and the temporal development of susceptibility to
pathogens. Some of the theoretical conceptualisation could, in principle,
be transferred to the haplotype. However, the \emph{intra}-locus interactions
that generate positive epistasis in haplotypes largely erode a theory
of independent \emph{inter}-allelic HA, since the other allelic partners
in \emph{cis} determine the value of a particular allelic contribution
to fitness. Moreover, heterozygote advantage at allelic level is unlikely
to be a major driver of HLA polymorphism unless all alleles have near-identical
fitnesses \citep{Lewontin1978,DeBoer2004,Stefan2019-wu}, a requirement
that appears to be at odds with the evidence.

Rare allele advantage can also be criticized for a number of reasons.
Although it is capable of generating large numbers of alleles on Takahata and Nei's analytic theory, their turnover is rapid
and their persistence times are insufficient to account for trans-species
polymorphism (TSP). Where parameter values are sufficient for TSP,
allele frequencies have a fairly even distribution. Again, an expectation
that is not in line with the evidence, either for allele or haplotype
distributions.

The principal objections to HA, NFDS, and divergent allele advantage
(DAA) as explanations of HLA polymorphism are that that they are purely
genetic explanations and they focus on alleles. They ignore the epidemiology
of infection, which has a strong density dependence for transmission,
including an all-important density-dependent threshold. Combining
genetics and epidemiology \citep{May1983-fk}, allows one to approach
HLA polymorphism from an alternative direction:
\begin{itemize}
\item introducing density as the source of raw data, rather than frequency,
avoids the risk that frequencies will be seen implicitly as coupled.
\item densities allow the introduction of rate equations based on principles
of mass action (or parallels), providing access to a range of demographic
models.
\item host densities can be governed by disease transmission, thereby coupling
capping of HLA haplotype population densities to their susceptibilities
in presenting pathogen-derived peptide.
\item capping of HLA haplotypes under selection automatically reduces their
effective fitness to zero, creating the potential for stable capped
populations at steady-state.
\item the models include the combination of first-order/exponential population
expansion with density-dependent mortality terms, as originally proposed
by Verhulst. These can cap or regulate populations in a density-dependent
manner, leading to use of logistic equations, the Lotka-Volterra competition
equations, much of the mathematical theory behind the stability or
otherwise of complex communities, etc. 
\item Lotka-Volterra competition equations provide access to a very large
body of theoretical work associated with complex communities of Linnean
species, through the simple expedient of treating genetic polymorphism
as a problem of molecular species such as haplotypes within a single
Linnean species.
\item the actions of Natural Selection then emerge as distinctive HLA haplotypes
that appear as capped silos through the action of density-limited
growth, competitive exclusion, and limiting similarity.
\end{itemize}
On the basis of the last bullet point, balancing selection could be
said to be an artefact of a focus on frequency since the density of
one pathogen-limited silo need have no connection to the size of others.
They would be established independently by the relationship between
a haplotype and its carrier's susceptibility to disease.

However, the analysis in this paper supports the view that balancing
selection is real and the direct result of selection and partitioning
on haplotype fitnesses as represented by their frequencies. The 350-1850
HLA haplotypes under selection form an inter-related network by permutation
of the 137 alleles that are themselves capable of demonstrating positive
epistasis. A feature of the network is the orderliness of the rank
distribution frequencies. There is an obvious mechanism by which this
might occur: through fine tuning both of antigen-binding sites and
their epistatic effects in haplotypes. The network or cloud then becomes
self-regulating, and balancing selection is reinstated. One could
then argue that the network is optimised to resist pathogens, with
a function of sexual reproduction being to generate the panmixis needed
for network equilibration.

There is a conceptual overlap between an HLA network of this type
and features of both Fisher's geometric model and a quasi-species
of the type proposed by Eigen, McCaskill and Schuster (the EMS) model
\citep{Eigen1988}.

Fisher's geometric model has its origin in a brief mention in \citep{Fisher1930},
pp. 38-41. An organism is characterised by a set of independent phenotypic
traits. All phenotypes are posited as being under stabilising selection,
with each having an optimal value. This creates the potential for
an optimum combination of phenotypic values.

Applied to an HLA network, HLA alleles can be regarded as mutants
with fitness optimisation within their respective silos. Haplotypes
are independent traits that combine to produce an optimum HLA network.

The EMS model, by contrast, focuses on the molecular or genetic inter-connectedness
of the components of the cloud or network, whose components differ
one from another by mutation. There are no independent traits. Both
Fisher's geometric model and the EMS model share a context of open-ended
population expansion, disguised to some extent by use of frequencies.
However, the EMS model explicitly acknowledges Fisher's fitness as
a first-order rate constant, and deals with the exponential increase
in density of the reactants by constant dilution and injection of
fresh reagents to create a steady-state. An inherent property of the
EMS is that natural selection acts rapidly to eliminate
variance. Variance is maintained by high mutation rates.

Our model obviates the need for the high mutation rates of the EMS
model by introducing negative density-dependent selection (NDDS) of
hosts imposed by a complex, reflexive, and antagonistic environment.
Obviating the need for rapid mutation allows room for a quasi-species
model with much slower rates of mutation and recombination.

An important feature of quasi-species is their connectedness. Connectedness
for DNA sequences is demonstrable at the nucleotide level by noting
the number of differences from a reference sequence; this can be calculated
as a Hamming distance. Strictly, there is no equivalent reference
sequence for an HLA quasi-species, because the major haplotypes are
independently stabilised into distinct silos. We can, however, infer
a degree of connectedness for alleles by noting distributions, as
illustrated in Fig.8. The top two panels show the distribution of
allele appearances in rank order to 37,000. We have argued that the
350-1850 haplotypes of lowest rank have the rank and frequency they
do because of positive selection on haplotypes with epistatic interactions
between alleles and extensive pleiotropy. These haplotypes then feed
all lower haplotype frequencies, mainly by recombination but also
by mutation. The presence of a single allele across hundreds of separate
haplotypes indicates a high degree of connectedness or pleiotropy.
By contrast, the third panel shows the distribution for an allele,
A{*}01:03g, that barely shows any epistatic contribution. It fails
to be represented in any haplotype of high frequency, and populates
weakly the lower frequency haplotype spectrum. The conclusion is that
connectedness and pleiotropy are both low in low-frequency alleles.
This is consistent with the conclusions summarised in 4.1 above.

In summary, there are three levels of positive epistatic selection
acting on the HLA transplantation loci.
\begin{itemize}
\item at the level of the polypeptide chain. All proteins that have biological
activity represent the effects of positive epistasis within their
gene product, since inter-amino-acid segregation would fundamentally
erode genetic transmission of successful mutants.
\item at the level of the allele. Positive epistasis is manifested in some
haplotypes by some combinations of the high-frequency allelic category
but not others. High allelic frequencies arise by selection on haplotypes.
However, such is the extent of their pleiotropy that high-frequency
alleles still have frequencies that fall well short of frequencies
needed to produce high-frequency haplotypes without epistasis and
linkage disequilibrium.
\item at the level of the haplotype. A relatively small number of alleles
generate a larger catalogue haplotypes. These haplotypes form a balanced
network that is optimised by shunting selection down onto alleles.
Selection is further shunted to the polypeptide sequences in the ABS.
\end{itemize}

\subsection{The pathogen environment}

It is axiomatic that the pathogen environment exercises selection
on human hosts. This generates HLA polymorphism as a signature; that
is, balkanisation of host genotypes. Balkanisation of hosts takes
individual genotypes below a transmission threshold, and this offers
an important epidemiological protection.

HLA polymorphism must equally exert selection on pathogens. In our
view, its principal effect is a reciprocal balkanisation of the pathogen
environment. Balkanisation reduces the fitness gains that can be made
by pathogens when they exploit genetic gaps in host defences. The
number of distinct HLA genotypes under selection (350-1850) gives
an approximate indication of the size of the pathogen environment
that is 'seen' by the host since, superficially, a host genotype that
is stably under selection requires a distinct selective agent. The
pathogen landscape shows complex epidemiological behaviours \citep{Gupta2024}
including development of strains. Collectively, these agencies may
greatly exceed the number of haplotypes under selection. On the other
hand, it may be that not all human pathogens are present simultaneously,
and that strains represent a numerical amplification to fill an available
niche.

Indeed, it is possible to account, in general terms, for much of pathogen
behaviour as being constrained by a finite density of hosts (even
if these include animals), within which the principles of competitive
exclusion and limiting displacement occur. It may also be that hosts
and pathogens show evidence of character displacement.

\subsection{Future directions}

There are a number of ways in which the ideas in this paper can be
explored further. Here are obvious ones:
\begin{itemize}
\item HLA allelic combinations and positive epistasis. A conclusion of this
paper is that the HLA haplotype distribution reflections selection
on positive epistasis within certain HLA allelic combinations. The
precise origin of the positive epistasis is currently unknown. It
may simply reflect greater breadth of pathogen-peptide binding. However,
we should note cautionary evidence, of which \citep{Rao2013} is an
example. Other explanations include differences in stimulating T-cells.
Alphafold appears capable of making major contributions \citep{McMaster2024}.
\item Functional overlaps of current haplotype networks. The NMDP databases
include HLA frequency data on a range of self-declared ethnicities
in addition to the Caucasian set. Preliminary inspection indicates
that the features identified in the Caucasian set - orderly rank order
distributions for alleles and haplotypes, evidence of positive epistasis,
etc. - hold true for other ethnicities. It should be possible rapidly
to establish their network characteristics and the extent to which
the HLA datasets for other ethnicities differ functionally. It is
theoretically possible that the HLA haplotype network was established
only once, and has co-evolved with the pathogen environment. Features
that would support long-term evolution of a single network include
the stabilisation offered by the pleiotropic properties of key alleles.
These potentially stabilise the trans-species polymorphism of class
II alleles, and the putative survival of Neanderthal/Denisovan class
I alleles.
\item Frequency stability of HLA networks. A key component of the current
proposal is the short-term frequency stability of the HLA haplotype
network. Data from historic population movements such as that into
Western Europe some 25,000 years ago may be able to clarify whether this
is the case. Nearer to the present day are population shocks such
as that caused by plague in Western Europe, notably the Black Death,
which would have profoundly damaged the prevailing HLA network of
the day. There is widespread distribution of plague pits in Western
Europe that can be dated. There are, of course, technical difficulties
in sequencing the HLA region, but we only really need the frequencies
of the binding sites of the transplantation alleles.
\item Discovery of infection matrices. Blanket pathogen sequencing, of the
kind described in \citep{Ozer2021}, should help define a library
of pathogen peptides that could be presented. It should be possible
to couple this with peptide-recognition properties of HLA haplotypes
to generate infection matrices that associate particular haplotypes
with particular pathogen genotypes \citep{Markle2024}.
\end{itemize}
\bibliographystyle{unsrtnat}
\bibliography{references}

\end{document}

\typeout{get arXiv to do 4 passes: Label(s) may have changed. Rerun}